\documentclass[twocolumn]{aastex631}
\usepackage{amsmath}
\usepackage{comment}
\usepackage{natbib}
\usepackage{booktabs, multirow}
\usepackage{changepage,threeparttable}
\usepackage{appendix}
\usepackage{graphicx}
\usepackage{hyperref}
\usepackage{float}
\usepackage{longtable}
\usepackage{times}
\usepackage{mathptmx}

\shortauthors{Moharana et al.}
\shorttitle{Filaments and Dense Cores in Aquila}

\begin{document}

\title{TRAO Survey of the Nearby Filamentary Molecular Clouds, the Universal Nursery of Stars (TRAO-FUNS). IV. Filaments and Dense Cores in the W40 and Serpens South Regions of Aquila}

\author[0009-0007-9411-0284]{Satyajeet Moharana}
\affiliation{Korea Astronomy and Space Science Institute, 776 Daedeokdae-ro, Yuseong-gu, Daejeon 34055, Republic of Korea; \href{mailto:satyajeet@kasi.re.kr}{satyajeet@kasi.re.kr}, \href{mailto:cwl@kasi.re.kr}{cwl@kasi.re.kr}}
\affiliation{University of Science and Technology, 217 Gajeong-ro, Yuseong-gu, Daejeon 34113, Republic of Korea}
\author[0000-0002-3179-6334]{Chang Won Lee}
\affiliation{Korea Astronomy and Space Science Institute, 776 Daedeokdae-ro, Yuseong-gu, Daejeon 34055, Republic of Korea; \href{mailto:satyajeet@kasi.re.kr}{satyajeet@kasi.re.kr}, \href{mailto:cwl@kasi.re.kr}{cwl@kasi.re.kr}}
\affiliation{University of Science and Technology, 217 Gajeong-ro, Yuseong-gu, Daejeon 34113, Republic of Korea}
\author[0000-0001-9333-5608]{Shinyoung Kim}
\affiliation{Korea Astronomy and Space Science Institute, 776 Daedeokdae-ro, Yuseong-gu, Daejeon 34055, Republic of Korea; \href{mailto:satyajeet@kasi.re.kr}{satyajeet@kasi.re.kr}, \href{mailto:cwl@kasi.re.kr}{cwl@kasi.re.kr}}
\author[0000-0003-0014-1527]{Eun Jung Chung}
\affiliation{Korea Astronomy and Space Science Institute, 776 Daedeokdae-ro, Yuseong-gu, Daejeon 34055, Republic of Korea; \href{mailto:satyajeet@kasi.re.kr}{satyajeet@kasi.re.kr}, \href{mailto:cwl@kasi.re.kr}{cwl@kasi.re.kr}}
\author[0000-0002-7497-2713]{Spandan Choudhury}
\affiliation{Korea Astronomy and Space Science Institute, 776 Daedeokdae-ro, Yuseong-gu, Daejeon 34055, Republic of Korea; \href{mailto:satyajeet@kasi.re.kr}{satyajeet@kasi.re.kr}, \href{mailto:cwl@kasi.re.kr}{cwl@kasi.re.kr}}
\author[0000-0002-2569-1253]{Mario Tafalla}
\affiliation{Observatorio Astronomico Nacional (IGN), Alfonso XII, 3, 28014 Madrid, Spain}
\author[0000-0002-1229-0426]{Jongsoo Kim}
\affiliation{Korea Astronomy and Space Science Institute, 776 Daedeokdae-ro, Yuseong-gu, Daejeon 34055, Republic of Korea; \href{mailto:satyajeet@kasi.re.kr}{satyajeet@kasi.re.kr}, \href{mailto:cwl@kasi.re.kr}{cwl@kasi.re.kr}}
\author[0000-0002-6386-2906]{Archana Soam}
\affiliation{Indian Institute of Astrophysics, II Block, Koramangala, Bengaluru 560034, India}
\author[0009-0001-5526-4615]{Donghyeok Koh}
\affiliation{Korea Astronomy and Space Science Institute, 776 Daedeokdae-ro, Yuseong-gu, Daejeon 34055, Republic of Korea; \href{mailto:satyajeet@kasi.re.kr}{satyajeet@kasi.re.kr}, \href{mailto:cwl@kasi.re.kr}{cwl@kasi.re.kr}}
\affiliation{University of Science and Technology, 217 Gajeong-ro, Yuseong-gu, Daejeon 34113, Republic of Korea}
\author[0000-0002-8614-0025]{Shivani Gupta}
\affiliation{Indian Institute of Astrophysics, II Block, Koramangala, Bengaluru 560034, India}
\affiliation{Pondicherry University, R.V. Nagar, Kalapet, 605014, Puducherry, India}
\author[0009-0007-0745-9147]{Maheswar Gopinathan}
\affiliation{Indian Institute of Astrophysics, II Block, Koramangala, Bengaluru 560034, India}
\author[0000-0003-4022-4132]{Woojin Kwon}
\affiliation{Department of Earth Science Education, Seoul National University, 1 Gwanak-ro, Gwanak-gu, Seoul 08826, Republic of Korea}
\affiliation{SNU Astronomy Research Center, Seoul National University, 1 Gwanak-ro, Gwanak-gu, Seoul 08826, Republic of Korea}
\affiliation{The Center for Educational Research, Seoul National University, 1 Gwanak-ro, Gwanak-gu, Seoul 08826, Republic of Korea}

\begin{abstract}
We present the results of molecular line observations toward the W40 and Serpens South regions of the Aquila molecular cloud complex, conducted as part of the TRAO-FUNS project to investigate the role of filamentary structures in the formation of dense cores and stars in molecular clouds. We performed a Gaussian decomposition of the C$^{18}$O spectra to disentangle multiple velocity components along the line-of-sight and a `Friends-of-Friends' algorithm on these decomposed components to identify 24 velocity-coherent filaments in the observed region. The `FellWalker' algorithm is applied on the N$_{2}$H$^{+}$ integrated intensity map to identify the dense cores embedded within the filaments. Many of the filaments previously identified from the Herschel survey are found to contain multiple velocity-coherent filaments. Virial analysis indicated that all of our identified filaments are thermally supercritical and gravitationally bound. Velocity gradients are observed along the filaments in the vicinity of embedded dense cores, indicating the presence of longitudinal flows that contribute to core formation. The median mass flow rate across the observed region is estimated to be $\sim$35 M$_{\odot}$ Myr$^{-1}$, with Serpens South showing a rate $\sim$40\% higher than W40. The analysis of non-thermal motions revealed that the dense cores mainly show subsonic to transonic motions, while their host filaments are mostly supersonic, suggesting that the turbulent motions in filaments may dissipate on smaller scales, allowing core formation. These findings highlight the essential role of the filaments' criticality, mass flow, and turbulent dissipation in the formation of dense cores within the filaments.
\end{abstract}


\section{Introduction} \label{sec:intro}
Infrared and submillimeter observations have revealed that molecular clouds are predominantly composed of filamentary structures, which play a fundamental role in the star formation process by channeling gas into the embedded dense cores and protostars \citep{2010A&A...518L.102A,2010A&A...518L.100M,2011A&A...529L...6A,2013A&A...554A..55H,2018A&A...610A..77H}. These filaments are ubiquitous, not only in active star-forming regions but also in quiescent, non-star-forming environments such as the Polaris cloud. In actively star-forming regions, the majority of prestellar cores are observed to be located within these filamentary networks \citep{2010A&A...518L.102A}. The Herschel Space Observatory has significantly advanced our understanding of filamentary molecular clouds and the initial conditions of star formation, revealing their presence across molecular clouds of varying mass scales, from low- to high-mass star-forming regions. Its high sensitivity enabled the first uniform characterization of filament properties—such as mass per unit length, radial density profiles, and characteristic widths—across the Milky Way star-forming regions \citep{2011A&A...529L...6A,2019A&A...621A..42A,2013A&A...550A..38P,2015A&A...584A..91K}.

While these surveys have significantly advanced our understanding of the star formation process in molecular clouds, it is important to recognize that the majority of the data are based on continuum emission. Consequently, many studies of the physical properties attributed to filamentary structures \citep[e.g.,][]{2014prpl.conf...27A,2019A&A...621A..42A} may represent a superposition of multiple distinct components that are merely aligned along the line-of-sight. If this is the case, it becomes essential to re-evaluate the derived physical properties of continuum filaments by decomposing them into velocity-coherent structures using spectroscopic data.

The TRAO-FUNS (Taeduk Radio Astronomy Observatory survey of Filaments, the Universal Nursery of Stars) project is a dedicated survey designed to investigate the velocity structure, kinematic behavior, and chemical properties of low-density filaments and the dense cores embedded within them, with the aim of understanding their formation mechanisms. Since the observed three-dimensional position-position-velocity (PPV) data can trace the multiple line-of-sight components in velocity, it provides a much more comprehensive understanding of the velocity-coherent structures in a molecular cloud. As part of this series, previous studies targeting L1478 in the California Molecular Cloud \citep{2019ApJ...877..114C}, IC 5146 \citep{2021ApJ...919....3C}, and the NGC 2068 and NGC 2071 regions in Orion B \citep{2023ApJ...957...94Y} successfully identified velocity-coherent filaments and dense cores, and derived their physical properties. These studies revealed that in low-mass (L1478), low- to intermediate-mass (IC 5146), and high-mass (Orion B) star-forming regions, the majority of dense cores reside within thermally supercritical filaments. Furthermore, analyses of non-thermal velocity dispersions using C$^{18}$O (1--0) and N$_{2}$H$^{+}$ (1--0) transitions consistently showed that dense cores exhibit lower non-thermal velocity dispersions compared to their surrounding filaments, suggesting the dissipation of turbulence during the core formation process.

This work, the fourth contribution of the TRAO-FUNS project, focuses on the molecular cloud complex Aquila Rift, located above the galactic plane ($b\simeq3\fdg5$) at galactic longitude of $l \simeq 28\fdg8$ \citep{2020ApJ...895..137S}. W40, Serpens South and MWC297 are the major star formation sites of this region that hosts a total of 651 starless, 292 prestellar and 58 protostellar cores as identified in the Herschel observations \citep{2015A&A...584A..91K}. A number of velocity-coherent structures and molecular outflows have been reported in this region from mapping observations by the Nobeyama 45 m telescope \citep{2019PASJ...71S...3N}. In this study, we have specifically focused on the W40 and Serpens South regions.

W40 is a young star cluster associated with an H{\sc ii} region also known as Sh 64 \citep{1959ApJS....4..257S}. Several high-mass O- and B-type main sequence stars have been identified in the central region of W40, responsible for illuminating the H{\sc ii} region \citep{1985ApJ...291..571S,2012AJ....144..116S}. Hundreds of YSOs have also been identified within this region by several studies over the years \citep{2013ApJS..209...31P,2013ApJ...779..113M,2022MNRAS.516.5244S}.

The Serpens South region, first identified by \cite{2008ApJ...673L.151G}, hosts a dense protostellar cluster embedded within a prominent hub-filamentary system. Multiple studies have provided evidence for filamentary accretion flows feeding the central cluster, as traced by N$_{2}$H$^{+}$ (1--0) emission \citep[e.g.,][etc.]{2013ApJ...766..115K,2014ApJ...790L..19F}. \cite{2014ApJ...791L..23N} has argued that the central cluster formation is triggered by filament collision. As reported by \cite{2009ApJS..184...18G} and \cite{2011A&A...535A..77M}, the high count of protostars compared to Class II YSOs suggests that the region is in an early stage of development and is highly active. Hence, it provides an ideal environment to study the morphology and gas kinematics of the natal filamentary structures as they are less affected by stellar feedbacks at this point in their evolution.

The distance to Aquila has been a subject of debate over the years. Earlier studies adopted a relatively closer distance of 260 pc \citep{2008ApJ...673L.151G,2010A&A...518L.102A,2011A&A...535A..77M,2013ApJ...766..115K,2015A&A...584A..91K}, based on photometric measurements of Serpens Main. However, more recent studies have favored a greater distance of $\sim$436 pc \citep{2018A&A...615A...9P,2022MNRAS.516.5244S,2024ApJ...969...70F}, as determined by \cite{2017ApJ...834..143O,2018ApJ...869L..33O,2023A&A...673L...1O} through a series of analyses incorporating multi-epoch Very Long Baseline Array (VLBA) observations, $Gaia$ Data Release 2 measurements, and VLBA observations of 22 GHz water masers toward the protostar CARMA-6, situated at the center of the Serpens South young cluster. In this study, we have adopted a distance of $455 \pm 50$ pc, as derived by \cite{2022ApJ...938...55A} by using $Gaia$ Early Data
Release 3 data for the Serpens South and W40 regions.

\cite{2019PASJ...71S...4S} have argued that both W40 and Serpens South regions are physically connected with each other, and the outer expanding shell of the W40 H{\sc ii} region has induced star formation in the central protocluster of Serpens South by interacting with the dense gas associated with it \citep[see also][]{2017ApJ...837..154N}. Such a complex environment, comprising a high-mass star-forming region and a hub-filamentary system, serves as an excellent target for a systematic investigation of the kinematic properties of filaments and embedded dense cores to get insights into their formation mechanisms. In this study, we present the physical properties of the filamentary structures identified by the C$^{18}$O (1--0) line and dense cores defined by N$_{2}$H$^{+}$ (1--0) emissions in these two regions.

This paper is organized in the following way. The observational details are described in Section \ref{sec:obs}. In Section \ref{sec:results}, we discuss the identification methods for filaments and dense cores and summarize their physical properties. Section \ref{sec:discussions} focuses on the kinematical properties of filaments and dense cores. In the end, a brief summary of the entire study is presented in Section \ref{sec:summary}.

\section{Observations and Data Reduction} \label{sec:obs}
\subsection{Observations}
On-the-fly (OTF) mapping observations toward W40 and Serpens South regions of Aquila were carried out using the multibeam receiver SEcond QUabbin Optical Imaging Array (SEQUOIA) of the Taeduk Radio Astronomy Observatory (TRAO)\footnote{\href{https://trao.kasi.re.kr}{https://trao.kasi.re.kr}}, which is a 14 m single-dish radio telescope located in South Korea. The receiver SEQUOIA-TRAO is equipped with 16-pixel Monolithic Microwave Integrated Circuit preamplifiers in a 4$\times$4 array and provides an observing frequency range of 85--115 GHz. The backend system, a fast Fourier transform spectrometer, has 8192 channels with a spectral resolution of $\sim$15 kHz, which corresponds to a velocity resolution of $\sim$0.04 km s$^{-1}$ at 110 GHz and a full spectral bandwidth of $\sim$60 MHz. The narrow-band second IF modules enable the observation of two frequencies simultaneously within the 85--100 or 100--115 GHz bands. The main beam efficiencies of the telescope are 0.48 at 98 GHz, and 0.46 at 110 GHz, respectively \citep{2019JKAS...52..227J}. The observations were made from 2016 December to 2018 October as part of the TRAO Key Science Program, TRAO-FUNS.

\begin{figure}[!htb]
    \centering
    \includegraphics[width=0.47\textwidth]{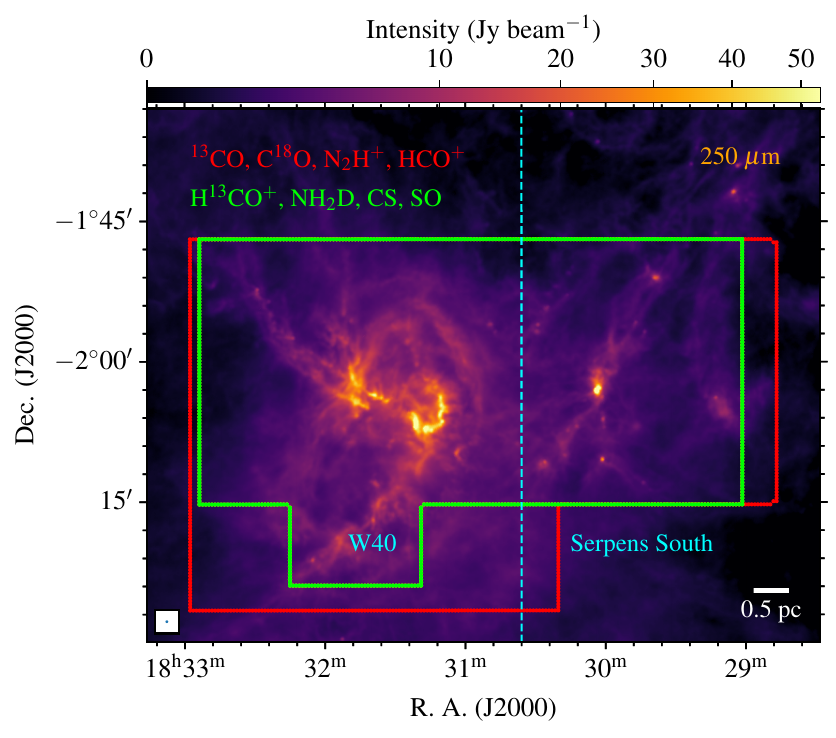}
    \caption{The survey areas of Aquila. The areas observed with various molecular lines are marked in red ($^{13}$CO, C$^{18}$O, N$_{2}$H$^{+}$ and HCO$^{+}$), and green (SO, CS, H$^{13}$CO$^{+}$, and NH$_{2}$D) boxes on the Herschel 250 $\mu$m continuum image. The cyan-colored vertical dashed line (RA = 18:30:37.2) roughly marks the boundary between the W40 region (eastern side) and the Serpens South region (western side). The blue dot at the bottom-left corner denotes the beam size of the Herschel 250 $\mu$m observations (18.2$^{\prime\prime}$). The scale bar at the lower-right corner represents a physical length of 0.5 pc, based on our adopted distance of 455 pc to Aquila.}
    \label{fig:boundary}
\end{figure}

To investigate the physical properties as well as chemical characteristics of dense cores and the surrounding filaments, we chose eight molecular lines. C$^{18}$O (1--0) and N$_{2}$H$^{+}$ (1--0) are selected to trace the relatively less dense filamentary structures and denser cores, respectively. $^{13}$CO (1--0) is observed simultaneously with C$^{18}$O (1--0) to reveal the large-scale bulk motion. Optically thick CS (2--1) and HCO$^{+}$ (1--0) along with optically thin H$^{13}$CO$^{+}$ (1--0) are observed to examine the evidence of infall motions in dense cores. SO (3$_{2}$--2$_{1}$) and NH$_{2}$D (1$_{11}$--1$_{01}$) are observed to investigate the chemical evolution of dense cores.  The observational details, along with the rest frequency, FWHM-beam size, observing area, pixel size, channel width, and rms noise level for each molecular line, are summarized in Table \ref{tab:obs}. Figure \ref{fig:boundary} shows the observed regions for each set of molecular lines over the Herschel 250 $\mu$m continuum image obtained from the Herschel Gould Belt Survey Archive\footnote{\href{http://gouldbelt-Herschel.cea.fr/archives}{http://gouldbelt-Herschel.cea.fr/archives}}.

\subsection{Data Reduction}
The raw data, obtained in the shape of tiles after OTF mapping, were re-gridded to a cell size of 20$^{\prime\prime}$ and converted into the CLASS format after a first-order baseline correction using the OTFTOOL \citep{2006JASS...23..269C}. Noise weighting was applied to obtain a better signal-to-noise ratio (SNR). After the pre-processing, the baselines of the spectra were removed by an iterative second-order polynomial fitting. The spectra were resampled with a channel width of 0.06 km s$^{-1}$. The datacube spans a velocity range of 60 km s$^{-1}$, with its central velocity set to 8 km s$^{-1}$, similar to the LSR velocity of Aquila. Then, all the tiles were combined to make one averaged scan map for each transition. The observation was continuously performed until the rms level of each average tile was $\lesssim$ 0.15 K $\left[T_{\rm A}^{*}\right]$ for C$^{18}$O, $^{13}$CO, and NH$_{2}$D, $\lesssim$ 0.1 K $\left[T_{\rm A}^{*}\right]$ for H$^{13}$CO$^{+}$, and $\lesssim$ 0.07 K $\left[T_{\rm A}^{*}\right]$ for the rest of the observed species at the 0.06 km s$^{-1}$ channel width.

\begin{deluxetable*}{l r@{.}l ccccc}
\tablecaption{Observational Details \label{tab:obs}}
\tablewidth{0pt}
\setlength{\tabcolsep}{12pt}
\tablehead{
\colhead{Molecule} &
\multicolumn{2}{c}{$\nu_{\rm ref}$} &
\colhead{$\theta_{\rm FWHM}$} &
\colhead{Area} &
\colhead{RMS\_I (Line)} &
\colhead{RMS\_II (Int. Intensity)} &
\colhead{$n_{\rm crit}$} \\
\colhead{} &
\multicolumn{2}{c}{(GHz)} &
\colhead{(arcsec)} &
\colhead{(arcmin$^{2}$)} &
\colhead{(K)} &
\colhead{(K km s$^{-1}$)} &
\colhead{(cm$^{-3}$)} \\
\colhead{(1)} &
\multicolumn{2}{c}{(2)} &
\colhead{(3)} &
\colhead{(4)} &
\colhead{(5)} &
\colhead{(6)} &
\colhead{(7)}
}
\startdata
$^{13}$CO (1--0) & 110 & 2013543 & 49 & 2255 & 0.144 & 0.121 & $1.9\times10^{3}$ \\
C$^{18}$O (1--0) & 109 & 7821734 & 49 & 2255 & 0.137 & 0.092 & $1.9\times10^{3}$ \\
N$_{2}$H$^{+}$ (1--0) & 93 & 1734020 & 52 & 2255 & 0.068 & 0.061 & $6.1\times10^{4}$ \\
HCO$^{+}$ (1--0) & 89 & 1885247 & 57 & 2255 & 0.068 & 0.065 & $6.8\times10^{4}$ \\
SO (3$_{2}$--2$_{1}$) & 99 & 2998700 & 52 & 1796 & 0.071 & 0.051 & $1.9\times10^{3}$ \\
CS (2--1) & 97 & 9809533 & 52 & 1796 & 0.070 & 0.055 & $1.3\times10^{5}$ \\
H$^{13}$CO$^{+}$ (1--0) & 86 & 7542884 & 57 & 1796 & 0.102 & 0.050 & $6.2\times10^{4}$ \\
NH$_{2}$D (1$_{11}$--1$_{01}$) & 85 & 9262630 & 57 & 1796 & 0.159 & 0.066 & $6.5\times10^{4}$ \\
\enddata
\tablecomments{
(1) The observed molecular lines. (2) Rest frequency of each molecular line \footnote{\href{https://splatalogue.online/\#/home}{https://splatalogue.online/\#/home}}. (3) FWHM of the telescope beam \citep{2019JKAS...52..227J}. (4) Total observed area. (5) Noise level in $T_{\rm A}^{*}$ of the final data cube. (6) Noise level of integrated intensity. (7) Critical density of the observed molecular lines at 10 K (60 K for SO). The references are: $^{13}$CO, C$^{18}$O, and SO: {\protect\citet{2021ApJ...919....3C}}; N$_{2}$H$^{+}$, HCO$^{+}$, H$^{13}$CO$^{+}$, and CS: {\protect\citet{2015PASP..127..299S}}; NH$_{2}$D: {\protect\citet{Wienen}}.}
\end{deluxetable*}

\section{Results} \label{sec:results}
\subsection{$^{\textit{13}}$CO and C$^{\textit{18}}$O Emissions}
\begin{figure*}[!htb]
    \centering
    \includegraphics[width=\textwidth]{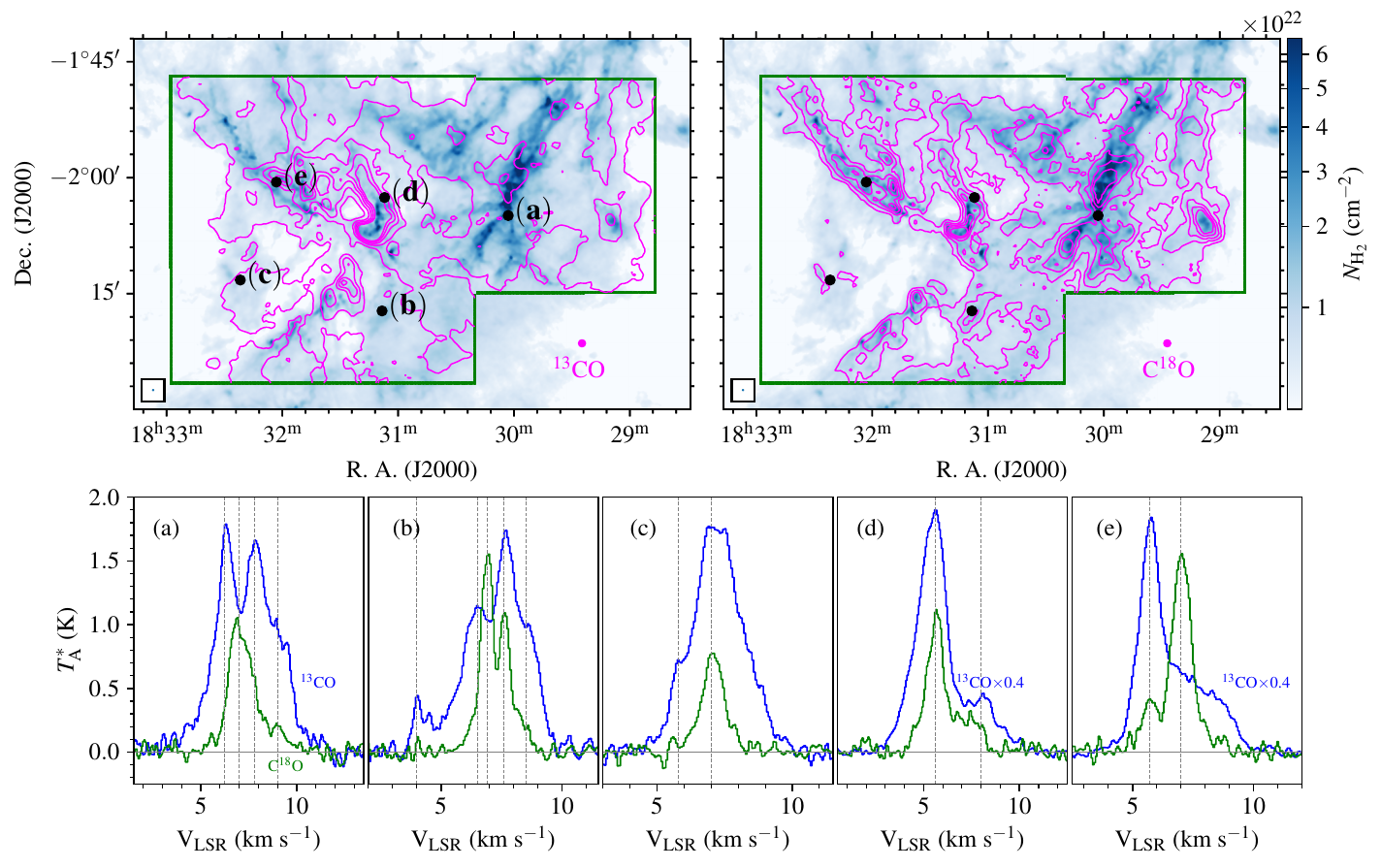}
    \caption{Integrated intensity maps of $^{13}$CO (1--0) (top left) and C$^{18}$O (1--0) (top right) with contours on the Herschel H$_{2}$ column density map and spectra of the selected positions (lower panel). The contour levels of are $\sigma \times \left[30, 45, 60, 75, 90, 105\right]$ K km s$^{-1}$, and $\sigma \times \left[10, 15, 20, 25, 30\right]$ K km s$^{-1}$ for $^{13}$CO and C$^{18}$O, respectively. The blue dots enclosed in square boxes at the lower-left corners of each panel indicate the beam size of the Herschel H$_{2}$ column density map (18.2$^{\prime\prime}$). The beam sizes corresponding to each molecular line are depicted as magenta-colored dots over their respective species labels. The blue ($^{13}$CO) and green (C$^{18}$O) colored spectra shown in the bottom panel are extracted from the positions marked by black-colored dots in the top panels. The vertical lines in each of the five lower panels indicate the peak positions of the $^{13}$CO and C$^{18}$O spectra.}
    \label{fig:c18o13co}
\end{figure*}

Figure \ref{fig:c18o13co} shows the integrated intensity contours of $^{13}$CO (top left) and C$^{18}$O (top right) over the 18.2$^{\prime\prime}$ high-resolution H$_{2}$ column density map derived from the 160-500 $\mu$m Herschel data\footnote{See Section 4.1 of \cite{2015A&A...584A..91K} for a detailed description of derivation of the H$_{2}$ column density map for Aquila region.}, suggesting that $^{13}$CO emission primarily traces the extended, large-scale distribution of molecular gas, while C$^{18}$O emission highlights the dense filamentary structures within it.

Panels (a)-(e) show representative spectra illustrating the diversity of features identified in the $^{13}$CO and C$^{18}$O line profiles across the mapped region. In panel (a), the C$^{18}$O line has one prominent peak at $\sim$7 km s$^{-1}$, but the $^{13}$CO line shows two prominent peaks at $\sim$6 and 8 km s$^{-1}$. However, it can be seen that the dip between the two peaks of $^{13}$CO is close to the peak position of the C$^{18}$O spectrum. This indicates that the double-peak feature of the $^{13}$CO line is caused by the self-absorption of the $^{13}$CO spectrum due to its high optical depth. There are also two additional peaks in $^{13}$CO at $\sim$9 and 9.5 km s$^{-1}$ along with a less prominent one in C$^{18}$O at $\sim$9 km s$^{-1}$. In panel (b), the C$^{18}$O spectral profile exhibits two distinct peaks at $\sim$7 and 7.5 km s$^{-1}$, indicative of multiple velocity components along the line-of-sight as C$^{18}$O is usually optically thin and position (b) is away from dense regions, as seen from the figure. The $^{13}$CO spectrum also shows a prominent peak near 7.5 km s$^{-1}$. However, relative to the $\sim$7 km s$^{-1}$ peak observed in the C$^{18}$O profile, there is a noticeable dip in the $^{13}$CO profile, suggestive of self-absorption in $^{13}$CO. Furthermore, the $^{13}$CO profile reveals two additional, less prominent peaks around 4 and 9 km s$^{-1}$, with no corresponding significant detection in C$^{18}$O above the 3$\sigma$ level, implying that these components trace lower-density molecular material primarily traced by $^{13}$CO. In panel (c), the C$^{18}$O spectrum shows a single peak at $\sim$7 km s$^{-1}$, indicative of a relatively simple velocity structure in that region of the cloud. In contrast, the $^{13}$CO profile displays a flattened top, which is suggestive of higher optical depth effects in $^{13}$CO. Additionally, there is also a less prominent peak in $^{13}$CO and C$^{18}$O at $\sim$6 km s$^{-1}$. Panel (d) shows a prominent peak at $\sim$6 km s$^{-1}$ followed by a less prominent peak at $\sim$7 and 8 km s$^{-1}$ in both the $^{13}$CO and C$^{18}$O profiles, indicating multiple components along the line-of-sight. In panel (e), there are two distinct peaks in the C$^{18}$O profile at $\sim$6 and 7 km s$^{-1}$ with the 6 km s$^{-1}$ peak being less prominent than the other, suggesting multiple components along the line-of-sight. The $^{13}$CO profile also shows a significant peak at 6 km s$^{-1}$. However, close to the 7 km s$^{-1}$ peak of C$^{18}$O profile, the $^{13}$CO profile looks like an extended shoulder indicative of self-absorption feature.

The five line profiles, chosen from various parts of the region, collectively show the overall complexity in the cloud. In all five cases, the $^{13}$CO spectra have larger velocity range and line widths than C$^{18}$O, which suggests that C$^{18}$O transition is tracing a more compact region with higher density than the $^{13}$CO transition and the $^{13}$CO emission is extended over a wider region, tracing the bulk motion of less dense molecular gas. It is also notable that $^{13}$CO is much brighter in the W40 region compared to Serpens South.

\subsection{HCO$^{\textit{+}}$, CS, SO, and N$_{\textit{2}}$H$^{\textit{+}}$ Emissions}
\begin{figure*}[!htb]
    \centering
    \includegraphics[width=\textwidth]{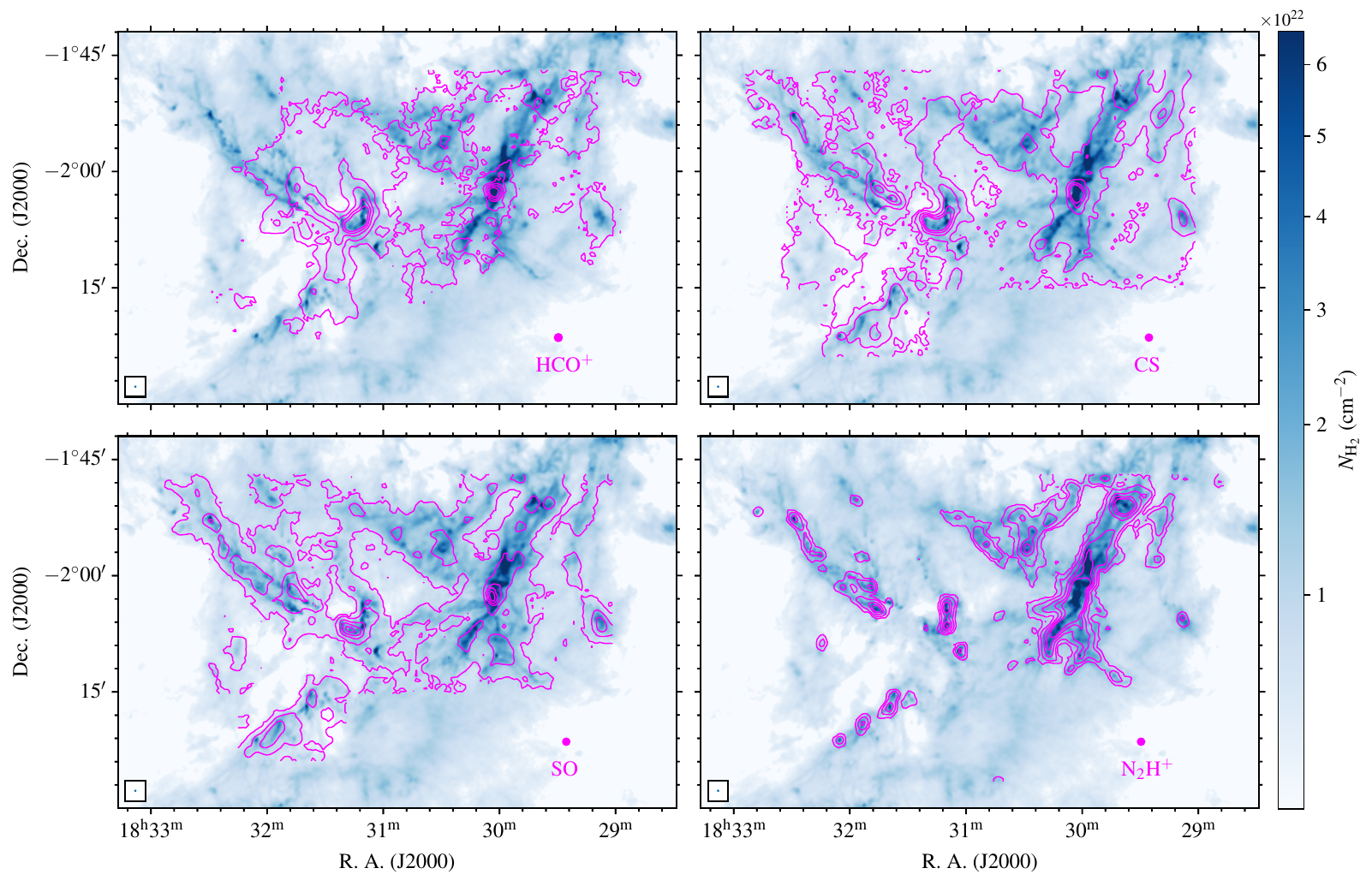}
    \caption{Distribution of HCO$^{+}$, CS, SO, and N$_{2}$H$^{+}$ integrated intensity contours over the Herschel H$_{2}$ column density map. The contour levels are $\sigma\times\left[3, 10, 20, 30, 40\right]$ K km s$^{-1}$ for all the species. The blue dots enclosed in square boxes at the lower-left corners of each panel indicate the beam size of the Herschel H$_{2}$ column density map. The beam sizes corresponding to each molecular line are depicted as magenta-colored dots over their respective species labels.}
    \label{fig:species_column}
\end{figure*}

Figure \ref{fig:species_column} presents the distributions of HCO$^{+}$, CS, SO, and N$_{2}$H$^{+}$ integrated intensity contours over the Herschel H$_{2}$ column density map. HCO$^{+}$, CS, and SO are observed to exhibit brighter emission in the W40 region compared to Serpens South, which may be due to the gas (and dust) being warmer in W40.

The distribution of HCO$^{+}$, CS, and SO closely follows the central high H$_{2}$ column density regions of W40; and it is well matched with that of $^{13}$CO (see top left panel of Figure \ref{fig:c18o13co}). However, in the case of Serpens South, there are three prominent peaks in the H$_{2}$ column density, but only the brightest one (central hub region) has the corresponding HCO$^{+}$, CS, and SO peaks, which may be due to the reported protostellar outflow activities \citep{2015Natur.527...70P} in this region. Though CS and SO are not widely regarded as outflow tracers, they have been reported to trace outflow activity in certain cases \citep[e.g.,][]{2010A&A...522A..91T}.

For N$_{2}$H$^{+}$ (bottom right panel of Figure \ref{fig:species_column}), it can be noted that the distribution traces the Serpens South region entirely, which suggests that the region consists of cold, dense gases and is in the early stages of star formation \citep[see][etc]{2013ApJ...766..115K,2016ApJ...833..204F,2024ApJ...969...70F}. However, in the W40 region, the emission shows a clumpy distribution in the bright C$^{18}$O and dense H$_{2}$ column density regions. 

The overall spatial distribution of these molecules highlights the diverse physical conditions and evolutionary stages present within the W40 and Serpens South regions. This paper focuses on the identification of velocity-coherent filamentary structures and the embedded dense cores and their kinematical properties. Discussions on the chemical differentiation between filaments and dense cores using the chemical tracers such as SO and NH$_{2}$D will be introduced in a forthcoming paper. In the following sections, we will explore the filamentary structures and the dense cores embedded within them, examining how these molecular features connect to the large-scale gas morphology and ongoing star formation activity.

\subsection{Filamentary Structures}
\subsubsection{Identification of Velocity-coherent Filaments} \label{subsec:velco}
 The $^{13}$CO (1--0) line is useful to trace the large-scale cloud but has limitations due to its large optical depth in the areas of high column density. Hence, the C$^{18}$O (1--0) line was used to resolve velocity components on the line-of-sight and identify velocity-coherent filamentary structures in the Aquila region. For this purpose, the C$^{18}$O spectrum at each pixel was subjected to Gaussian decomposition to disentangle the multiple velocity components using the automated Gaussian decomposing Python class \texttt{Decompose} from the package \texttt{funstools},\footnote{\href{https://github.com/radioshiny/funstools}{https://github.com/radioshiny/funstools}} which is an optimally developed Python-based program tailored for TRAO-FUNS data. The decomposition process follows the three steps described in \cite{2023ApJ...957...94Y}.

\texttt{Step 1} consists of velocity and spatial smoothing of the spectra using the initial smoothing parameters to improve SNR and spatial continuity, respectively. From the smoothed spectrum, the peak velocity positions of the expected Gaussian components are determined by applying the \texttt{find\_peaks} algorithm in the \texttt{scipy.signal} package. \texttt{Step 1} returns the initial guesses for each decomposed Gaussian component for the next step. \texttt{Step 2} fits a Gaussian profile to the spectra using the initial guesses determined from \texttt{Step 1} to find the best-fit spectrum and returns the resulting Gaussian fitting parameters for each velocity component of the spectrum. In \texttt{Step 3}, a final Gaussian fitting is performed to enhance the spatial continuity of the decomposed results. This is accomplished by modifying the initial guesses for each pixel based on the best-fit parameters derived from \texttt{Step 2} in its neighboring pixels. Then, a multiple Gaussian fit is repeated with the adjusted initial guesses.

\begin{figure*}[!htb]
    \centering
    \includegraphics[width=\textwidth]{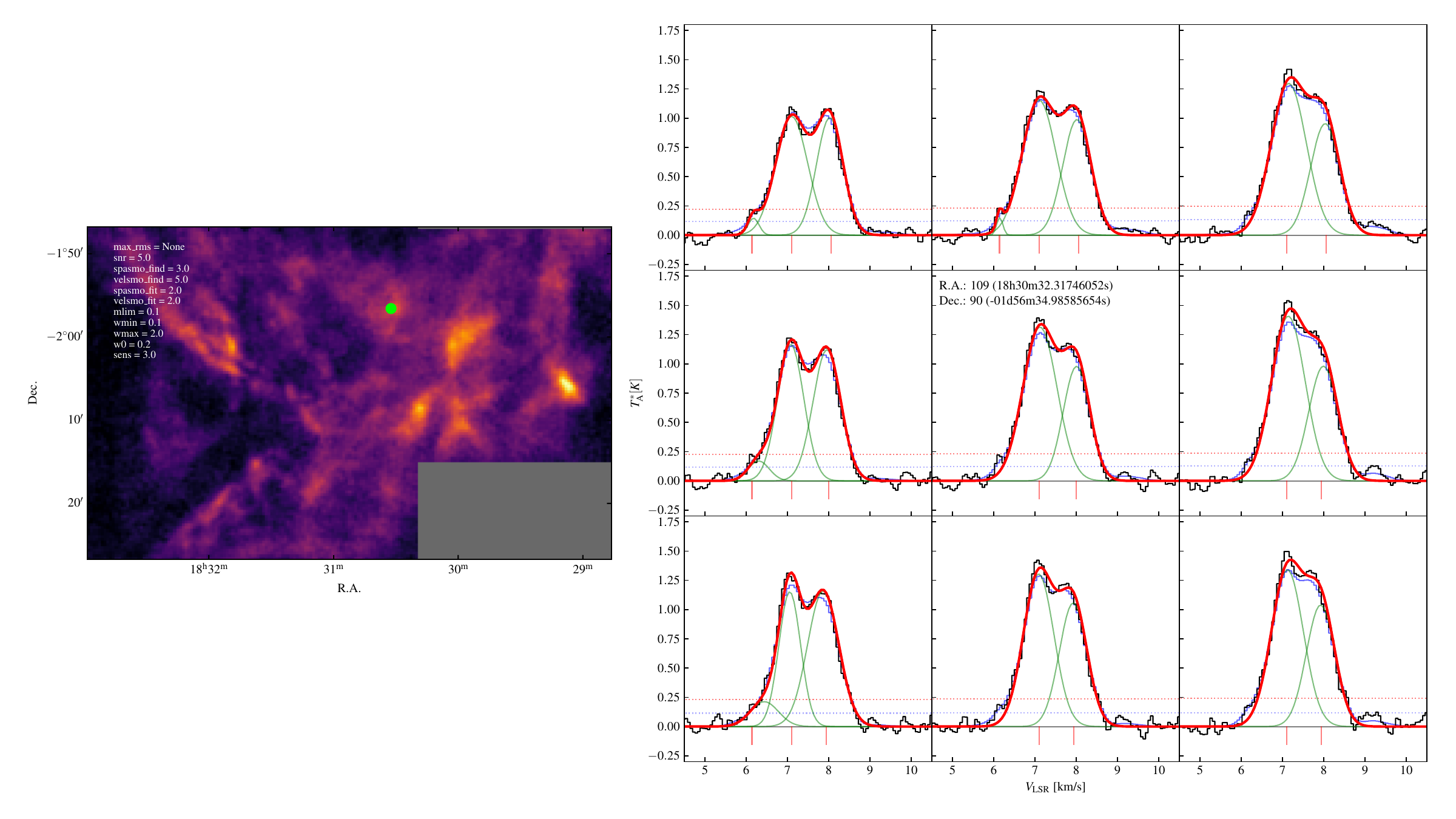}
    \caption{Left panel: Integrated intensity map of the C$^{18}$O (1–0) data cube. The green dot marks the pixel location whose spectrum is shown in the central panel on the right. Right panel: The $3\times3$ layout shows the C$^{18}$O spectrum at the pixel position marked by the green dot and those of its eight surrounding pixels located within $\sqrt{2}$ pixel units, along with the corresponding step-by-step Gaussian decomposition results. Black solid line: original spectra. Blue solid line: smoothed profiles with the initial smoothing parameters. Red vertical lines: initial guesses of velocity components found in \texttt{Step 1}. Green solid line: decomposed Gaussian profiles determined in \texttt{Step 3} with modified initial guesses from \texttt{Step 2}. Red thick solid line: total Gaussian fit. Red and blue horizontal dotted lines: $5\times\sigma$ levels of original and smoothed profiles, respectively.}
    \label{fig:decompose}
\end{figure*}

\begin{figure}[!htb]
    \centering
    \includegraphics[width=0.47\textwidth]{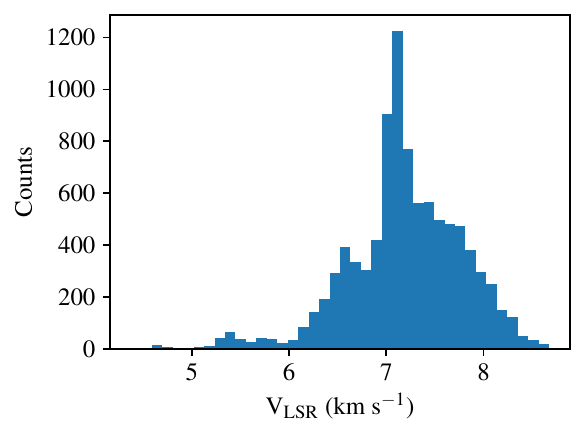}
    \caption{Distribution of C$^{18}$O velocity centroid values obtained by the decomposition of Gaussian-fitted line profiles throughout the entire observed region of 2255 arcmin$^{2}$.}
    \label{fig:component_vel}
\end{figure}

During the decomposition of multiple Gaussian components from the observed line profiles, we employed two different combinations of smoothing factors. In \texttt{Step 1}, where the number and velocities of Gaussian components are identified from the shape of the line profile and used to generate the initial guesses for fitting, we applied aggressive smoothing--5-channel velocity smoothing and 3-pixel spatial smoothing--to minimize the impact of noise and to ensure large-scale spatial continuity. In contrast, for \texttt{Steps 2} and \texttt{3}, where multiple Gaussian components are fitted to the observed line profiles, we applied minimal smoothing--2-channel velocity smoothing and 2-pixel spatial smoothing--to preserve the velocity and spatial resolution. Figure \ref{fig:decompose} presents an example of the Gaussian decomposition around the position marked with a green dot in the left panel. Using the automated Gaussian decomposing code, 10,673 Gaussian components were identified with their peak intensities of an SNR $>$ 5. Figure \ref{fig:component_vel} illustrates the velocity distribution of the decomposed components.

The decomposed velocity components are then used to identify the velocity-coherent filaments using the friends-of-friends (FoF) algorithm. Note that FoF has been extensively used to identify velocity-coherent filaments in other regions as part of TRAO-FUNS study, except for L1478 of the California Molecular Cloud (CMC), where filament identification was carried out using the \texttt{astrodendro} Python package. The detailed processes of identifying the filaments are summarized as follows: first, all the decomposed velocity components are sorted based on their intensity values. Then, the component having the highest intensity is treated as the seed. If the velocity difference ($v_{\rm diff}$) between a component and the seed component is less than the velocity dispersion ($\sigma$) of the seed component and the positional distance between the two components is $\leq$ $\sqrt{2}$ pixel distance (maximum distance between the seed pixel and one of the neighboring 8 pixels), they become a part of a velocity-coherent structure. The minimum number of pixels for a velocity-coherent structure was set to be 28, which corresponds to the number of pixels in a $5 \times \theta_{\rm FWHM}$ area\footnote{Note that this threshold was determined through an iterative procedure. During the execution of the FoF algorithm, we observed the presence of small, localised structures comprising 5--10 pixels. To avoid such small structure, we imposed a minimum size criterion and found that adopting a threshold of 28 pixels, corresponding to $\sim 5\times \theta_{\rm FWHM}$, provides a roubust choice.}. This process was iteratively done until no component was left to be compared.

\begin{figure*}[!htb]
    \centering
    \includegraphics[width=\textwidth]{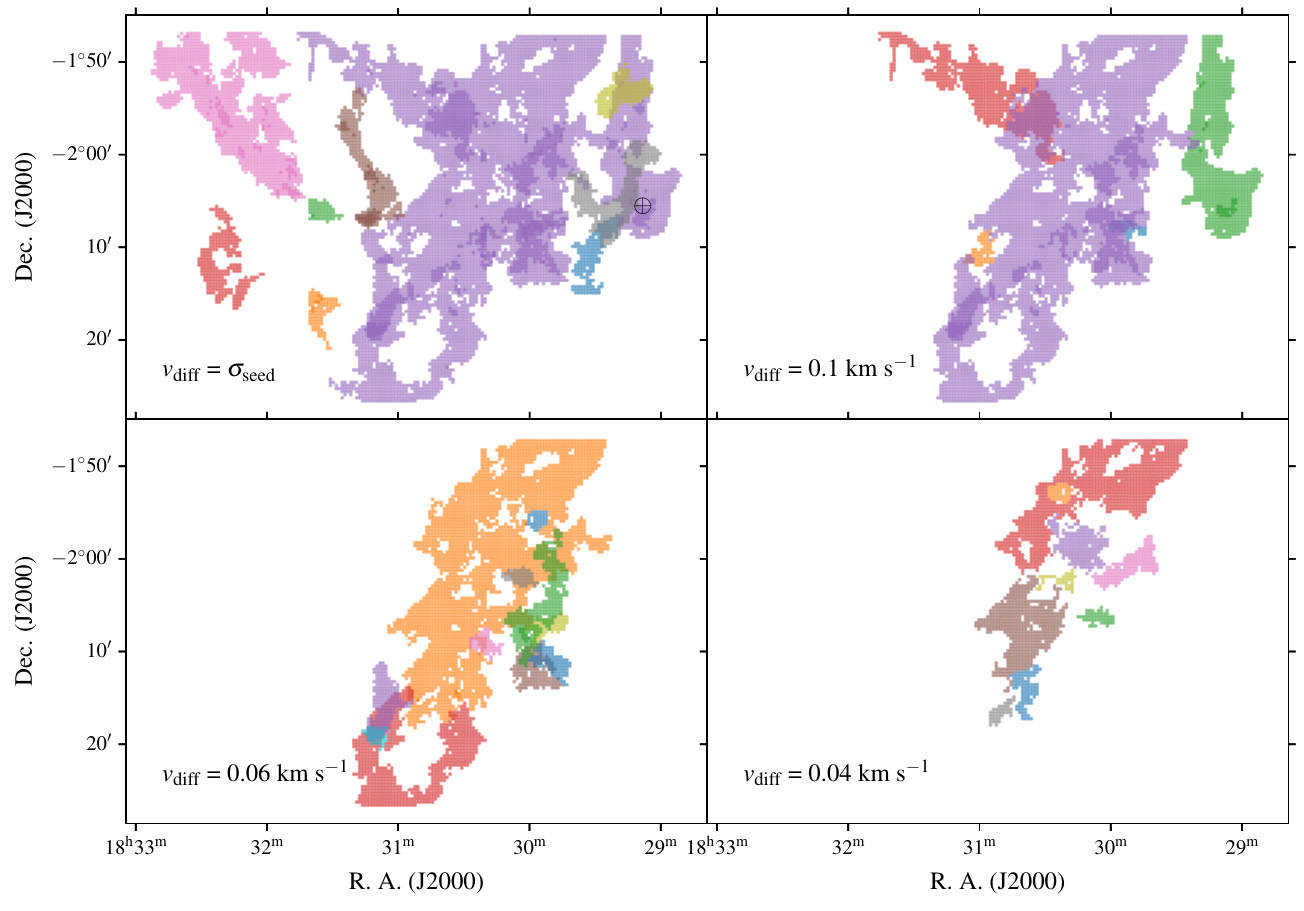}
    \caption{Procedure for identifying velocity-coherent filamentary structures using the FoF algorithm applied to the decomposed velocity components of the C$^{18}$O data toward the Aquila region. The top left panel shows the identified filamentary structures with the velocity difference set as the velocity dispersion of the seed component (marked with the `$\oplus$' symbol). The top right panel shows the structures disintegrated from the large structure (violet colored, top left panel) when the velocity difference was set to be 0.1 km s$^{-1}$. The bottom left panel shows the structures disintegrated from the large structure (violet-colored, top right panel) when the velocity difference was set to be 0.06 km s$^{-1}$. The bottom right panel shows the structures disintegrated from the large structure (orange colored, bottom left panel) when the velocity difference was set to be 0.04 km s$^{-1}$.}
    \label{fig:fil}
\end{figure*}

In the starting run of the FoF algorithm, when the velocity difference was set to be less than the velocity dispersion of the seed component ($\sigma_{\rm seed} \simeq 0.38$ km s$^{-1}$), 9 structures were identified, as shown in Figure \ref{fig:fil} (top left panel). However, one of the structures in violet, despite having multiple velocity components, was identified as a single velocity-coherent structure and can be seen spreading across a wide range. Hence, for this particular structure, the FoF algorithm was again run with a smaller velocity difference, 0.1 km s$^{-1}$, which was able to identify multiple independent structures having a single velocity component (colored regions in Figure \ref{fig:fil}, top right panel). However, even in the second run of the FoF, we find several regions where multiple velocity components still exist (the violet-colored structure, top right panel). So, a third run for the FoF algorithm was performed with an even smaller velocity difference of 0.06 km s$^{-1}$ for this structure. This trial further resulted in multiple independent structures (Figure \ref{fig:fil}, bottom left panel). However, there still remained a very big and complex structure shown in orange, which, despite having no multiple velocity components, was seen to contain multiple independent velocity coherent structures in the position-position-velocity (PPV) space that were somehow connected by a small number of components having intermediate velocity values. Hence, a final trial of the FoF algorithm was performed with a velocity difference of 0.04 km s$^{-1}$ to separate these independent structures after multiple tests with the differences of 0.055, 0.05, 0.045, 0.04, and 0.035 km s$^{-1}$, indicating that the previously identified big cloud structure was finally decomposed (see bottom right panel of Figure \ref{fig:fil}). Note that for other identified structures, if a FoF run with a given velocity difference revealed multiple velocity components, we also performed additional runs using a smaller velocity difference to try and separate those components. Lastly, by visual inspection, it was checked whether any Gaussian components with significant emission ($\sim5\sigma_{\rm rms}$) were missed during the FoF run. These components were recovered in the final filamentary structures.

Through all these steps, a total of 37 velocity-coherent structures were identified in the Aquila region. However, due to multiple trials of the FoF algorithm and with minimal velocity differences, many of these structures were found to be small, rounded clumps rather than typical elongated filamentary structures. Hence, after a careful visual inspection of their distribution in the 3D PPV space, several structures were merged with each other, satisfying continuity in the PPV space. In some cases, a small number of pixels were also discarded that seemed like outliers to the structures but got connected due to the adopted velocity threshold. Finally, a total of 23 velocity-coherent structures were identified in the Aquila region and named as F1, F2, ..., F23 for further analysis (see Figure \ref{fig:filvel}).

After the identification of velocity-coherent structures, their skeletons were defined using the Python package \texttt{FilFinder} \citep{2015MNRAS.452.3435K}, which extracts their skeletons from the 2D spatial image via a medial axis transform. To perform this, first we produced the integrated intensity maps of each structure from their resultant parameters (amplitude, velocity centroid, and velocity dispersion) derived from the Gaussian decomposition. The \texttt{FilFinder} algorithm not only provides information on the skeleton of the structures (see the left panel in Figure \ref{fig:filvel}) but also the ones for the branches stretched out from the skeleton.

The structures F14A and F14B were identified as a single structure, F14, while running the FoF algorithm. However, \texttt{FilFinder} provided two different skeletons for the structure; hence, we visually divided the structure into two structures as F14A and F14B. Note that structures F10 and F14A exhibit complex velocity distributions that could not be effectively separated using the FoF algorithm. Therefore, Principal Component Analysis (PCA) and Density-Based Spatial Clustering of Applications with Noise (DBSCAN) were collectively employed to separate them into smaller velocity-coherent structures (see Figure \ref{fig:pca_db},  Appendix \ref{app:fil14a} for details). F14A was separated into F14a, F14c, F14d, and F14e, and F10 was separated into F10a and F10b. For continuity in naming, F14B was renamed to F14b.

\subsubsection{Lengths and Widths of the Filaments}
The length of the structures ($l$) is adopted as the length of the respective skeletons obtained using the \texttt{FilFinder} algorithm. The minimum and maximum lengths obtained were 0.6 pc (F22) and 4.0 pc (F12), respectively, with a median value of 1.4 pc. However, note that these derived lengths may be affected by projection effects and should be treated as lower limits to the true physical lengths of the filaments.

The \texttt{FilFinder} algorithm was also used to find the widths of each structure by creating radial profiles centered on the structure's skeleton. Then, the averaged radial profile is fitted with the 1D-Gaussian model to determine the FWHM as the width of the structure ($W$). The widths range from 0.03 pc (F10b) to 0.32 pc (F6) with a median value of 0.13 pc. Note that our median width is comparable to that of the characteristic width of filaments (0.1 pc) revealed by the Herschel observations \citep{2011A&A...529L...6A} and to the median value of $0.12\pm0.04$ pc obtained by \cite{2015A&A...584A..91K}. However, out of 23 identified velocity-coherent structures in this study, 9 have widths larger than 0.15 pc. This can be attributed to our adopted distance to the target (455 pc), compared to that of \cite{2015A&A...584A..91K} (260 pc). Also, our derived widths may be affected by the coarser spatial resolution of C$^{18}$O (49$^{\prime\prime}$) compared to that of Herschel (18.2$^{\prime\prime}$).

Table \ref{tab:prop} lists the lengths and widths of each velocity-coherent structure determined using the \texttt{FilFinder} algorithm. However, though F2, F8, and F16 were identified as velocity coherent structures by the FoF algorithm and F14e by the DBSCAN algorithm (see Appendix \ref{app:fil14a}), \texttt{FilFinder} was not able to provide skeletons for these structures. Accordingly, all other structures have been considered as filaments. Table \ref{tab:prop} also lists the aspect ratio (AR), defined as the ratio of length $l$ and width $W$ of the structure (AR = $l/W$), that suggests how the velocity-coherent structures are elongated \citep{2014prpl.conf...27A}. Note that ARs for all of the filaments are $>3$, signifying their elongated morphology.

\begin{figure*}[!htb]
    \centering
    \includegraphics[width=\textwidth]{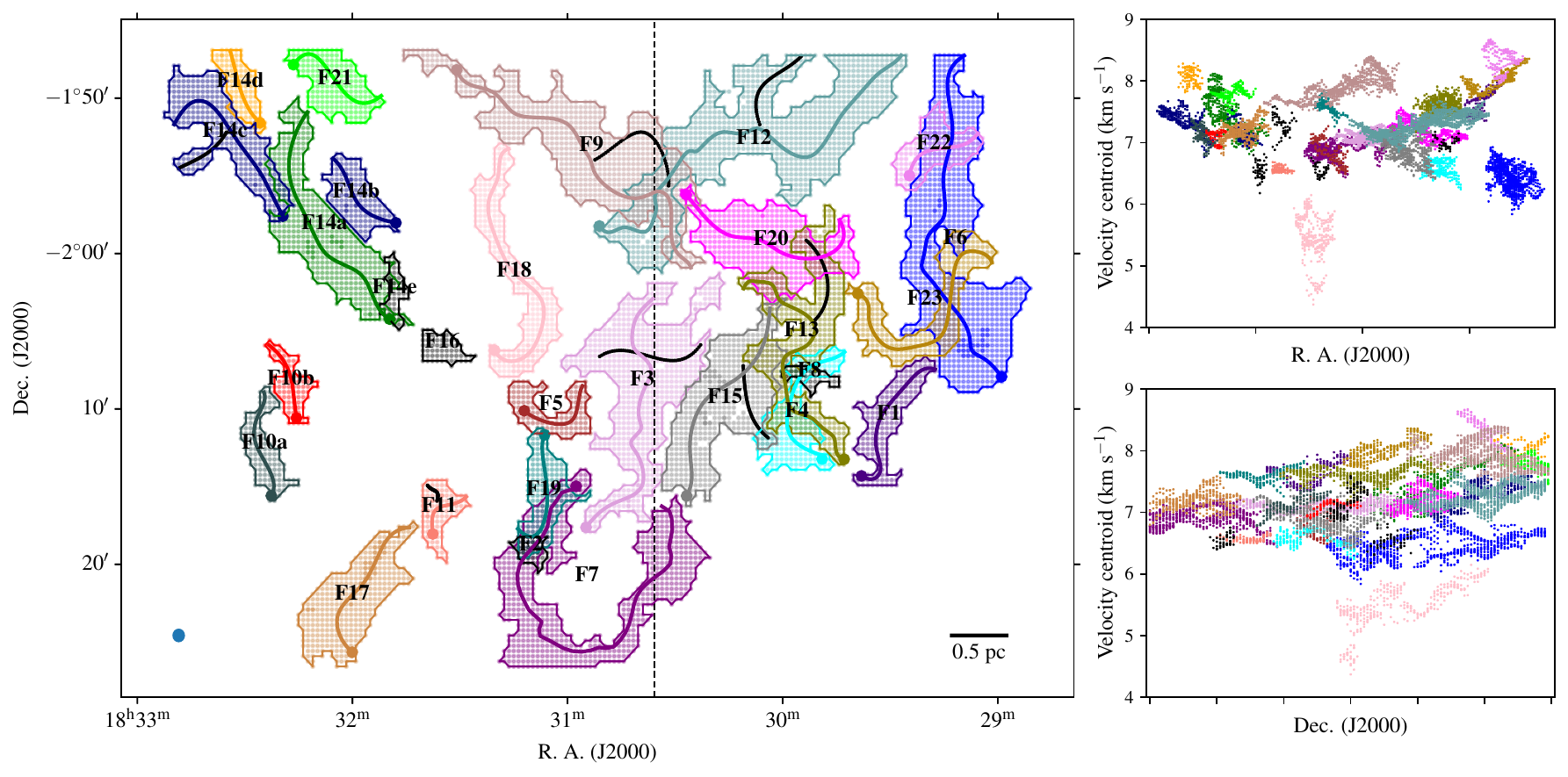}
    \caption{Left: Morphologies of the identified velocity-coherent filaments and their skeletons (thick lines) in the Aquila region. The thick black lines drawn over some filaments are the branches extended from the main skeleton. The dots over the skeletons represent the 0 pc offset position along the skeleton. The velocity-coherent structures F2, F8, F14e, and F16 are also shown without skeletons. The vertical black dashed line (RA = 18:30:37.2) roughly marks the boundary between W40 and Serpens South regions. The blue dot in the lower-left corner represents the C$^{18}$O beam size. Right: Position-Velocity (PV) plots of the filaments along R.A. (east to west, top panel) and Dec. (south to north, bottom panel) directions.}
    \label{fig:filvel}
\end{figure*}

\begin{figure*}[!htb]
    \centering
    \includegraphics[width=\textwidth]{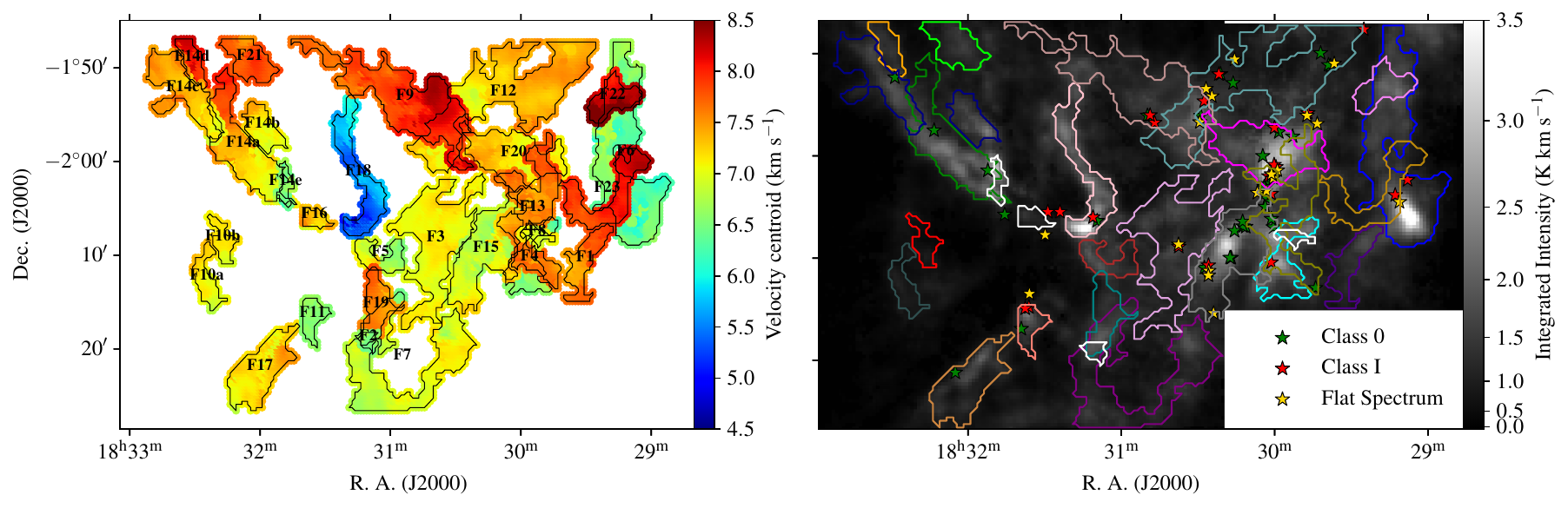}
    \caption{Left: Distribution of Gaussian-decomposed velocity centroid values within the filaments along with the velocity-coherent structures F2, F8, F14e, and F16. Right: Morphology of the filaments over the C$^{18}$O integrated intensity image. F2, F8, F14e, and F16 are also shown with white-colored boundaries. The green, red, and yellow-colored stars correspond to the positions of the class 0, I, and flat spectrum YSOs, respectively, over the W40 and Serpens South region, identified by \cite{2023ApJS..266...32P}.}
    \label{fig:fil_details}
\end{figure*}

\begin{figure*}[!htb]
    \centering
    \includegraphics[width=\textwidth]{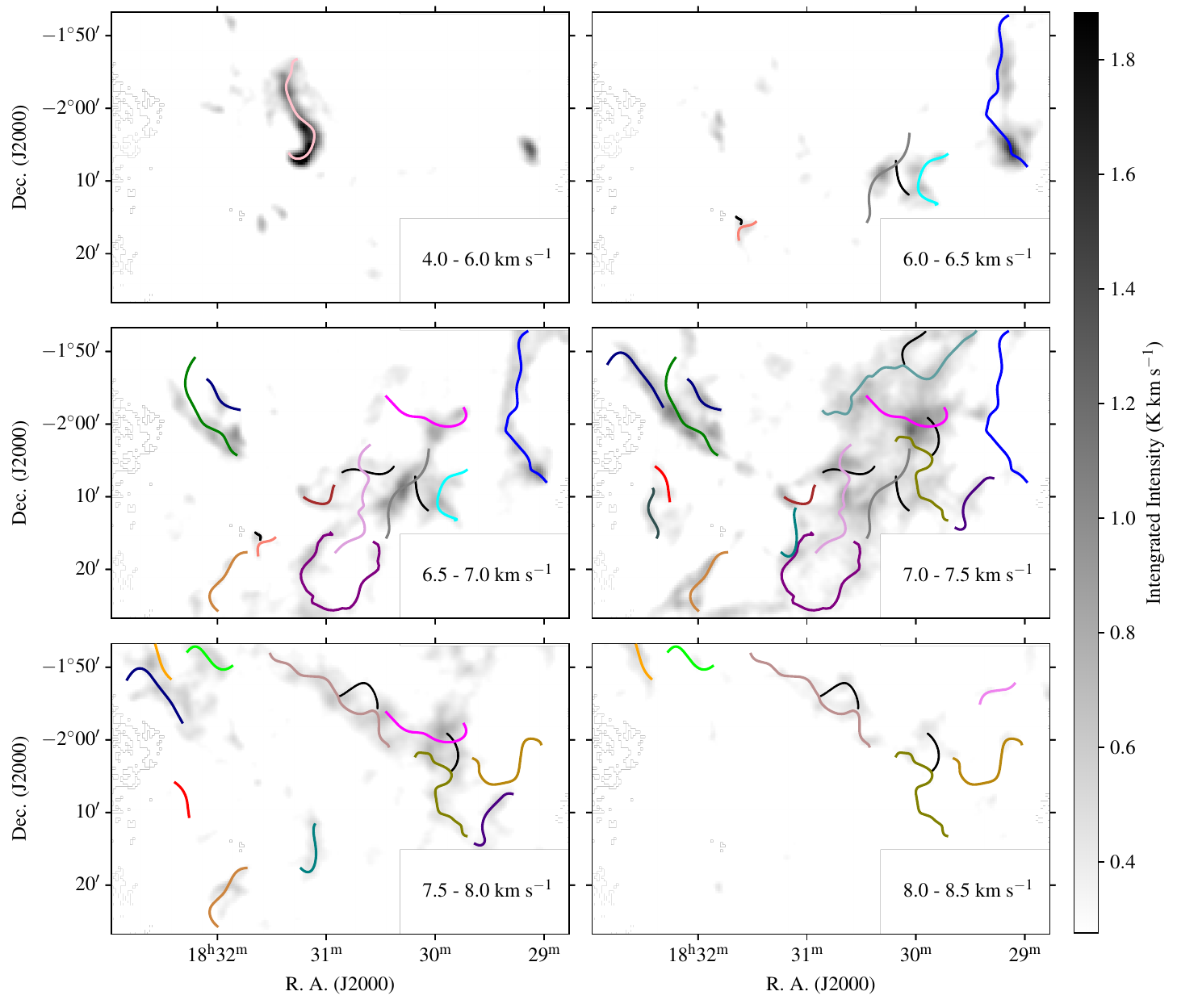}
    \caption{Velocity channel maps of the identified filamentary structures. The colored lines denote the filament skeletons, following the same color scheme as in Figure \ref{fig:filvel}, and the black-colored lines signify the branches corresponding to the main skeleton. The lower limit to the color scale is set at the 3$\times\sigma_{\rm RMS\_I}$ level for better visualisation of the filamentary structures. A filament skeleton is selected in a specific channel map if 70\% or more of its velocity centroid values fall within the corresponding velocity range.}
    \label{fig:velch}
\end{figure*}

Figure \ref{fig:filvel} illustrates the morphologies, skeleton structures, and spatial distribution along the R.A. and Dec. axes for all the identified velocity-coherent filaments. The 2D distribution of Gaussian-decomposed velocity centroid values, providing an insight to their 3D distribution\footnote{The distribution of velocity centroids in the 3D PPV space for each filament is available at \href{https://imbunty575.github.io/Aquila-Filaments/Filaments_PPV.html}{https://imbunty575.github.io/Aquila-Filaments/Filaments\_PPV.html}.} (left panel), and the spatial distribution of YSOs (right panel), as identified by \cite{2023ApJS..266...32P}, over the filaments is presented in Figure \ref{fig:fil_details}. Figure \ref{fig:velch} contains the velocity channel maps of the filamentary structures with intervals of 0.5 km s$^{-1}$ from 4--8.5 km s$^{-1}$, which show the velocity distribution in different structures in detail. Around 40\% of the total identified filaments exhibit mean velocities between 4 and 7 km s$^{-1}$, whereas the rest lie within the 7--8.5 km s$^{-1}$ interval. The first panel of Figure \ref{fig:velch} suggests that Filament 18 (F18) has the lowest velocity component among all the filaments. The average velocity centroids and the total velocity dispersions for each filament are tabulated in Table \ref{tab:prop}.

\subsubsection{Column Densities, Masses, and Line Masses of Filaments} \label{sec:column}
The C$^{18}$O column densities of the identified filaments were derived using the formula obtained under the assumption of local thermodynamic equilibrium \citep{1991ApJ...374..540G,2015MNRAS.450.1094P},
\begin{align}
    N_{\rm C^{18}O} = \ & \frac{3k_{\rm B}}{8\pi^{3}B\mu^{2}}\frac{e^{hBJ(J+1)/k_{\rm B}T_{\rm ex}}}{J+1} \nonumber \\
    & \times \frac{T_{\rm ex} + \frac{hB}{3k_{\rm B}}}{1 - e^{-h\nu/k_{\rm B}T_{\rm ex}}} \int \tau \, {\rm d}v, 
    \label{eq:c18o_column}
\end{align}
where $B$ is the rotational constant, $\mu$ is the permanent dipole moment of C$^{18}$O molecule, $J$ is the lower rotational level, $\tau$ is the optical depth, $k_{\rm B}$ is the Boltzmann constant, and $h$ is the Planck constant. The excitation temperature ($T_{\rm ex}$) is calculated using the relation given by \cite{2008ApJ...679..481P}:
\begin{equation}
T_{\rm ex} = \frac{T_{0}}{\ln{\left(1+T_{0}\left(\frac{T_{\rm R}}{1-e^{-\tau}}+\frac{T_{0}}{e^{T_{0}/T_{\rm bg}}-1}\right)^{-1}\right)}},
\end{equation}
where $T_{\rm R}$ is the radiation temperature, $T_{0}=h\nu/k_{\rm B}$ and $T_{\rm bg}$ is the cosmic microwave background temperature (2.73 K). We used the radiation temperature of $^{13}$CO for $T_{\rm R}$ under the assumption that $^{13}$CO and C$^{18}$O have similar excitation temperatures and $^{13}$CO is optically thick compared to C$^{18}$O ($\tau_{\rm ^{13}CO}\gg\tau_{\rm C^{18}O}
$). \cite{2019PASJ...71S...3N} reported that $\tau_{\rm ^{13}CO}$ typically exceeds 1 across the dense filamentary structures in Aquila and reaches values greater than 2–3 toward the Serpens South region. These findings support the assumption of high optical depth for $^{13}$CO in these environments. $\tau_{\rm C^{18}O}$ is derived by adopting the abundance ratio of $\rm [^{13}CO/C^{18}O] = 5.5$ \citep{1982ApJ...262..590F} and the relation
\begin{equation}
\frac{T_{\rm max, \:C^{18}O}}{T_{\rm max, \:^{13}CO}} \approx \frac{1-e^{-\tau_{\rm C^{18}O}}}{1-e^{-\tau_{\rm ^{13}CO}}},
\end{equation}
where $T_{\rm max, \:C^{18}O}$ and $T_{\rm max, \:^{13}CO}$ are the maximum intensities of C$^{18}$O and $^{13}$CO, respectively. Our derived excitation temperatures ($T_{\rm ex}$) exhibit a strong one-to-one correlation with the dust temperatures ($T_{\rm dust}$) obtained from Herschel observations, maintaining an average ratio of $T_{\rm ex}/T_{\rm dust} \approx 1$ across all identified velocity-coherent filaments. However, in the W40 region, specifically for filament F18, we observed a significantly higher ratio ($T_{\rm ex}/T_{\rm dust} > 2$). A similar trend was reported by \cite{2019PASJ...71S...3N}, who derived an excitation temperature map for the entire Aquila region using $^{12}{\rm CO}$ (1--0). 

The integral term at the right side of Equation \ref{eq:c18o_column} can be approximated as
\begin{equation}
    \int\tau(\nu) \:{ d\nu} \approx \frac{1}{J(T_{\rm ex})-J(T_{\rm bg})}\frac{\tau(\nu_{0})}{1-e^{-\tau(\nu_{0})}}\int T_{\rm mb}\:  d\nu,
\end{equation}
where $\nu_{0}$ is the central frequency, $T_{\rm mb}$ is the observed main beam temperature of the spectral line, and $J(T)$ is the source function, $J(T) = T_{0}/(e^{T_{0}/T}-1)$. $\int T_{\rm mb}\:d\nu$ was adopted as the area under the fitted Gaussian profile. In this way, the C$^{18}$O column densities for the filaments were estimated and found to be in the wide range between 10$^{14}$ cm$^{-2}$ to 10$^{16}$ cm$^{-2}$ with a median value of $8.7\times10^{14}$ cm$^{-2}$.

\begin{figure}[!htb]
    \centering
    \includegraphics[width=0.47\textwidth]{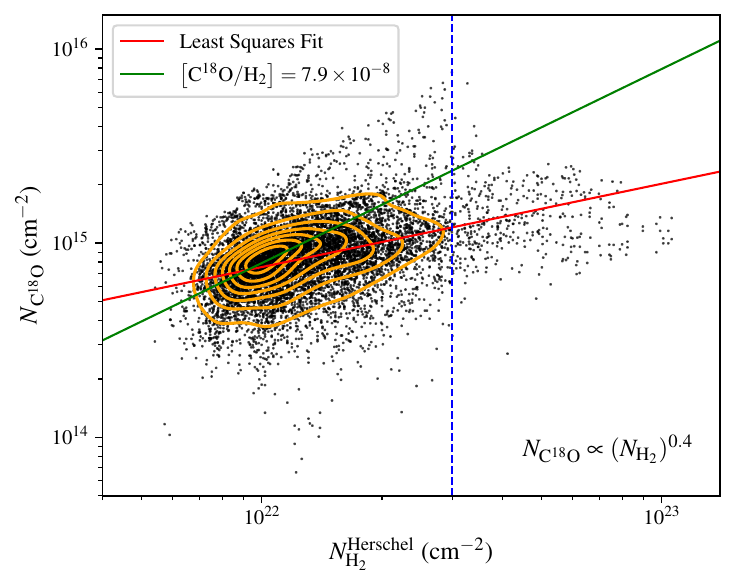}
    \caption{Comparison of C$^{18}$O and H$_{2}$ column densities of identified filaments. The red solid line represents the least squares fit to the data, resulting in the power law relation $N_{\rm C^{18}O} \propto (N_{\rm H_{2}})^{0.4}$. The obtained average abundance ratio $\left[\rm C^{18}O/H_{2}\right]=7.9\times10^{-8}$ is presented with the green solid line. The orange contours represent kernel density estimates (KDE), highlighting the regions where the majority of the column density measurements are concentrated. The vertical blue-dashed line indicates the approximate position in $N_{\rm H_{2}}^{ \rm Herschel}$ beyond which C$^{18}$O depletion is significant.}
    \label{fig:nh2_compare}
\end{figure}

We smoothed the Herschel high-resolution H$_{2}$ column density map (18.2$^{\prime\prime}$) to TRAO C$^{18}$O beam size ($\theta_{\rm FWHM} = 49^{\prime\prime}$) and divided the H$_{2}$ column density value with our derived C$^{18}$O column density value at those regions of the filaments where there is only a single velocity component ($\sim88\%$ of the pixels associated with the filaments exhibit a single velocity component) to derive the fractional abundance of C$^{18}$O at each pixel position. The average fractional abundance\footnote{Note that pixels (positions) where multiple filaments overlap are not considered during the calculation of this abundance ratio.} obtained is $[\rm C^{18}O/H_{2}]=(7.9\pm3.7)\times10^{-8}$. Figure \ref{fig:nh2_compare} shows the comparison between C$^{18}$O and H$_{2}$ column densities obtained from C$^{18}$O emissions and Herschel data, respectively. A least squares fit to the data gives a power law relation of the form $N_{\rm C^{18}O} \propto (N_{\rm H_{2}})^{0.4}$. However, this relation was not adopted for the derivation of C$^{18}$O fractional abundance, as the fit was strongly biased towards the regions where $N_{\rm H_{2}}^{\rm  Herschel}$ goes high, but $N_{\rm C^{18}O}$ does not show a proportional rise ($\gtrsim 3\times10^{22}$ cm$^{-2}$, vertical blue-dashed line of Figure \ref{fig:nh2_compare}). This behavior likely arises due to CO depletion, as at high densities of $n_{\rm H_{2}}>10^{5}$ cm$^{-3}$ and low dust temperatures of $<20$ K, CO is usually depleted and freezes onto the surfaces of dust grains \citep{1999ApJ...523L.165C}.  If the filament depth is assumed to be comparable to its width ($\sim0.1$ pc), a volume density of 10$^{5}$ cm$^{-3}$ corresponds to a column density of $\sim 3 \times 10^{22}$ cm$^{-2}$, which aligns with the threshold above which the increase in $N_{\rm C^{18}O}$ becomes less pronounced.

$N_{\rm H_{2}}^{\rm Herschel}$ values of the filaments range from $5.4\times10^{21}$ to $1\times10^{23}$ cm$^{-2}$ with a median of $1.2\times10^{22}$ cm$^{-2}$. The masses of the filaments were derived from the $N_{\rm H_{2}}^{\rm Herschel}$ data instead of the derived $N_{\rm C^{18}O}$ data to avoid any CO depletion effect (masses from $N_{\rm C^{18}O}$ are on average $\sim10$\% lower). At positions where multiple velocity-coherent filaments overlap (F6 and F23, F6 and F22, F9 and F12, etc.), $N_{\rm C^{18}O}$ is converted to $N_{\rm H_{2}}$ using the derived $\left[\rm C^{18}O/H_{2}\right]$ abundance ratio, and subsequently used to derive the mass at those positions. The minimum and maximum masses for our identified filaments are 33 M$_{\odot}$ (F10b) and 726 M$_{\odot}$ (F12), respectively, with a median value of 154 M$_{\odot}$. The line masses ($M_{\rm line}$) are calculated by dividing the mass of a filament by its length and range from 41 M$_{\odot}$ pc$^{-1}$ (F23) to 216 M$_{\odot}$ pc$^{-1}$ (F14a) with a median value of 107 M$_{\odot}$ pc$^{-1}$. Table \ref{tab:prop} lists the average values of $T_{\rm dust}$, $N_{\rm H_{2}}^{\rm Herschel}$ and $N_{\rm C^{18}O}$, masses and line masses of the identified filaments. Note that the uncertainties in the masses and line masses are propagated from the uncertainty in the adopted distance to Aquila and the standard deviation of H$_{2}$ column density values within each of the filaments. Also, the line mass maybe affected due to the projection effect in the derived lengths of the filaments.

\subsubsection{Velocity Dispersion and Critical Line Mass of Filaments} \label{crit_mass}
Velocity dispersion plays a crucial role in understanding the kinematic properties of filaments. The observed velocity dispersion ($\sigma_{\rm obs}$) is related to the FWHM of the line profile with the relation $\sigma_{\rm obs} = {\rm FWHM}/\sqrt{8 \ln{2}}$. The thermal velocity dispersion ($\sigma_{\rm th}$) is defined as $\sigma_{\rm th} = [k_{\rm B}T_{\rm gas}/m_{\rm obs}]^{1/2}$ for the observed molecule with mass $m_{\rm obs}$. The non-thermal velocity dispersion ($\sigma_{\rm nth}$), originating from the turbulent motions of gas, is related to $\sigma_{\rm obs}$ and, $\sigma_{\rm th}$ as follows,
\begin{equation}
\begin{aligned}
\sigma_{\rm nth} &= \sqrt{\sigma_{\rm obs}^{2}-\sigma_{\rm th}^{2}} \\
                 &= \sqrt{\sigma_{\rm obs}^{2} - \frac{k_{\rm B}T_{\rm gas}}{m_{\rm obs}}}.
\end{aligned}
\end{equation}
For C$^{18}$O line, $m_{\rm obs}$ is calculated as $m_{\rm obs} = \mu_{\rm C^{18}O}\times m_{\rm H}$, where $\mu_{\rm C^{18}O}=30$ is the molecular weight of C$^{18}$O and $m_{\rm H} = 1.67\times10^{-27}$ kg is the mass of the hydrogen atom. Note that we have used $T_{\rm dust}$ to infer $T_{\rm gas}$ throughout our analysis.

We define the total velocity dispersion ($\sigma_{\rm tot}$) of the gas, including both thermal and non-thermal effects, as 
\begin{equation}\label{eq:tot_disp}
\sigma_{\rm tot} = \sqrt{\sigma_{\rm nth}^{2}+c_{s}^{2}},
\end{equation}
where $c_{s}$ is the isothermal speed of sound at temperature $T_{\rm gas}$, and defined as $c_{s} = [k_{\rm B}T_{\rm gas}/m_{\rm H}\mu_{\rm mean}]^{1/2}$. Here $\mu_{\rm mean}$ is the mean molecular weight per free particle ($=2.37$) assuming the mean mass of gas with the composition consisting of 71\% hydrogen, 27\% helium, and 2\% metal \citep{2008A&A...487..993K}. The average total velocity dispersion ($\left<\sigma_{\rm tot}\right>$) ranges from 0.35 km s$^{-1}$ (F19) to 0.45 km s$^{-1}$ (F20) with a median value of 0.41 km s$^{-1}$ across the filaments.

Assuming the filament is a self-gravitating isothermal cylinder with infinite length, the thermal critical line mass of the filament in an equilibrium under thermal and gravitational pressure and the non-thermal critical line mass supported by turbulent motions can be defined as (\citealp{1963AcA....13...30S,1964ApJ...140.1056O}):
\begin{equation} \label{eq:th_crit}
M_{\rm line}^{\rm th,\: crit} = 2\left<c_{s}\right>^{2}/G,
\end{equation}
\begin{equation} \label{eq:nth_crit}
M_{\rm line}^{\rm nth, \:crit} = 2\left<\sigma_{\rm nth}\right>^{2}/G,
\end{equation}
where $\left<c_{s}\right>$ and $\left<\sigma_{\rm nth}\right>$ are the average sound speed and average non-thermal velocity dispersion of a filament. $M_{\rm line}^{\rm th,\: crit}$ ranges from 22 M$_{\odot}$ pc$^{-1}$ (F12) to 34 M$_{\odot}$ pc$^{-1}$ (F18) with a median value of 25 M$_{\odot}$ pc$^{-1}$ and $M_{\rm line}^{\rm nth,\: crit}$ ranges from 30 M$_{\odot}$ pc$^{-1}$ (F19) to 71 M$_{\odot}$ pc$^{-1}$ (F20) with a median value of 53 M$_{\odot}$ pc$^{-1}$ across the filaments.

The contributions from thermal and non-thermal motions can be collectively attributed by the total critical line mass (virial line mass), defined as
\begin{equation} \label{eq:total_crit}
    M_{\rm line}^{\rm tot,\:crit} = 2\left<\sigma_{\rm tot}\right>^{2}/G,
\end{equation}
where $\left<\sigma_{\rm tot}\right>$ is the average total velocity dispersion of a filament. $M_{\rm line}^{\rm tot,\:crit}$ is found to be in the range of 58 M$_{\odot}$ pc$^{-1}$ (F19) to 93 M$_{\odot}$ pc$^{-1}$ (F20) with a median value of 79 M$_{\odot}$ pc$^{-1}$ across the filaments. The virial mass of the filaments ($M_{\rm vir}$) can be calculated from $M_{\rm line}^{\rm tot,\:crit}$ as follows:

\begin{equation}
M_{\rm vir} = M_{\rm line}^{\rm tot,\:crit} \times l,
\end{equation}
where $l$ is the length of the filament.

Table \ref{tab:prop} lists the average values of $\sigma_{\rm tot}$, the thermal and total critical line masses along with their uncertainties for all the identified filaments. The uncertainties in $M_{\rm line}^{\rm th,\: crit}$, $M_{\rm line}^{\rm nth,\: crit}$ and $M_{\rm line}^{\rm tot,\: crit}$ are estimated from the uncertainties in their respective velocity dispersions. Criticality plays a crucial role in the gravitational stability of a filament and the formation of dense cores within it (explained in further sections).

\begin{deluxetable*}{lccrcccrrrrrrcc}
\tablecaption{Physical Properties of the Filaments in the Aquila Region \label{tab:prop}}
\tabletypesize{\footnotesize}
\tablehead{
\colhead{Fil. ID} & \colhead{$l$} & \colhead{$W$} & \colhead{AR} & 
\colhead{$\left<V_{\rm LSR}\right>$} & 
\colhead{$\left<T_{\rm dust}\right>$} & \colhead{$\left<\sigma_{\rm tot}\right>$} & 
\colhead{$M$} & \colhead{$M_{\rm line}$} & 
\colhead{$M_{\rm line}^{\rm tot,\: crit}$} & 
\colhead{$M_{\rm line}^{\rm th,\: crit}$} & 
\colhead{$N_{\rm H_{2}}^{\rm Herschel}$} & 
\colhead{$N_{\rm C^{18}O}$} & 
\colhead{Cores} & \colhead{YSOs} \\
\colhead{} & 
\colhead{{\tiny (pc)}} & 
\colhead{{\tiny (pc)}} & 
\colhead{} & 
\colhead{{\tiny (km s$^{-1}$)}} & 
\colhead{{\tiny (K)}} & 
\colhead{{\tiny (km s$^{-1}$)}} & 
\colhead{{\tiny (M$_{\odot}$)}} & 
\colhead{{\tiny (M$_{\odot}$ pc$^{-1}$)}} & 
\colhead{{\tiny (M$_{\odot}$ pc$^{-1}$)}} & 
\colhead{{\tiny (M$_{\odot}$ pc$^{-1}$)}} & 
\colhead{{\tiny ($\times10^{21}$ cm$^{-2}$)}} & 
\colhead{{\tiny ($\times10^{14}$ cm$^{-2}$)}} & 
\colhead{} & 
\colhead{}\\
\colhead{(1)} & \colhead{(2)} & \colhead{(3)} & \colhead{(4)} & 
\colhead{(5)} & \colhead{(6)} & 
\colhead{(7)} & \colhead{(8)} & 
\colhead{(9)} & 
\colhead{(10)} & 
\colhead{(11)} & 
\colhead{(12)} & 
\colhead{(13)} & 
\colhead{(14)} & \colhead{(15)}
}
\startdata
F1 &1.4 &0.05& 28.8& 7.4 $\pm$ 0.2& 14.8& 0.39 $\pm$ 0.06& 78 $\pm$ 17& 54 $\pm$ 12& 74 $\pm$ 22& 24 $\pm$ 1& 9.5& 7.1& 0& 0 \\ 
F2 &$\cdots$ &$\cdots$ &$\cdots$& 6.5 $\pm$ 0.1& 17.0& 0.35 $\pm$ 0.07& 20 $\pm$ 5& $\cdots$& $\cdots$& $\cdots$& 11.4& 5.2& 0& 0 \\ 
F3 &1.9 &0.19& 9.9& 7.1 $\pm$ 0.1& 16.6& 0.40 $\pm$ 0.06& 328 $\pm$ 73& 174 $\pm$ 39& 76 $\pm$ 23& 27 $\pm$ 2& 10.8& 8.3& 4& 2 \\ 
F4 &1.1 &0.12& 8.8& 6.6 $\pm$ 0.1& 14.4& 0.38 $\pm$ 0.06& 138 $\pm$ 31& 131 $\pm$ 29& 69 $\pm$ 22& 23 $\pm$ 1& 13.4& 9.3& 2& 0 \\ 
F5 &0.9 &0.06& 15.3& 6.8 $\pm$ 0.2& 18.9& 0.42 $\pm$ 0.06& 72 $\pm$ 16& 78 $\pm$ 18& 85 $\pm$ 23& 31 $\pm$ 3& 12.3& 8.1& 0& 0 \\ 
F6 &3.3 &0.32& 9.7& 6.4 $\pm$ 0.2& 14.9& 0.39 $\pm$ 0.05& 500 $\pm$ 111& 161 $\pm$ 36& 71 $\pm$ 19& 24 $\pm$ 2& 12.7& 13.7& 4& 3 \\ 
F7 &3.9 &0.17& 22.6& 6.9 $\pm$ 0.2& 15.9& 0.38 $\pm$ 0.06& 312 $\pm$ 69& 81 $\pm$ 18& 70 $\pm$ 21& 26 $\pm$ 2& 10.1& 6.7& 0& 0 \\ 
F8 &$\cdots$ &$\cdots$ &$\cdots$& 7.0 $\pm$ 0.1& 14.2& 0.40 $\pm$ 0.07& 35 $\pm$ 8& $\cdots$& $\cdots$& $\cdots$& 17.1& 7.8& 0& 0 \\ 
F9 &3.1 &0.19& 16.5& 7.9 $\pm$ 0.2& 14.4& 0.41 $\pm$ 0.05& 472 $\pm$ 105& 151 $\pm$ 34& 80 $\pm$ 18& 23 $\pm$ 2& 13.6& 8.7& 7& 5 \\ 
F10a &1.0 &0.09& 11.6& 7.1 $\pm$ 0.1& 17.3& 0.40 $\pm$ 0.05& 46 $\pm$ 10& 44 $\pm$ 10& 79 $\pm$ 18& 28 $\pm$ 2& 8.4& 7.4& 0& 0 \\ 
F10b &0.7 &0.03& 24.3& 7.1 $\pm$ 0.1& 17.3& 0.37 $\pm$ 0.05& 33 $\pm$ 7& 45 $\pm$ 10& 66 $\pm$ 19& 28 $\pm$ 1& 8.3& 5.8& 1& 0 \\ 
F11 &0.6 &0.08& 7.0& 6.6 $\pm$ 0.1& 18.0& 0.43 $\pm$ 0.05& 63 $\pm$ 14& 113 $\pm$ 25& 87 $\pm$ 20& 29 $\pm$ 1& 13.9& 9.3& 2& 0 \\ 
F12 &4.0 &0.26& 15.3& 7.3 $\pm$ 0.2& 13.6& 0.42 $\pm$ 0.04& 726 $\pm$ 162& 183 $\pm$ 41& 85 $\pm$ 15& 22 $\pm$ 2& 16.3& 10.3& 11& 11 \\ 
F13 &2.4 &0.16& 15.2& 7.6 $\pm$ 0.1& 14.1& 0.41 $\pm$ 0.06& 301 $\pm$ 67& 124 $\pm$ 28& 81 $\pm$ 22& 23 $\pm$ 1& 13.4& 8.2& 5& 13 \\ 
F14a &1.4 &0.26& 5.4& 7.4 $\pm$ 0.3& 16.4& 0.43 $\pm$ 0.05& 302 $\pm$ 67& 216 $\pm$ 48& 86 $\pm$ 20& 27 $\pm$ 4& 14.3& 10.5& 5& 2 \\ 
F14b &0.9 &0.08& 10.6& 7.1 $\pm$ 0.1& 16.5& 0.45 $\pm$ 0.04& 82 $\pm$ 18& 96 $\pm$ 21& 93 $\pm$ 17& 27 $\pm$ 2& 12.2& 9.3& 1& 0 \\ 
F14c &1.7 &0.12& 14.3& 7.4 $\pm$ 0.3& 14.5& 0.38 $\pm$ 0.06& 171 $\pm$ 38& 100 $\pm$ 22& 70 $\pm$ 22& 23 $\pm$ 1& 11.6& 8.6& 4& 1 \\ 
F14d &0.7 &0.04& 16.5& 8.1 $\pm$ 0.1& 14.3& 0.44 $\pm$ 0.05& 46 $\pm$ 10& 70 $\pm$ 16& 89 $\pm$ 18& 23 $\pm$ 2& 10.1& 8.7& 0& 0 \\ 
F14e &$\cdots$ &$\cdots$ &$\cdots$& 6.5 $\pm$ 0.1& 22.2& 0.46 $\pm$ 0.06& 57 $\pm$ 13& $\cdots$& $\cdots$& $\cdots$& 22.6& 12.1& 0& 0 \\ 
F15 &2.1 &0.18& 11.5& 6.9 $\pm$ 0.2& 14.7& 0.44 $\pm$ 0.05& 352 $\pm$ 79& 170 $\pm$ 38& 89 $\pm$ 20& 24 $\pm$ 2& 15.7& 10.4& 4& 12 \\ 
F16 &$\cdots$ &$\cdots$ &$\cdots$& 7.3 $\pm$ 0.2& 23.4& 0.45 $\pm$ 0.04& 22 $\pm$ 5& $\cdots$& $\cdots$& $\cdots$& 8.5& 6.5& 0& 0 \\ 
F17 &1.3 &0.15& 8.9& 7.2 $\pm$ 0.1& 17.4& 0.42 $\pm$ 0.04& 170 $\pm$ 38& 128 $\pm$ 28& 84 $\pm$ 17& 28 $\pm$ 2& 10.8& 8.2& 3& 1 \\ 
F18 &2.5 &0.08& 30.9& 5.5 $\pm$ 0.4& 21.2& 0.43 $\pm$ 0.05& 307 $\pm$ 69& 124 $\pm$ 28& 89 $\pm$ 22& 34 $\pm$ 4& 21.3& 26.9& 4& 2 \\ 
F19 &1.2 &0.10& 12.4& 7.6 $\pm$ 0.1& 17.2& 0.35 $\pm$ 0.05& 65 $\pm$ 15& 52 $\pm$ 12& 58 $\pm$ 17& 28 $\pm$ 2& 8.2& 6.1& 0& 0 \\ 
F20 &2.0 &0.19& 10.6& 7.2 $\pm$ 0.1& 13.8& 0.45 $\pm$ 0.04& 419 $\pm$ 94& 208 $\pm$ 47& 93 $\pm$ 15& 22 $\pm$ 1& 21.4& 13.1& 3& 6 \\ 
F21 &1.1 &0.12& 9.0& 7.8 $\pm$ 0.1& 14.3& 0.40 $\pm$ 0.05& 98 $\pm$ 22& 91 $\pm$ 20& 74 $\pm$ 18& 23 $\pm$ 2& 11.9& 7.9& 1& 0 \\
F22 &0.6 &0.15& 4.1& 8.3 $\pm$ 0.2& 13.9& 0.39 $\pm$ 0.06& 51 $\pm$ 11& 82 $\pm$ 18& 72 $\pm$ 20& 23 $\pm$ 1& 7.1& 4.8& 0& 0 \\ 
F23 &2.4 &0.13& 18.4& 8.0 $\pm$ 0.2& 15.6& 0.40 $\pm$ 0.06& 99 $\pm$ 22& 41 $\pm$ 9& 77 $\pm$ 22& 25 $\pm$ 1& 7.4& 4.9& 0& 0 \\
\enddata
\tablecomments{\footnotesize 
(1) Filament IDs. (2) Filament lengths. (3) Filament widths. (4) Aspect ratio of filaments: defined as the ratio between lenght and width. (5) Average velocity centroid. (6) Average dust temperature, obtained from the Herschel dust temperature map, smoothed to TRAO C$^{18}$O resolution. (7) Average total velocity dispersion, derived using Equation \ref{eq:tot_disp}. (8) Mass of filaments. (9) Line mass of filaments, defined as $M/l$. (10) Total critical line mass of filaments, derived using Equation \ref{eq:total_crit}. (11) Thermal critical line mass of filaments, derived using Equation \ref{eq:th_crit}. (12) Average H$_{2}$ column density of filaments, obtained from the Herschel H$_{2}$ column density map, smoothed to TRAO C$^{18}$O resolution. (13) Average C$^{18}$O column density of filaments, derived using Equation \ref{eq:c18o_column}. (14) Number of identified N$_{2}$H$^{+}$ dense cores present in each filament. (15) Number of YSOs present in each filament, identified by \cite{2023ApJS..266...32P}.
}
\end{deluxetable*}

\subsection{Identification of N$_{\textit{2}}$H$^{\textit{+}}$ Dense Cores}
\begin{figure*}[!htb]
    \centering
    \includegraphics[width=\textwidth]{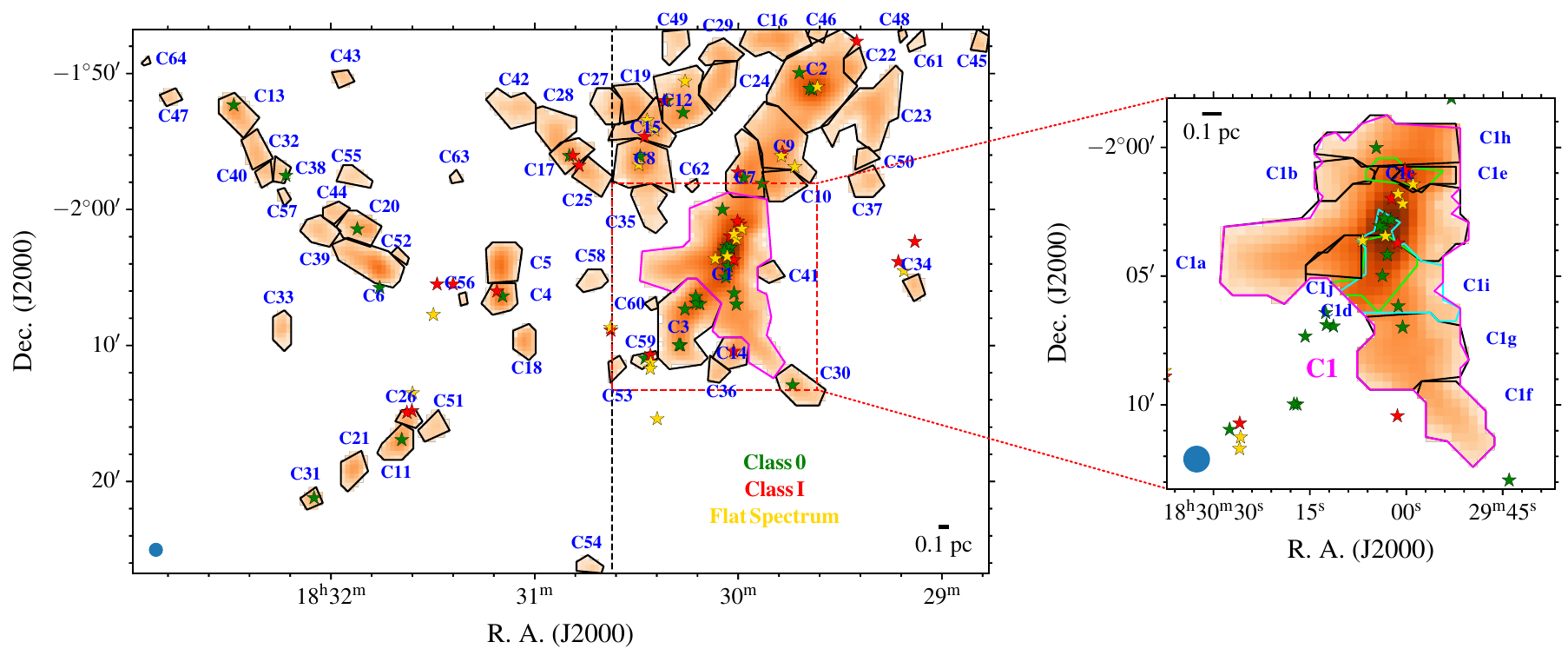}
    \caption{Left: The distribution of N$_{2}$H$^{+}$ dense cores identified by the FellWalker algorithm in the Aquila region over the integrated intensity image of N$_{2}$H$^{+}$. The vertical black dashed line (RA = 18:30:37.2) roughly marks the boundary between W40 and Serpens South regions. Right: Zoomed-in figure of the region inside the red-dashed box containing C1 in the left panel. C1a to C1j were identified by applying the FoF method to the $F_{2,\: 3\:\rightarrow\: 1,\: 2}$ hyperfine component of N$_{2}$H$^{+}$ spectra present within the C1 region. Core C1j and C1c (lime-colored borders) overlap with cores C1d and C1e, respectively, in the position space but have different velocities than those. Some parts of core C1i (cyan-colored border) also spatially overlap with core C1a. The green, red, and yellow-colored stars correspond to the positions of the class 0, I, and flat spectrum YSOs, respectively, over the W40 and Serpens South region, identified by \cite{2023ApJS..266...32P} in both the figures. The blue dot in the lower-left corner of both the figures represents the N$_{2}$H$^{+}$ beam size.}
    \label{fig:cores}
\end{figure*}

The N$_{2}$H$^{+}$ (1--0) molecular line, typically optically thin, serves as an effective tracer of dense cores in nearby star-forming regions. The FellWalker algorithm \citep{2015A&C....10...22B} in the \texttt{CUPID} package \citep{2007ASPC..376..425B} within the Starlink\footnote{\href{https://starlink.eao.hawaii.edu/}{https://starlink.eao.hawaii.edu/}} software was applied to the N$_{2}$H$^{+}$ integrated intensity map to identify the dense cores. Prior to running the algorithm, the map was smoothed using a Gaussian kernel to suppress small-scale noise and prevent spurious localised features from being identified as cores. The rms noise ($\sigma_{\rm smooth}$) of this smoothed map was estimated to be 0.017 K km s$^{-1}$. Pixels with intensities $>$$2\times\sigma_{\rm smooth}$ were allowed to be associated with a peak. Two neighboring peaks are considered separate if the difference between the peak values and the minimum value between the peaks is larger than $1.5\times\sigma_{\rm smooth}$\footnote{The default value for this parameter is $3\times\sigma_{\rm smooth}$ \citep{2015A&C....10...22B} using which we were getting 56 dense cores. However, this led to the merging of some cores that were visually separable (e.g., C1 and C3, C19 and C27, etc.). Hence, we adopted the parameter as $1.5\times\sigma_{\rm smooth}$.}. In total, we identified 64 N$_{2}$H$^{+}$ dense cores in the observed region (see Left panel of Figure \ref{fig:cores}). Note that almost all of the Herschel-identified protostellar cores and a large number of prestellar cores \citep{2015A&A...584A..91K} spatially overlap with our identified N$_{2}$H$^{+}$ dense cores (see Figure \ref{fig:herschel_cores}; Appendix \ref{app:n2hp_herschel}).

While running the FellWalker algorithm, the identified core 1 (C1) was significantly larger compared to the typical dense core radius ($\sim$ 0.1 pc). Despite multiple iterations with various sets of parameters, FellWalker was unable to separate C1 into further segments. The N$_{2}$H$^{+}$ velocity distribution, derived from hyperfine fitting at each pixel within C1, shows that the structure is not velocity-coherent (see Figure \ref{fig:nh2_compare}(a); Appendix \ref{app:core_c1}), so it cannot be considered a single dense core. Hence, we separated C1 into 10 smaller regions (C1a, C1b, $\cdots$, C1j) based on their velocity distribution (see Appendix \ref{app:core_c1} for a detailed description of the separation process). The right panel of Figure \ref{fig:cores} shows the spatial distribution of dense cores C1a to C1j over the N$_{2}$H$^{+}$ integrated intensity image. 

Note that there are several other cores, such as C2, C3, C9, etc., with large radii, which could not be divided into further segments by the FellWalker algorithm as well. However, the velocity centroid distributions within these cores remain coherent. Hence, we did not attempt to separate them into smaller segments. Also, there are a number of dense cores with radius of $\gtrsim 0.1$ pc. This large size of the cores can be attributed to the coarse spatial resolution (52$^{\prime\prime}$) of our N$_{2}$H$^{+}$ data.

\subsection{Physical Properties of the Dense Cores} \label{sec:core_properties}
The N$_{2}$H$^{+}$ column densities of the dense cores were derived using the N$_{2}$H$^{+}$ integrated intensity following the equation given by \cite{2002ApJ...572..238C} as follows:
\begin{align}
    N_{\rm N_{2}H^{+}} =\ & \frac{8\pi W_{\rm int}}{\lambda^{3}A}\frac{{\rm g}_{{ l}}}{{\rm g}_{{ u}}}
    \frac{1}{J_{\nu}(T_{\rm ex}) - J_{\nu}(T_{\rm bg})} \nonumber \\
    & \times \frac{1}{1 - e^{-h\nu / k_{\rm B} T_{\rm ex}}}
    \frac{Q_{\rm rot}}{{\rm g}_{l}e^{-E_{l}/k_{\rm B} T_{\rm ex}}},
    \label{eq:n2hp_column}
\end{align}
where $W_{\rm int}$ is the integrated intensity of the N$_{2}$H$^{+}$ emission, $A$ is the Einstein coefficient \citep[$=3.628\times10^{-5}\:\rm s^{-1}$;][]{2006ApJ...648..461D}, $\lambda$ and $\nu$ are the wavelength and frequency of the observed transition, respectively. $Q_{\rm rot}$ is the partition function, derived based on the $T_{\rm ex}$ values at each pixel and $E_{l}$ is the energy of the lower level. g$_{l}=1$ and g$_{u}=3$ are the statistical weights of the lower and upper levels, respectively. Note that for the C1 region, where multiple cores overlap spatially, the integrated intensities at each spatial position of the cores were calculated using the results from the hyperfine fitting.

Using the derived N$_{2}$H$^{+}$ column density map, where the integrated intensity ($W_{\rm int}$) has a signal-to-noise ratio (SNR) greater than 7, along with the Herschel H$_{2}$ column density map, we calculated the mean fractional abundance of N$_{2}$H$^{+}$ as $[\rm N_{2}H^{+}/H_{2}] = (9.4 \pm 3.6) \times 10^{-10}$. Based on this fractional abundance, we estimated the H$_{2}$ column densities ($N_{\rm H_{2}}$) for cores C1a to C1j and determined their masses ($M_{\rm core}$) by integrating the $N_{\rm H_{2}}$ values across their spatial extent. For the rest of the cores, the masses are derived from Herschel $\rm H_{2}$ column density values. The derived core masses ($M_{\rm core}$) range from 1 M$_{\odot}$ (C64) to 249 M$_{\odot}$ (C2), with a median mass of 31 M$_{\odot}$.

The diameter of each dense core was determined by equating its area to that of a circle. This calculated diameter was then deconvolved by subtracting the N$_{2}$H$^{+}$ beam size ($\theta_{\rm FWHM}$) in quadrature, as follows:
\begin{equation} \label{eq:radius}
d_{\rm core} = \sqrt{d^{2}-\theta_{\rm FWHM}^{2}} 
\end{equation}
The radius of each dense core was then calculated from this deconvolved diameter. However, for cores C56, C60, C62, C63, and C64, the deconvolution could not be performed because $d < \theta_{\rm FWHM}$ in these cases. For these cores, we adopted $\theta_{\rm FWHM}$ as an upper limit for the diameter. The hyperfine N$_{2}$H$^{+}$ spectra at each pixel position within each dense core were then fitted with hyperfine profiles to determine the systematic velocities and total velocity dispersions (see Equation \ref{eq:tot_disp}). 

\cite{2023ApJS..266...32P} presented a catalog of 172 protostars in the Aquila region, identified through their 1-850 $\mu$m SEDs, compiled from multiple surveys, including the Two Micron All Sky Survey, $Spitzer$, Herschel, the Wide-field Infrared Survey Explorer, and JCMT/SCUBA-2 data, as part of the eHOPS survey \citep{eHOPS_dataset}. In the present study, we utilize this catalog to investigate whether the dense cores we identified are protostellar or starless. Among the eHOPS sources, 33 class 0 protostars, 21 class I protostars, and 19 flat-spectrum YSOs are located within the boundaries of our identified dense cores (marked by asterisks in Figure \ref{fig:cores} with green, red, and yellow symbols, respectively). A dense core associated with any of these protostars is classified as a dense core with YSOs, while a core without such an association is considered starless. Out of the 73 identified N$_{2}$H$^{+}$ dense cores, 25 ($\sim$35\%) of them are found to harbor YSOs.

Table \ref{tab:cores} lists the estimated physical quantities for all the identified dense cores as well as the number of YSOs present within each dense core. 

\startlongtable
\begin{deluxetable*}{lccrcccrrclrc}
\tablecaption{Physical Properties of the Dense Cores in the Aquila Region. \label{tab:cores}}
\tablewidth{0pt}
\tabletypesize{\footnotesize}
\tablehead{
\colhead{Core ID} & \colhead{R. A. (J2000)} & \colhead{Dec. (J2000)} & \colhead{$R_{\rm core}$} & \colhead{$\left<T_{\rm dust}\right>$} & \colhead{V$_{\rm LSR}$} & \colhead{$\sigma_{\rm tot}$} & \colhead{$N_{\rm H_{2}}^{\rm Herschel}$} & \colhead{$M_{\rm core}$} & \colhead{$\alpha_{\rm vir}^{\rm core}$} & \colhead{Fil. ID} & \colhead{$X$} & \colhead{YSOs}\\
\colhead{} & \colhead{(hh:mm:ss)} & \colhead{(dd:mm:ss)} & \colhead{(pc)} & \colhead{(K)} & \colhead{(km s$^{-1}$)} & \colhead{(km s$^{-1}$)} & \colhead{{\tiny ($\times10^{21}$ cm$^{-2}$)}} & \colhead{(M$_{\odot}$)} & \colhead{} & \colhead{} & \colhead{} \\
\colhead{(1)} & \colhead{(2)} & \colhead{(3)} & \colhead{(4)} & \colhead{(5)} & \colhead{(6)} & \colhead{(7)} & \colhead{(8)} & \colhead{(9)} & \colhead{(10)} & \colhead{(11)} & \colhead{(12)} & \colhead{(13)}
}
\startdata
C1a$^{*}$ & 18:30:02.30 & -02:02:04.94 & 0.36 & 14.6 & 7.40 & 0.43 $\pm$ 0.14 & 77.8 & 221 $\pm$ 50 & 0.33 $\pm$ 0.22 & F20 & 0.19 & 7 \\
C1b & 18:30:06.30 & -02:01:04.94 & 0.11 & 14.4 & 7.72 & 0.41 $\pm$ 0.14 & 28.3 & 19 $\pm$ 4 & 1.38 $\pm$ 0.94 & $\cdots$ & $\cdots$ & 0 \\
C1c$^{*}$ & 18:29:59.63 & -02:01:04.93 & 0.09 & 13.3 & 6.89 & 0.33 $\pm$ 0.05 & 22.0 & 15 $\pm$ 3 & 1.19 $\pm$ 0.37 & $\cdots$ & $\cdots$ & 2 \\
C1d$^{*}$ & 18:30:02.30	& -02:03:44.93 & 0.15 & 13.6 & 6.92 & 0.37 $\pm$ 0.06 & 40.4 & 42 $\pm$ 10 & 0.35 $\pm$ 0.11 & F15 & 0.28 & 5 \\
C1e$^{*}$ & 18:30:00.96 & -02:01:04.93 & 0.08 & 13.2 & 7.74 & 0.40 $\pm$ 0.05 & 41.6 & 21 $\pm$ 5 & 1.22 $\pm$ 0.31 & $\cdots$ & $\cdots$ & 1 \\
C1f & 18:29:56.96 & -02:09:04.92 & 0.17 & 14.1 & 7.62 & 0.31 $\pm$ 0.14 & 22.1 & 37 $\pm$ 8 & 0.37 $\pm$ 0.33 & F13 & 0.79 & 0 \\
C1g$^{*}$ & 18:29:59.63	& -02:08:04.92 & 0.23 & 13.8 & 7.43 & 0.41 $\pm$ 0.13 & 28.9 & 87 $\pm$ 20 & 0.45 $\pm$ 0.28 & F13 & 0.33 & 1 \\
C1h$^{*}$ & 18:29:56.96	& -01:59:04.92 & 0.21 & 13.2 & 7.31 & 0.42 $\pm$ 0.08 & 67.0 & 113 $\pm$ 26 & 0.31 $\pm$ 0.12 & F20 & 0.74 & 1 \\
C1i$^{*}$ & 18:30:02.30	& -02:03:24.93 & 0.22 & 13.9 & 7.69 & 0.44 $\pm$ 0.17 & 51.8 & 65 $\pm$ 15 & 0.65 $\pm$ 0.50 & F13 & 0.23 & 1 \\
C1j & 18:30:06.30	& -02:03:44.94 & 0.11 & 13.8 & 7.19 & 0.45 $\pm$ 0.05 & 39.7 & 27 $\pm$ 6 & 1.19 $\pm$ 0.27 & F15 & 0.22 & 0 \\
C2$^{*}$ & 18:29:36.96 & -01:50:44.88 & 0.41 & 12.9 & 7.45 & 0.35 $\pm$ 0.02 & 54.1 & 249 $\pm$ 56 & 0.22 $\pm$ 0.03 & F12 & 0.22 & 4 \\
C3$^{*}$ & 18:30:11.63 & -02:06:24.95 & 0.33 & 14.3 & 7.07 & 0.39 $\pm$ 0.02 & 54.0 & 178 $\pm$ 40 & 0.31 $\pm$ 0.04 & F15 & 0.31 & 6 \\
C4$^{*}$ & 18:31:09.01 & -02:06:25.00 & 0.15 & 21.6 & 4.64 & 0.45 $\pm$ 0.02 & 46.3 & 42 $\pm$ 9 & 0.49 $\pm$ 0.12 & F18 & 0.46 & 2 \\
C5 & 18:31:09.01 & -02:04:05.00 & 0.19 & 22.5 & 5.34 & 0.40 $\pm$ 0.02 & 39.6 & 56 $\pm$ 13 & 0.48 $\pm$ 0.08 & F18 &  0.43 & 0 \\
C6 & 18:31:45.03 & -02:04:04.96 & 0.24 & 22.4 & 7.08 & 0.48 $\pm$ 0.01 & 37.5 & 73 $\pm$ 16 & 0.74 $\pm$ 0.05 & F14a & 0.25 & 0 \\
C7$^{*}$ & 18:29:55.63 & -01:58:24.92 & 0.23 & 12.6 & 7.65 & 0.41 $\pm$ 0.02 & 50.9 & 149 $\pm$ 34 & 0.26 $\pm$ 0.04 & F12 & 0.57 & 3 \\
C8$^{*}$ & 18:30:27.65 & -01:56:24.98 & 0.26 & 13.4 & 7.53 & 0.36 $\pm$ 0.02 & 39.2 & 111 $\pm$ 25 & 0.32 $\pm$ 0.05 & F9 & 0.35 & 2 \\
C9$^{*}$ & 18:29:48.96 & -01:56:04.91 & 0.30 & 13.1 & 7.45 & 0.39 $\pm$ 0.01 & 27.6 & 135 $\pm$ 30 & 0.35 $\pm$ 0.02 & F12 & 0.56 & 3 \\
C10 & 18:29:50.29 & -01:57:24.91 & 0.16 & 13.0 & 7.53 & 0.41 $\pm$ 0.02 & 40.8 & 58 $\pm$ 13 & 0.36 $\pm$ 0.08 & $\cdots$ & $\cdots$ & 0 \\
C11$^{*}$ & 18:31:38.36 & -02:16:44.96 & 0.15 & 17.3 & 6.59 & 0.40 $\pm$ 0.01 & 17.9 & 30 $\pm$ 7 & 0.55 $\pm$ 0.07 & F17 & 0.56 & 1 \\
C12$^{*}$ & 18:30:15.64 & -01:52:44.96 & 0.27 & 13.6 & 6.87 & 0.33 $\pm$ 0.02 & 27.4 & 76 $\pm$ 17 & 0.40 $\pm$ 0.06 & F12 & 0.44 & 4 \\
C13$^{*}$ & 18:32:27.71 & -01:52:24.86 & 0.17 & 14.0 & 7.33 & 0.38 $\pm$ 0.02 & 37.2 & 41 $\pm$ 9 & 0.51 $\pm$ 0.10 & F14c & 0.73 & 1 \\
C14$^{*}$ & 18:29:59.62 & -02:10:04.92 & 0.15 & 14.5 & 6.77 & 0.35 $\pm$ 0.01 & 26.6 & 38 $\pm$ 8 & 0.36 $\pm$ 0.04 & F4 & 0.97 & 1 \\
C15$^{*}$ & 18:30:24.98 & -01:54:04.98 & 0.21 & 13.4 & 7.27 & 0.38 $\pm$ 0.02 & 35.2 & 65 $\pm$ 14 & 0.44 $\pm$ 0.07 & F12 & 0.68 & 3 \\
C16 & 18:29:47.63 & -01:47:24.91 & 0.21 & 12.0 & 7.56 & 0.41 $\pm$ 0.02 & 27.0 & 54 $\pm$ 12 & 0.63 $\pm$ 0.10 & F12 & 0.63 & 0 \\
C17$^{*}$ & 18:30:48.99 & -01:55:45.00 & 0.16 & 14.7 & 8.05 & 0.35 $\pm$ 0.02 & 29.9 & 36 $\pm$ 8 & 0.42 $\pm$ 0.09 & F9 & 0.63 & 3 \\
C18 & 18:31:02.33 & -02:09:25.00 & 0.12 & 16.9 & 6.09 & 0.34 $\pm$ 0.01 & 28.3 & 27 $\pm$ 6 & 0.15 $\pm$ 0.05 & $\cdots$ & $\cdots$ & 0 \\
C19 & 18:30:30.32 & -01:52:24.98 & 0.16 & 13.6 & 6.61 & 0.28 $\pm$ 0.02 & 19.2 & 31 $\pm$ 7 & 0.31 $\pm$ 0.07 & F12 & 0.52 & 0 \\
C20$^{*}$ & 18:31:49.03 & -02:01:24.95 & 0.17 & 19.6 & 7.01 & 0.40 $\pm$ 0.02 & 31.9 & 46 $\pm$ 10 & 0.45 $\pm$ 0.09 & F14a & 0.92 & 1 \\
C21 & 18:31:53.04 & -02:19:04.92 & 0.15 & 18.5 & 7.03 & 0.33 $\pm$ 0.02 & 29.0 & 30 $\pm$ 7 & 0.36 $\pm$ 0.10 & F17 & 0.69 & 0 \\
C22 & 18:29:27.62 & -01:49:24.85 & 0.12 & 13.5 & 7.57 & 0.30 $\pm$ 0.01 & 24.4 & 15 $\pm$ 3 & 0.94 $\pm$ 0.10 & $\cdots$ & $\cdots$ & 0 \\
C23 & 18:29:19.62 & -01:52:04.82 & 0.31 & 13.7 & 6.73 & 0.41 $\pm$ 0.02 & 17.3 & 88 $\pm$ 20 & 0.61 $\pm$ 0.07 & F6 & 0.68 & 0 \\
C24 & 18:30:03.64 & -01:50:04.95 & 0.19 & 13.5 & 7.35 & 0.36 $\pm$ 0.02 & 18.5 & 38 $\pm$ 8 & 0.58 $\pm$ 0.10 & F12 & 0.53 & 0 \\
C25 & 18:30:44.99 & -01:57:05.00 & 0.15 & 14.4 & 7.71 & 0.43 $\pm$ 0.02 & 20.0 & 33 $\pm$ 7 & 0.61 $\pm$ 0.14 & F9 & 0.44 & 0 \\
C26$^{*}$ & 18:31:37.03 & -02:15:24.96 & 0.08 & 18.1 & 6.91 & 0.33 $\pm$ 0.02 & 29.6 & 14 $\pm$ 3 & 1.24 $\pm$ 0.24 & F11 & 1.02 & 2 \\
C27 & 18:30:34.32 & -01:52:24.99 & 0.15 & 13.3 & 7.37 & 0.35 $\pm$ 0.02 & 19.5 & 34 $\pm$ 8 & 0.38 $\pm$ 0.09 & F12 & 1.17 & 0 \\
C28 & 18:30:53.00 & -01:54:45.00 & 0.18 & 13.8 & 8.12 & 0.37 $\pm$ 0.02 & 24.4 & 50 $\pm$ 11 & 0.44 $\pm$ 0.08 & F9 & 1.17 & 0 \\
C29 & 18:30:03.64 & -01:48:44.95 & 0.13 & 13.5 & 7.38 & 0.31 $\pm$ 0.01 & 19.7 & 20 $\pm$ 5 & 0.31 $\pm$ 0.06 & F12 & 1.08 & 0 \\
C30$^{*}$ & 18:29:42.28 & -02:12:44.87 & 0.19 & 14.4 & 7.65 & 0.31 $\pm$ 0.01 & 18.2 & 33 $\pm$ 7 & 0.47 $\pm$ 0.05 & F13 & 0.55 & 1 \\
C31$^{*}$ & 18:32:05.05 & -02:21:04.88 & 0.08 & 16.8 & 7.13 & 0.33 $\pm$ 0.02 & 28.0 & 12 $\pm$ 3 & 1.40 $\pm$ 0.27 & F17 & 0.91 & 1 \\
C32 & 18:32:21.05 & -01:55:44.88 & 0.14 & 14.4 & 7.23 & 0.32 $\pm$ 0.01 & 21.1 & 28 $\pm$ 6 & 0.35 $\pm$ 0.05 & F14c & 0.71 & 0 \\
C33 & 18:32:13.05 & -02:08:24.89 & 0.12 & 17.3 & 7.09 & 0.32 $\pm$ 0.02 & 10.9 & 12 $\pm$ 3 & 1.34 $\pm$ 0.26 & F10b & 0.54 & 0 \\
C34 & 18:29:07.59 & -02:05:24.75 & 0.10 & 14.7 & 6.02 & 0.31 $\pm$ 0.02 & 34.2 & 23 $\pm$ 5 & 0.66 $\pm$ 0.13 & F6 & 4.06 & 0 \\
C35 & 18:30:26.31 & -01:58:04.98 & 0.17 & 14.2 & 7.88 & 0.33 $\pm$ 0.01 & 20.0 & 37 $\pm$ 8 & 0.40 $\pm$ 0.04 & F9 & 1.96 & 0 \\
C36 & 18:30:04.96 & -02:11:44.93 & 0.08 & 14.2 & 6.36 & 0.33 $\pm$ 0.01 & 20.5 & 13 $\pm$ 3 & 1.27 $\pm$ 0.13 & F4 & 1.25 & 0 \\
C37 & 18:29:19.61 & -01:57:44.81 & 0.15 & 14.1 & 7.34 & 0.31 $\pm$ 0.02 & 18.3 & 25 $\pm$ 6 & 0.39 $\pm$ 0.10 & F6 & 0.64 & 0 \\
C38$^{*}$ & 18:32:14.38 & -01:57:04.90 & 0.08 & 14.8 & 7.47 & 0.32 $\pm$ 0.02 & 29.4 & 15 $\pm$ 3 & 1.10 $\pm$ 0.21 & F14a & 2.04 & 1 \\
C39 & 18:32:02.37 & -02:01:24.92 & 0.15 & 16.7 & 7.21 & 0.33 $\pm$ 0.02 & 26.1 & 31 $\pm$ 7 & 0.33 $\pm$ 0.09 & F14a & 2.72 & 0 \\
C40 & 18:32:18.38 & -01:57:04.89 & 0.08 & 14.6 & 7.12 & 0.29 $\pm$ 0.02 & 23.0 & 13 $\pm$ 3 & 1.04 $\pm$ 0.21 & F14c & 1.77 & 0 \\
C41 & 18:29:50.29 & -02:04:24.90 & 0.09 & 14.1 & 7.73 & 0.38 $\pm$ 0.02 & 19.6 & 15 $\pm$ 3 & 1.51 $\pm$ 0.27 & F13 & 1.69 & 0 \\
C42 & 18:31:03.67 & -01:52:25.00 & 0.19 & 14.8 & 7.74 & 0.30 $\pm$ 0.01 & 19.3 & 43 $\pm$ 10 & 0.35 $\pm$ 0.03 & F9 & 1.51 & 0 \\
C43 & 18:31:55.69 & -01:50:04.95 & 0.08 & 14.3 & 7.78 & 0.28 $\pm$ 0.01 & 25.6 & 12 $\pm$ 3 & 1.00 $\pm$ 0.10 & F21 & 1.24 & 0 \\
C44 & 18:31:59.70 & -02:00:04.93 & 0.08 & 16.7 & 6.92 & 0.31 $\pm$ 0.02 & 23.8 & 14 $\pm$ 3 & 1.02 $\pm$ 0.20 & F14a & 1.53 & 0 \\
C45 & 18:28:47.61 & -01:47:44.70 & 0.08 & 11.7 & 6.86 & 0.23 $\pm$ 0.01 & 9.2 & 6 $\pm$ 1 & 1.34 $\pm$ 0.17 & $\cdots$ & $\cdots$ & 0 \\
C46 & 18:29:35.63 & -01:46:44.88 & 0.06 & 11.0 & 7.30 & 0.25 $\pm$ 0.01 & 10.7 & 6 $\pm$ 1 & 1.69 $\pm$ 0.20 & F12 & 2.00 & 0 \\
C47 & 18:32:46.39 & -01:51:44.80 & 0.07 & 14.2 & 7.47 & 0.27 $\pm$ 0.01 & 16.6 & 7 $\pm$ 2 & 1.59 $\pm$ 0.17 & F14c & 0.76 & 0 \\
C48 & 18:29:11.62 & -01:46:44.80 & 0.07 & 11.1 & 6.77 & 0.29 $\pm$ 0.02 & 15.9 & 2 $\pm$ 1 & 5.37 $\pm$ 1.13 & F6 & 1.90 & 0 \\
C49 & 18:30:16.98 & -01:47:04.97 & 0.13 & 13.0 & 7.53 & 0.30 $\pm$ 0.01 & 10.1 & 12 $\pm$ 3 & 0.40 $\pm$ 0.09 & F12 & 2.05 & 0 \\
C50 & 18:29:22.28 & -01:56:04.82 & 0.09 & 14.0 & 6.98 & 0.31 $\pm$ 0.02 & 12.9 & 11 $\pm$ 3 & 1.35 $\pm$ 0.26 & $\cdots$ & $\cdots$ & 0 \\
C51 & 18:31:27.69 & -02:15:44.97 & 0.13 & 18.6 & 6.29 & 0.33 $\pm$ 0.02 & 15.6 & 17 $\pm$ 4 & 0.29 $\pm$ 0.15 & F11 & 1.37 & 0 \\
C52 & 18:31:38.36 & -02:03:04.97 & 0.03 & 24.4 & 7.75 & 0.38 $\pm$ 0.01 & 17.5 & 6 $\pm$ 1 & 3.58 $\pm$ 0.33 & $\cdots$ & $\cdots$ & 0 \\
C53 & 18:30:35.65 & -02:11:24.98 & 0.07 & 16.1 & 6.80 & 0.26 $\pm$ 0.01 & 15.2 & 8 $\pm$ 2 & 1.22 $\pm$ 0.13 & F3 & 2.41 & 0 \\
C54 & 18:30:42.32 & -02:26:24.95 & 0.08 & 12.7 & 7.77 & 0.26 $\pm$ 0.02 & 8.6 & 9 $\pm$ 2 & 1.23 $\pm$ 0.26 & $\cdots$ & $\cdots$ & 0 \\
C55 & 18:31:53.03 & -01:57:24.95 & 0.12 & 17.7 & 7.16 & 0.31 $\pm$ 0.01 & 18.9 & 20 $\pm$ 4 & 0.77 $\pm$ 0.08 & F14b & 2.73 & 0 \\
C56 & 18:31:19.68 & -02:06:44.99 & $\lesssim$ 0.07 & 27.9 & 7.94 & 0.36 $\pm$ 0.02 & 22.1 & 5 $\pm$ 1 & 3.90 $\pm$ 0.72 & $\cdots$ & $\cdots$ & 0 \\
C57 & 18:32:14.38 & -01:59:04.90 & 0.03 & 15.2 & 7.39 & 0.29 $\pm$ 0.02 & 19.8 & 7 $\pm$ 2 & 1.82 $\pm$ 0.36 & F14a & 4.17 & 0 \\
C58 & 18:30:43.66 & -02:05:04.99 & 0.11 & 16.3 & 7.08 & 0.26 $\pm$ 0.01 & 13.7 & 15 $\pm$ 3 & 0.68 $\pm$ 0.07 & F3 & 2.95 & 0 \\
C59$^{*}$ & 18:30:27.64 & -02:10:44.97 & 0.07 & 16.3 & 6.99 & 0.30 $\pm$ 0.02 & 18.9 & 10 $\pm$ 2 & 1.33 $\pm$ 0.27 & F15 & 6.25 & 3 \\
C60 & 18:30:23.64 & -02:07:04.97 & $\lesssim$ 0.07 & 15.4 & 6.88 & 0.27 $\pm$ 0.02 & 13.7 & 4 $\pm$ 1 & 2.92 $\pm$ 0.61 & F3 & 3.30 & 0 \\
C61 & 18:29:06.28 & -01:47:24.78 & 0.06 & 13.1 & 7.28 & 0.29 $\pm$ 0.02 & 13.5 & 7 $\pm$ 2 & 1.99 $\pm$ 0.40 & $\cdots$ & $\cdots$ & 0 \\
C62 & 18:30:12.97 & -01:58:04.96 & $\lesssim$ 0.07 & 14.2 & 7.77 & 0.26 $\pm$ 0.01 & 13.4 & 3 $\pm$ 1 & 3.06 $\pm$ 0.33 & F20 & 5.77 & 0 \\
C63 & 18:31:23.68 & -01:57:44.99 & $\lesssim$ 0.07 & 19.0 & 6.23 & 0.29 $\pm$ 0.01 & 16.3 & 4 $\pm$ 1 & 3.34 $\pm$ 0.33 & F18 & 4.97 & 0 \\
C64 & 18:32:53.05 & -01:49:04.78 & $\lesssim$ 0.07 & 14.3 & 7.43 & 0.24 $\pm$ 0.01 & 9.5 & $\sim$1 & 7.46 $\pm$ 0.84 & $\cdots$ & $\cdots$ & 0 \\
\enddata
\tablecomments{\footnotesize (1) Dense core IDs. Core IDs with asterisk have YSOs. (2) \& (3) Coordinates of peak intensity positions of dense cores in sexagesimal units. (4) Radius of dense cores, derived using Equation \ref{eq:radius}. (5) Average dust temperature, obtained from the Herschel dust temperature map, smoothed to TRAO N$_{2}$H$^{+}$ resolution. (6) \& (7) Velocity centroid and total velocity dispersion, obtained from hyperfine structure fitting to the average N$_{2}$H$^{+}$ spectra. (8) H$_{2}$ column density of dense cores at peak intensity position, obtained from the Herschel H$_{2}$ column density map, smoothed to TRAO N$_{2}$H$^{+}$ resolution. (9) Mass of dense cores. (10) Virial parameter, derived using Equation \ref{eq:virial}. (11) Name of the filaments surrounding the dense cores. (12) Relative abundance of dense cores defined in Equation \ref{eq:rel_ab}. (13) Number of YSOs present in each dense core, identified by \cite{2023ApJS..266...32P}.}
\end{deluxetable*}

\section{Discussions} \label{sec:discussions}
\subsection{Comparison of C$^{\textit{18}}$O Filaments with Herschel Continuum Filaments}
\begin{figure*}[!htb]
    \centering
    \includegraphics[width=0.8\textwidth]{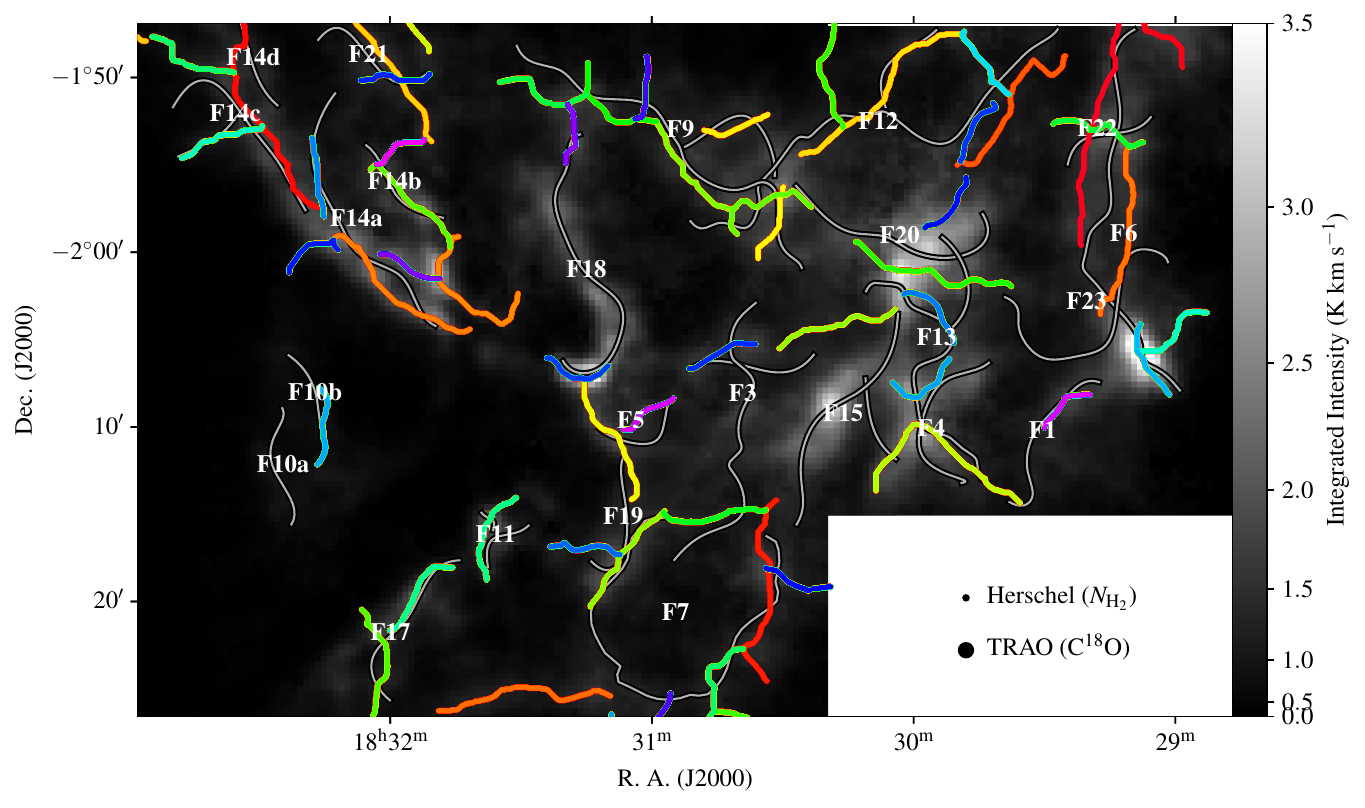}
    \caption{The velocity-coherent filaments (this study) vs. the Herschel continuum filaments \citep{2015A&A...584A..91K}. Gray lines show the skeletons and branches of the velocity-coherent filaments, while the colored lines trace the Herschel continuum filament skeletons. The beam sizes of Herschel H$_{2}$ column density and TRAO $\rm C^{18}O$ maps are presented with black dots at the lower-right corner of the figure.}
    \label{fig:Herschel}
\end{figure*}

Using the continuum data obtained by the Herschel Gould Belt Survey, \cite{2015A&A...584A..91K} identified filament networks in the Aquila region from the H$_{2}$ column density map. Figure \ref{fig:Herschel} shows the comparison between these Herschel continuum filaments and the velocity-coherent filaments identified in this study. Although both the identification methods and the data are different, with significant differences in spatial resolution, they are found to be coincident, especially in high-intensity regions.

Figure \ref{fig:Herschel} shows that the Herschel filaments exist mostly within regions where the identified velocity-coherent filaments are defined. However, several filaments that appear as single structures in the Herschel data exhibit multiple velocity components indicative of being complex and thus are separated into a number of velocity-coherent filaments in this study (e.g., the regions of filaments F4, F6, F9, F12, F13, F20, F22, and F23). From the analysis of the H$_2$ column densities of individual velocity-coherent filaments (see Section \ref{sec:column}), identified through Gaussian decomposition of the C$^{18}$O spectra, we find that in regions where multiple components spatially overlap, the H$_2$ column densities inferred from Herschel dust emission are, on average, overestimated by $\sim$55\% relative to those derived for the individual decomposed components. Similar comparisons have been done by \cite{2021ApJ...919....3C} in the IC 5146 molecular cloud and by \cite{2023ApJ...957...94Y}
in the Orion B molecular cloud, finding similar results. These results highlight the critical importance of incorporating velocity information for the identification of velocity-coherent structures and for assessing their dynamical state.

However, note that there are several cases in which a single velocity-coherent filament is found to contain multiple Herschel filaments (F9, F14a, F14c, etc.). This can be attributed to the coarser spatial resolution of our C$^{18}$O data (49$^{\prime\prime}$) compared to that of Herschel (18.2$^{\prime\prime}$).

\subsection{Kinematic Information of Filaments and Dense Cores}  \label{kinematics}
We determined which dense core is related to which filament by comparing the velocity centroid of the filament from the decomposed C$^{18}$O spectra and the velocity centroid of the dense core from hyperfine structure fitting to the average N$_{2}$H$^{+}$ spectra. For the dense cores that share positional coincidence with more than one filament, we assign the core to that filament only that has the closest velocity centroid value (see Figure \ref{fig:fil_core_spectra}). In this way, we define two groups of filaments, such as `filaments with N$_{2}$H$^{+}$ dense cores' and `filaments without N$_{2}$H$^{+}$ dense cores.' Note that, for several dense cores, we could not find any surrounding C$^{18}$O filaments (e.g., C18, C22, C45, C50, C52, C54, C61, C64). In most of these regions, the C$^{18}$O emission is weak and was not detected above the $5\sigma_{\rm RMS\_I}$ threshold level and was thus excluded during the FoF algorithm processing. Also, in some cases, we could not find any significant velocity coherent filamentary structure with area $\geq5\times\theta_{\rm FWHM}$ while running the FoF algorithm.

\subsubsection{Systematic Velocities}
\begin{figure}[!htb]
    \centering
    \includegraphics[width=0.47\textwidth]{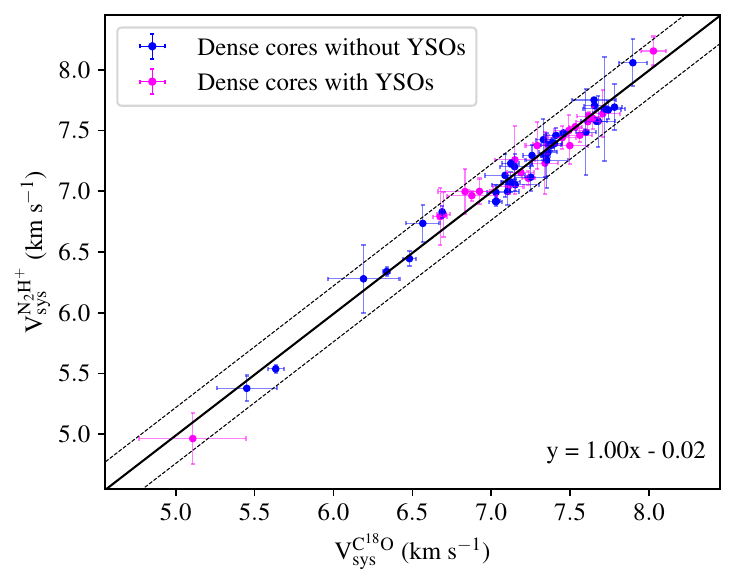}
    \caption{Systematic velocities of dense cores measured from hyperfine fitting to the N$_{2}$H$^{+}$ spectra in comparison with those of surrounding filaments obtained from the Gaussian decomposition of C$^{18}$O spectra. Magenta and blue colors represent dense cores with and without YSOs. The black solid line represents the least-squares fit of the data points, and the dashed lines show the range from the fit result by the sound speed of 0.23 km s$^{-1}$, derived from the median dust temperature of 15 K.}
    \label{fig:sys}
\end{figure}

The systematic velocities of the filaments are derived from the decomposed C$^{18}$O spectra corresponding to the pixel positions overlapping with the dense cores. A core-to-envelope velocity difference less than the speed of sound may indicate that both the core and the surrounding filament are kinematically connected. Figure \ref{fig:sys} shows that the systematic velocities of dense cores (see Section \ref{sec:core_properties}) and surrounding filaments have a tight correlation with a difference less than 0.23 km s$^{-1}$ (speed of sound at median dust temperature of 15 K). This similarity suggests that the dense cores and the surrounding filaments move together with a close kinematical relation between the two objects. Also note that, we could not find any difference in this analysis for dense cores with (magenta dots) and without (blue dots) YSOs. 

\subsubsection{Velocity Variation along the Filaments and the Embedded Dense Cores}
To understand how the velocity changes along the filaments, particularly near the embedded dense cores, it is important to examine the distribution of velocity centroid values obtained from the decomposed C$^{18}$O spectra as a function of distance along the filament's skeleton. The distance of each pixel position is measured as a position offset in parsecs along the filament's skeleton starting from the zero offset position specified for that filament in Figure \ref{fig:filvel} (dot mark over the skeleton). Note that there are multiple pixel positions perpendicular to the skeleton, in which case we assume that all these pixel positions share the same offset value. In this way the distribution of velocity centroid values (green dots) along the skeleton of filament F9 (taken as an example) is shown in the right-middle panel of Figure \ref{fig:analysis}. The distribution of integrated intensity along the filament skeleton is also shown in the right-top panel of Figure \ref{fig:analysis}. The centroid positions of the embedded dense cores are indicated by vertical magenta lines, while the magenta dots represent their average centroid velocities, as listed in Table \ref{tab:cores}. A comparison between the average centroid velocity of the cores and the surrounding filament's velocity centroid values confirms the tight correlation discussed in the previous section.

In the case of F9, there is a change of systematic velocity along the filament with an end-to-end velocity gradient of $\sim$0.2 km s$^{-1}$ pc$^{-1}$, with local fluctuations at the sub-parsec scale near the centroid position of embedded dense cores, indicative of longitudinal gas flow motions along the filaments. To quantify the gas flow rate, we determined local velocity gradients ($\nabla V_{\parallel}$) in the vicinity\footnote{Note that the offset range used to derive the velocity gradient was determined by visually inspecting the velocity variations in the immediate vicinity of each dense core. Hence, the derived values might be slightly underestimated or overestimated accordingly.} of core centroids by performing least-squares fits to the velocity distributions (black lines in the right-middle panel of Figure \ref{fig:analysis}). For this purpose, velocity measurements within a radial distance of 0.05 pc from the filament skeleton were included in the fits. Figure \ref{fig:velocity} (Appendix \ref{app:velocity}) presents the velocity variations along the skeletons of the remaining filaments containing dense cores, with fitted gradients indicated in black solid lines. The derived $\nabla V_{\parallel}$ values, corresponding to the slopes of the linear fits, are listed in Table \ref{tab:vel_grad} (Appendix \ref{app:velocity}) and span from 0.2 km s$^{-1}$ pc$^{-1}$ (C1f; F13) to 1.5 km s$^{-1}$ pc$^{-1}$ (C42; F9), with a median of 0.6 km s$^{-1}$ pc$^{-1}$. The uncertainties in $\nabla V_{\parallel}$ were obtained from the least-squares fitting procedure. However, it should be noted that all velocity gradients were derived under the assumption of an inclination angle of $45^{\circ}$; within a plausible inclination range of $30^{\circ}$--$60^{\circ}$, the inferred gradients would vary by a factor of approximately 1.7 to 0.6.

A number of previous studies have reported the presence of a $\lambda/4$ shift between the sinusoidal oscillations of density (traced by integrated intensity here) and velocity distributions \citep[e.g.,][etc.]{2011A&A...533A..34H,2022ApJ...940..112K}, which is interpreted as the evidence of gravitational fragmentation. In our case, however, we do not observe any clear $\lambda/4$ shift, possibly due to the coarse spatial resolution of our data. Nevertheless, in certain cases, such as cores C28 and C8 in Figure \ref{fig:analysis}, and several others shown in Figure \ref{fig:velocity} (Appendix \ref{app:velocity}), the velocity centroids along the filament skeleton exhibit a local minimum near the core center. Such a feature can arise when a core lies closer to the observer than the filament and gas is flowing toward the core.

\subsubsection{Mass Flow Rates along the Filaments}
Over the past decade, both single-dish and interferometric observations have consistently revealed velocity gradients along filaments in a variety of molecular clouds spanning different physical and evolutionary conditions \citep[see, e.g.,][]{2013ApJ...766..115K,2017A&A...606A.123H,2019A&A...629A..81T,2019ApJ...877..114C,2021ApJ...919....3C,2022MNRAS.514.6038Z,2022ApJ...940..112K,2025A&A...696A.202S}. These velocity gradients serve as quantitative tracers of the longitudinal mass flow rate along the filaments, thereby providing a direct measure of the dynamical accretion processes feeding the dense cores.

The mass flow rates ($\dot{M}_{\parallel}$) corresponding to the derived velocity gradients can be estimated by the relation given by \cite{2013ApJ...766..115K} for axial flow motions with the assumption of a simple cylindrical model as folows:
\begin{equation} \label{eq:mass_flow}
    \dot{M_{\parallel}} = \frac{\nabla V_{\parallel}M_{\rm fil}}{\tan \alpha},
\end{equation}

where $M_{\rm fil}$ denotes the filament mass within the range in length where the velocity gradient ($\nabla V_{\parallel}$) is measured, and $\alpha$ is the inclination angle between the filament axis and the plane of the sky. For simplicity, we adopted $\alpha = 45^{\circ}$ for all calculations. The resulting values of $\dot{M}_{\parallel}$, with uncertainties propagated from those of $M_{\rm fil}$ and $\nabla V_{\parallel}$, are listed in Table \ref{tab:vel_grad} (Appendix \ref{app:velocity}). Across the filaments, the average mass flow rates range from 7 M$_{\odot}$ Myr$^{-1}$ (F11) to 178 M$_{\odot}$ Myr$^{-1}$ (F20), with a median of 35 M$_{\odot}$ Myr$^{-1}$. Notably, filament F15, which roughly corresponds to the southern filament of Serpens South as defined by \cite{2013ApJ...766..115K}, has a flow rate of $\sim 52$ M$_{\odot}$ Myr$^{-1}$, which is in excellent agreement with the reported value of $\sim 49$ M$_{\odot}$ Myr$^{-1}$ (re-scaled to our adopted distance of 455 pc) based on N$_2$H$^{+}$ (1--0) observations \citep{2013ApJ...766..115K}. A regional comparison further indicates that the median $\dot{M}_{\parallel}$ in Serpens South is about 40\% higher than in W40, consistent with the enhanced mass flows expected in hub-filament systems.

\subsubsection{Variation of Non-thermal Velocity Dispersion}
The non-thermal velocity dispersions ($\sigma_{\rm nth}$) at each pixel position for both the filaments and the embedded N$_{2}$H$^{+}$ dense cores were calculated as described in Section \ref{crit_mass}. The Mach numbers ($\mathcal{M}$), defined as the ratio between $\sigma_{\rm nth}$ and $c_{s}$ (sound speed), provide a measure of the level of turbulence relative to thermal motions within the gas. Motions are classified as subsonic when $\mathcal{M} < 1$, and as supersonic when $\mathcal{M} > 1$ \citep{2011A&A...533A..34H}. If $\mathcal{M}$ lies within the range $1 \pm \delta\mathcal{M}$, the motions are considered transonic, where $\delta\mathcal{M}$ is the uncertainty in $\mathcal{M}$, usually propagated from uncertainty in the adopted gas temperature and uncertainty in the observed velocity dispersion ($\sigma_{\rm obs}$).

To derive $\delta\mathcal{M}$, we adopted an uncertainty of 2 K corresponding to our adopted Herschel $T_{\rm dust}$ as suggested by \cite{2015A&A...584A..91K}. Since, $\sigma_{\rm obs}$ corresponds to the velocity dispersion of the decomposed components obtained by Gaussian decomposition of the C$^{18}$O spectra, we were unsure about the uncertainty for each individual component in case of multiple components along the line-of-sight. Hence, we selected only those pixels where a single Gaussian component was identified and used an average uncertainty of 0.03 km s$^{-1}$ in $\sigma_{\rm obs}$, derived from the diagonal elements of the covariance matrices of the respective fits. In this way, the average propagated uncertainty in $\mathcal{M}$ was obtained to be 0.17.

\begin{figure*}[!htb]
    \centering
    \includegraphics[width=\textwidth]{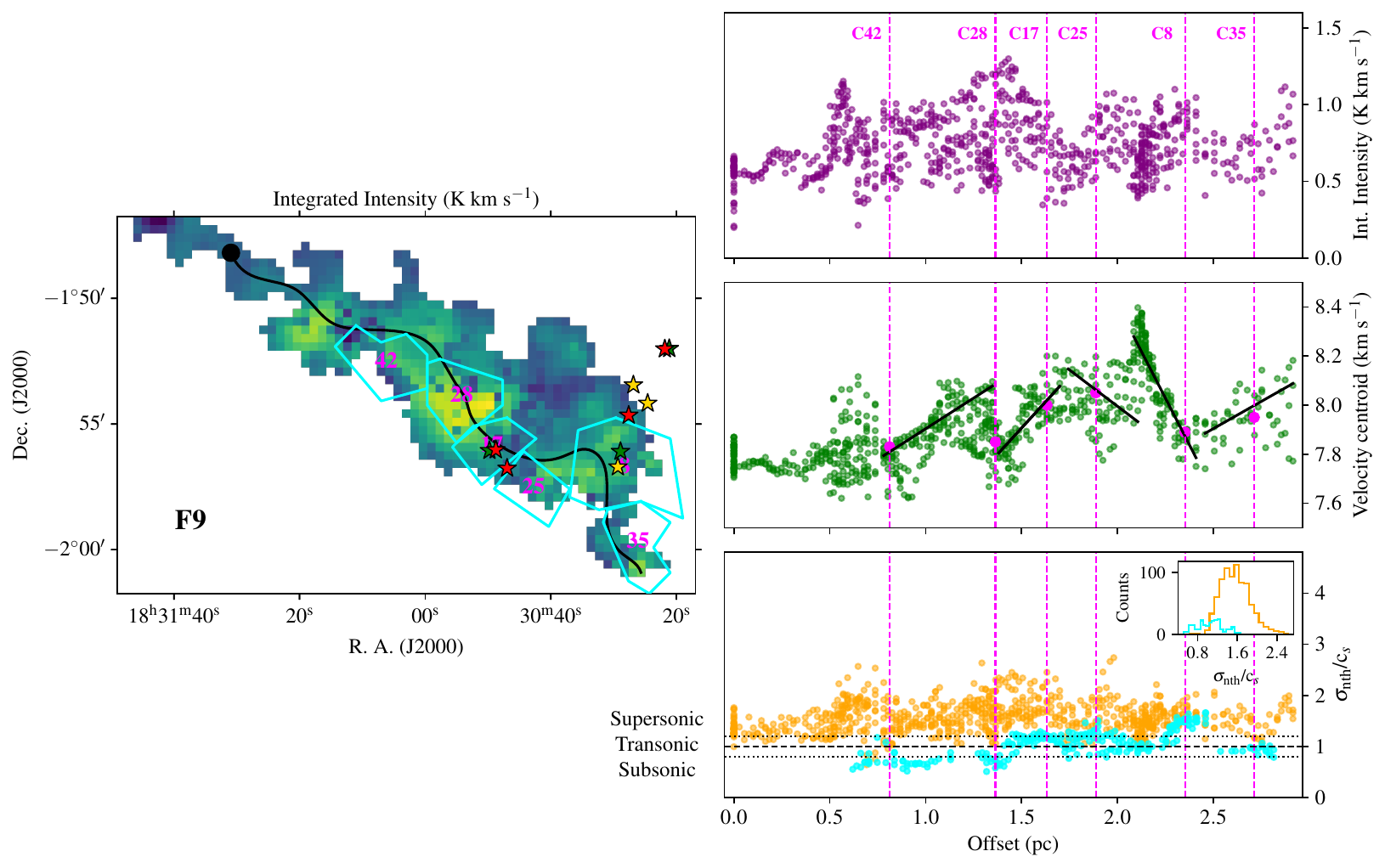}
    \caption{Left panel: C$^{18}$O integrated intensity map of filament F9. The black solid line is the skeleton of the filament obtained from \texttt{FilFinder}. The black dot over the skeleton marks the 0 offset position along the skeleton. The cyan polygons are the dense cores embedded in the filament with core IDs labeled in magenta color. The green, red, and yellow colored stars denote the positions of class 0, I, and flat-spectrum YSOs, respectively, identified by \cite{2023ApJS..266...32P}. Right panel: Distribution of filament's integrated intensity (purple; top panel), velocity centroid (green; middle panel), and Mach numbers ($\sigma_{\rm nth}$/$c_{s}$) (orange; lower panel) along the filament skeleton. The core centroid positions are marked with magenta vertical lines in all three panels. The cores' average centroid velocities as listed in Table \ref{tab:cores} are marked in magenta dots along the magenta-colored vertical lines in the middle panel. The least-squares fit to the velocity distribution near each core position is also presented with black solid lines in the middle panel. The cores' Mach numbers at each pixel position are presented with cyan-colored points in the lower panel. The horizontal dashed line ($\sigma_{\rm nth}$/$c_{s}=1$) in the lower panel marks the boundary between subsonic and supersonic motions, while the two horizontal dotted lines provide the range for transonic motions. The inset figure in the third panel shows the histograms for Mach number distribution for both filament and dense cores.}
    \label{fig:analysis}
\end{figure*}

The right-lower panel of Figure \ref{fig:analysis} presents the distribution of $\mathcal{M}$ along the skeleton of filament F9 (orange dots) and the embedded dense cores (cyan dots). This distribution reveals that the $\mathcal{M}$ for the filament are predominantly $>1$, indicating that the motions within the filament are mostly supersonic. In contrast, $\mathcal{M}$ for the dense cores embedded within the filament are $\lesssim$ 1 and primarily within the transonic limit (except for core C8), suggesting that these cores are primarily subsonic or transonic. Figure \ref{fig:mach} (Appendix \ref{app:mach}) shows the variation of Mach numbers for all the filaments with their corresponding N$_{2}$H$^{+}$ dense cores. A similar trend to that of F9 is observed for the rest of the filaments as well as their embedded dense cores. Considering the insignificant velocity shift between the filaments and the embedded dense cores and the lower Mach numbers in dense cores than those in the surrounding filaments, turbulence dissipation may have been processed to form the dense cores in the respective filaments. 

However, there are few exceptions where the $\mathcal{M}$ of dense cores are $>1$, particularly those located near the hub-filamentary region of Serpens South (filaments F13, F15, and F20) and in the F12 filament. These dense cores are located in a very complex location with a large number of YSOs with possible infall and outflow motions that may have effectively caused spectral broadening, leading to supersonic Mach numbers.

\subsection{Gravitational Stability in Filaments}
\begin{figure*}[!htb]
    \centering
    \includegraphics[width=\textwidth]{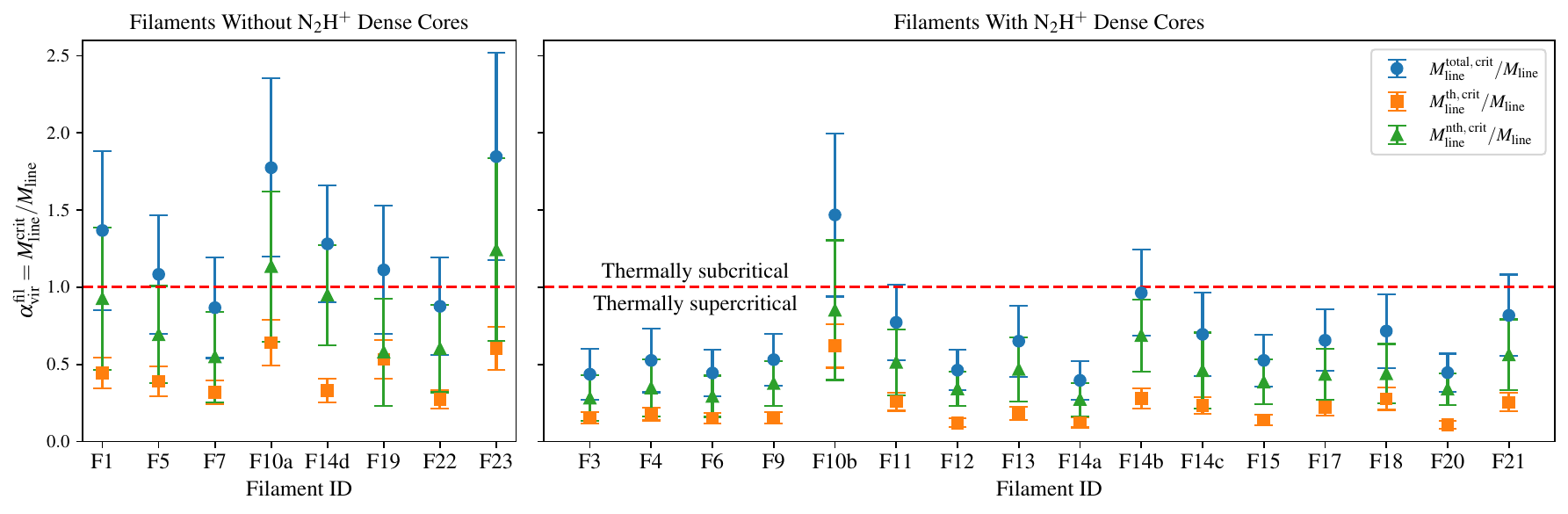}
    \caption{Virial parameters for the filaments without (left panel) and with (right panel) dense cores. The virial parameters are derived considering thermal support only (orange boxes), non-thermal support only (green triangles), and both thermal plus non-thermal support (blue circles), respectively. The horizontal red-dashed line represents $\alpha_{\rm vir}^{\rm fil}=1$.}
    \label{fig:alpha_filament}
\end{figure*}

To understand how dense cores form within the filaments, it is important to analyze the gravitational stability of the filaments. Here, we have tried to look at the gravitational stability of the filaments by analyzing the relationship between their line mass and critical line mass (see Section \ref{crit_mass}). The virial parameter for filaments is defined as the ratio of the total critical line mass to the line mass ($\alpha_{\rm vir}^{\rm fil}=M_{\rm line}^{\rm tot,\: crit}/M_{\rm line}$). We have also calculated the ratio of thermal and non-thermal critical line masses ($M_{\rm line}^{\rm th,\:crit}$ and $M_{\rm line}^{\rm nth,\:crit}$, respectively) to $M_{\rm line}$ to check for their individual contributions. For filaments with and without N$_{2}$H$^{+}$ dense cores defined in Section \ref{kinematics}, Figure \ref{fig:alpha_filament} presents $\alpha_{\rm vir}^{\rm fil}$ (blue circles) as well as $M_{\rm line}^{\rm th,\:crit}/M_{\rm line}$ (orange boxes) and $M_{\rm line}^{\rm nth,\:crit}/M_{\rm line}$ (green triangles) values for each filament. The estimated uncertainties in these ratios are calculated from the propagation of uncertainties of observed velocity dispersion and line mass.

The critical value $\alpha_{\rm vir}^{\rm fil}=2$ is the lowest critical virial parameter proposed for non-magnetized clouds (\citealt{2013ApJ...779..185K, 2013A&A...553A.119A}), which suggests that filaments having $\alpha_{\rm vir}^{\rm fil}\gg2$ are gravitationally unbound, whereas filaments with $\alpha_{\rm vir}^{\rm fil}\simeq2\pm1$ are gravitationally bound and stable (e.g., the solar system), and filaments with $\alpha_{\rm vir}^{\rm fil}\ll2$ are gravitationally bound but unstable. Similarly, filaments with $M_{\rm line}$ $<$ $M_{\rm line}^{\rm th,\:crit}$ are treated as thermally subcritical, where the structure is gravitationally stable as thermal kinetic energy is larger than the gravitational potential energy and provides support against gravitational collapse. Filaments with  $M_{\rm line}$ $>$ $M_{\rm line}^{\rm th,\:crit}$ are treated as thermally supercritical, where the structure is gravitationally unstable and either under radial collapse or fragmentation.

A close inspection of Figure \ref{fig:alpha_filament} indicates that all the filaments, irrespective of the presence of N$_{2}$H$^{+}$ dense core, have $\alpha_{\rm vir}^{\rm fil}<2$, and the ratio $M_{\rm line}^{\rm th,\:crit}/M_{\rm line}<1$, suggesting that all the filaments are gravitationally bound and thermally supercritical. However, filaments having dense cores show an overall trend of lesser $\alpha_{\rm vir}^{\rm fil}$ and $M_{\rm line}^{\rm th,\:crit}/M_{\rm line}$ ratios compared to those without dense cores, suggesting that these filaments are more gravitationally bound and thermally supercritical, leading to fragmentation and thus formation of dense cores. The gravitationally bound and thermally supercritical status of the filaments without dense cores suggests that these filaments may have an environment potentially capable of forming dense cores. Note that $M_{\rm line}^{\rm nth,\:crit}/M_{\rm line}^{\rm th,\: crit}$ $\sim$2 across the filaments. This suggests that the effects of non-thermal motions are significant throughout the entire molecular cloud. Given the large YSO population in the region, feedback from protostellar outflows represents a plausible source of this turbulence. In addition, large-scale gas flows along the filaments are likely to further sustain and enhance these non-thermal motions.

\begin{figure}[!htb]
    \centering
    \includegraphics[width=0.47\textwidth]{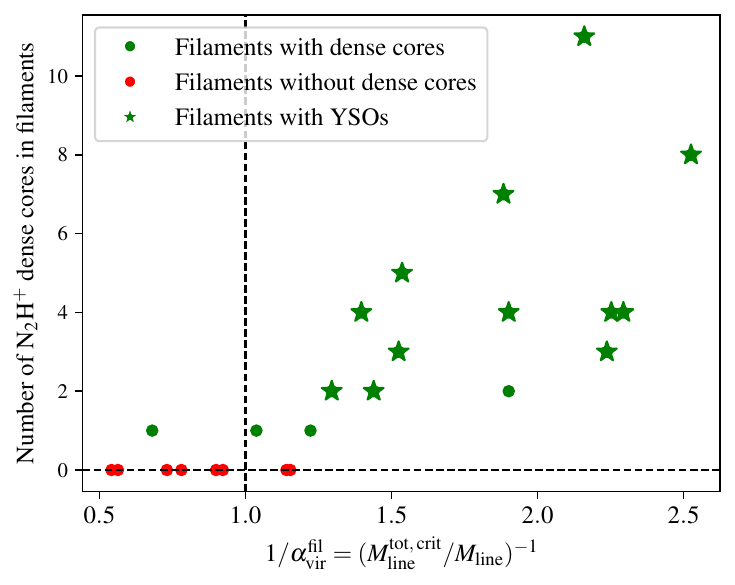}
    \caption{Number of dense cores in each filament as a function of the ratio between $M_{\rm line}$ and $M_{\rm line}^{\rm tot,\:crit}$. The vertical black-dashed line shows the location of $M_{\rm line}/M_{\rm line}^{\rm tot,\:crit}=1$ and the horizontal black-dashed line shows the location where filaments do not contain any dense core.}
    \label{fig:line_to_th}
\end{figure}

Figure \ref{fig:line_to_th} shows the number of dense cores per filament as a function of $1/\alpha_{\rm vir}^{\rm fil}$, i.e., the ratio $M_{\rm line}/M_{\rm line}^{\rm tot,\:crit}$. From the figure, it can be seen that filaments with larger $M_{\rm line}/M_{\rm line}^{\rm tot,\:crit}$ values tend to contain a greater number of dense cores, whereas those with relatively smaller ratios host fewer or none. This trend supports the interpretation that a higher degree of criticality enhances a filament’s ability to fragment and produce multiple dense cores. Similar findings have also been reported by \cite{2015A&A...574A.104T} and \cite{2023ApJ...957...94Y}, underscoring the importance of filament criticality in governing fragmentation and core formation.

\subsection{Virial State of the Dense Cores}
In order to evaluate the stability of the identified dense cores, we derived their virial parameter using the equation given by \cite{1992ApJ...395..140B} as follows:
\begin{equation} \label{eq:virial}
    \alpha_{\rm vir}^{\rm core} = \frac{5R_{\rm core}}{GM_{\rm core}}\sigma^{2}_{\rm tot,\: N_{2}H^{+}},
\end{equation}
where $R_{\rm core}$ is the deconvolved radius of the core obtained using Equation \ref{eq:radius}, $M_{\rm core}$ is the mass of the cores, and $\sigma^{2}_{\rm tot,\: N_{2}H^{+}}$ is the total velocity dispersion estimated from Equation \ref{eq:tot_disp}, using the non-thermal contributions from the averaged $\rm N_{2}H^{+}$ line profile over the dense cores.

A system is said to be in virial equilibrium if its kinetic energy is comparable to its gravitational potential energy. For a non-magnetic sphere with uniform density, the equilibrium between gravitational and kinetic energy approaches the critical value of $\alpha_{\rm vir}^{\rm core} = 2$ \citep{2007ApJ...657..870V,2011MNRAS.411...65B,2022MNRAS.515.2822R}. When $\alpha_{\rm vir}^{\rm core} < 2$, gravitational forces dominate over kinetic energy, indicating that the dense core is gravitationally unstable and may collapse. Conversely, for cores with $\alpha_{\rm vir}^{\rm core} > 2$, kinetic energy exceeds gravitational energy, suggesting that such cores are unlikely to undergo collapse and form stars. 

\begin{figure}[!htb]
    \centering
    \includegraphics[width=0.47\textwidth]{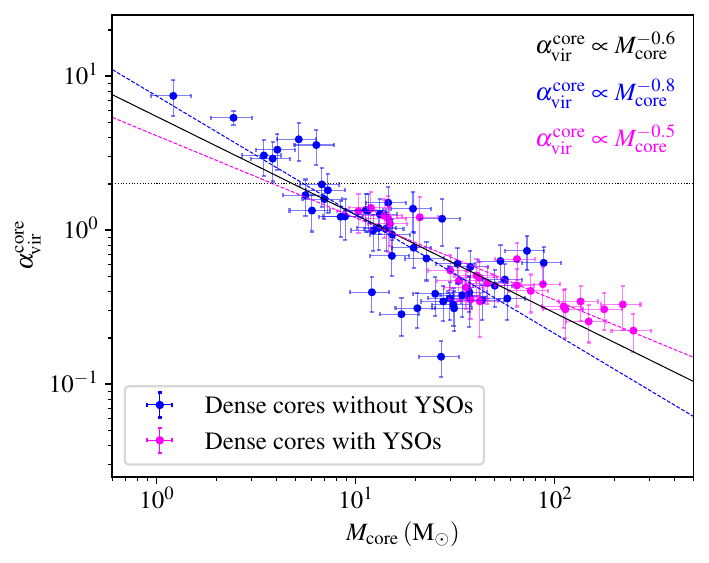}
    \caption{Virial parameters for dense cores with (magenta) and without (blue) YSOs. The horizontal black-dotted line represents $\alpha_{\rm vir}^{\rm core}=2$. The black solid line represents the least-squares fit to all data points, corresponding to a power-law index of -0.6. The magenta and blue dashed lines indicate the least-squares fits to data points corresponding to dense cores with and without YSOs, with power-law indices of -0.8 and -0.5, respectively.}
    \label{fig:alpha_core}
\end{figure}

In our analysis, we found that almost 90\% of dense cores have $\alpha_{\rm vir}^{\rm core} < 2$, with all the dense cores with YSOs falling within this limit, indicating their gravitationally bound status. Those small number of cores with $\alpha_{\rm vir}^{\rm core} > 2$ are not gravitationally bound and may not form stars in the future. The derived virial parameters of dense cores ($\alpha_{\rm vir}^{\rm core}$) are plotted against the core mass ($M_{\rm core}$) in Figure \ref{fig:alpha_core}, showing that $\alpha_{\rm vir}^{\rm core}$ is inversely proportional to $M_{\rm core}$ with a power-law index of -0.6 for our observed region, suggesting that massive cores tend to be gravitationally unstable. Also, note that dense cores without YSOs have a smaller power-law index of -0.8, compared to that of dense cores with YSOs (-0.5). This smaller index suggests that the gravitational boundedness of dense cores without YSOs increases more rapidly with mass, indicating that more massive ones may be close to forming stars. In the case of dense cores containing YSOs, the feedback effects from the YSOs may increase the non-thermal linewidths of the N$_{2}$H$^{+}$ emission. As $\alpha_{\rm vir}^{\rm core} \propto \sigma^{2}_{\rm tot,\:N_{2}H^{+}}$, this feedback likely results in larger $\alpha^{\rm core}_{\rm vir}$ values, thereby producing a shallower slope.

However, it should be noted that the derived virial parameters in this study have an average uncertainty of $\sim30\%$ and several dense cores are not fully resolved due to the coarse spatial resolution of TRAO ($\theta_{\rm FWHM}=52^{\prime\prime}$). Therefore, high-resolution observations are essential for a more detailed analysis of the core properties using virial analysis.

\subsection{Infall Signature in Dense Cores}
To investigate the evolutionary status of the identified dense cores, we examined the presence of inward gas motions toward their central regions. Such motions are crucial for core evolution, as they facilitate the accumulation of gaseous material within the core. This requires selecting an appropriate molecular tracer--one that is optically thick enough to distinguish infalling gas on the far side of the core from that moving inward on the near side (the characteristic `blue asymmetry' profile), while also probing the densest inner regions with minimal contamination from chemical depletion. Also, to ensure that the observed asymmetric double-peaked line profiles are not influenced by multiple velocity components along the line-of-sight, it is necessary to compare the optically thick line with an optically thin counterpart. CS (2--1) and HCO$^{+}$ (1--0), two optically thick molecular lines and well-known infall tracers \citep[e.g.,][etc.]{1999ApJ...526..788L,2001ApJS..136..703L,2005A&A...442..949F,2009MNRAS.392..170S}, and H$^{13}$CO$^{+}$ (1--0) line, the optically thin isotopologue of HCO$^{+}$, are collectively chosen to identify these blue asymmetries over the dense cores. 

\begin{figure*}[!htb]
    \centering
    \includegraphics[width=\textwidth]{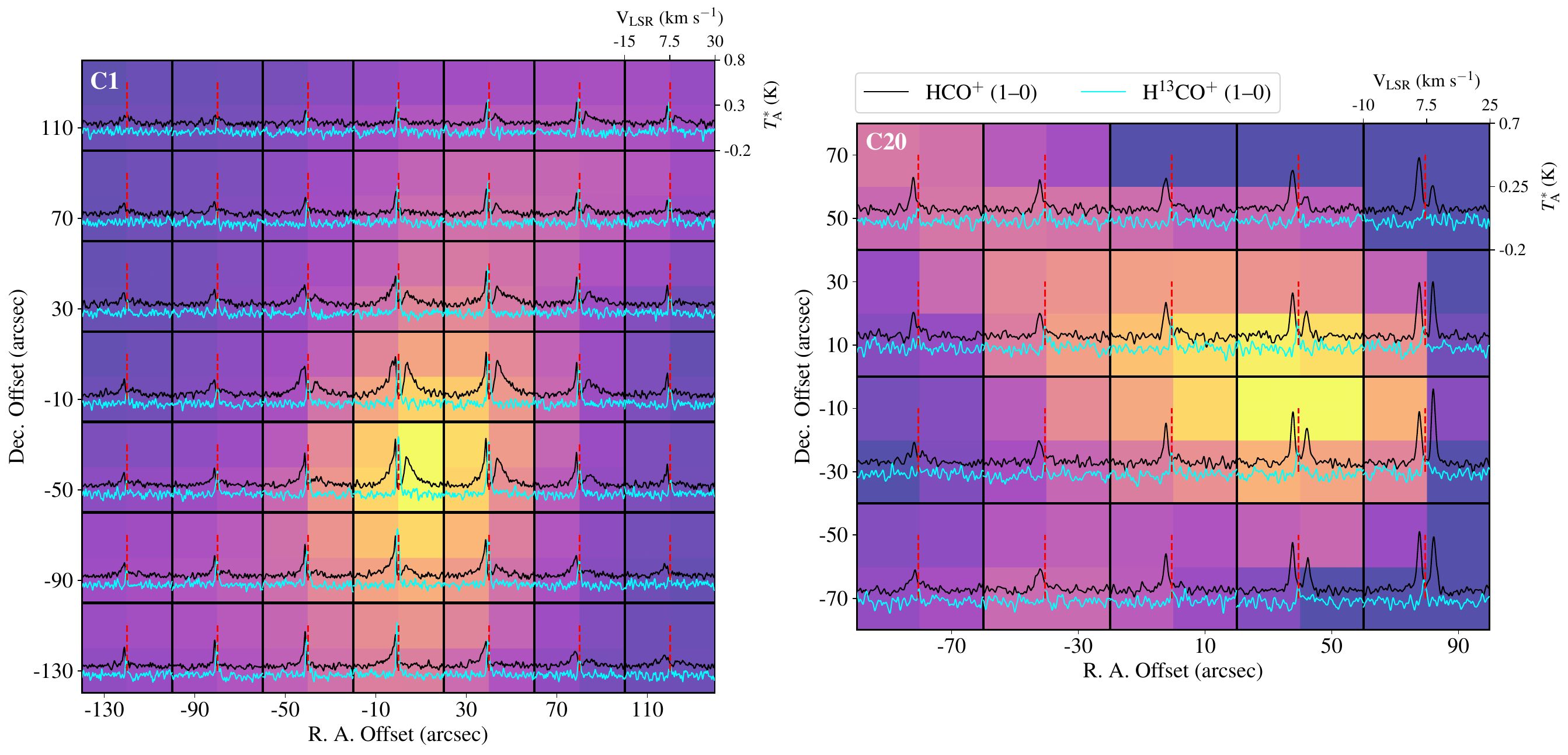}
    \caption{Infall asymmetry towards the central regions of dense cores C1 (left) and C20 (right) as seen from the line profile of optically thick HCO$^{+}$ (black). The spectra of optically thin H$^{13}$CO$^{+}$ (cyan) is also shown to check for the presence of any multiple velocity components along the line-of-sight. The dashed red vertical line marks the systematic velocity of the dense cores obtained from hyperfine fitting of N$_{2}$H$^{+}$ spectra. The color tones in the background indicate N$_{2}$H$^{+}$ integrated intensity. Each spectrum is the average of the spectra for the 2$\times$2 pixels shown in the background of each spectrum. HCO$^{+}$ spectra are shown with an offset of 0.2 K in the $y$-axis. The (0, 0) offset positions refer to (18:30:04, -2:02:36) and (18:31:52, -2:01:36) for C1 and C20, respectively.}
    \label{fig:infall}
\end{figure*}

Figure \ref{fig:infall} shows HCO$^{+}$ and H$^{13}$CO$^{+}$ spectra over the central regions of two dense cores, C1 and C20, with N$_{2}$H$^{+}$ integrated intensity map as the background. Blue asymmetry in HCO$^{+}$ profiles can be clearly seen over both the dense cores, while H$^{13}$CO$^{+}$ spectra peak right at the dip between blue and red HCO$^{+}$ profiles, which coincides with the systematic velocity obtained from hyperfine structure fitting of N$_{2}$H$^{+}$ spectra. Figures \ref{fig:spectral_profiles1}, \ref{fig:spectral_profiles2}, \ref{fig:spectral_profiles3}, and \ref{fig:spectral_profiles4} (Appendix \ref{app:profiles}) present the spectral profiles of various species averaged over 30\% of the peak intensity contour of each of the identified dense cores. A close inspection of the spectral profiles of CS, HCO$^{+}$ and H$^{13}$CO$^{+}$ from these figures reveals that $\sim60\%$ of the identified dense cores exhibit this characteristic blue asymmetry profiles in CS and $\sim45\%$ in HCO$^{+}$, indicating the ongoing inward gas motions toward their central regions. Note that almost all cores that show blue asymmetry in HCO$^{+}$ also display a similar feature in CS. However, in a few cases, cores with blue profiles in CS do not show reliable evidence in HCO$^{+}$, primarily because HCO$^{+}$ emission was not detected above the 3$\sigma_{\rm RMS\_I}$ level in those cores. Furthermore, we also noticed that $\sim80\%$ of the dense cores that contain YSOs also showed this blue asymmetry. Notably, all of these identified dense cores with blue asymmetry have virial parameter $\lesssim$ 1 suggesting the dominance of gravitational energy in these cores.

\begin{figure}
    \centering
    \includegraphics[width=0.47\textwidth]{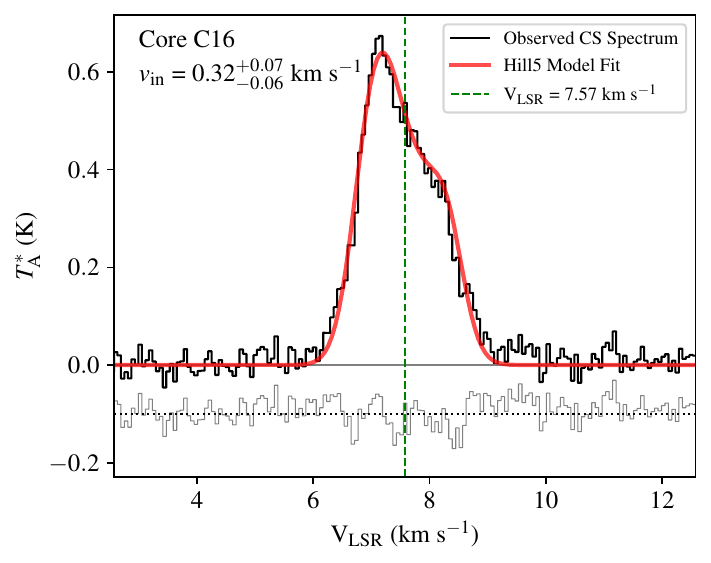}
    \caption{Hill5 model fit (red line) to the CS (2--1) spectrum (black line) of core C16 averaged over the 30\% peak intensity contour. The systemic velocity (V$_{\rm LSR}$) of the core, derived from the hyperfine fitting of the N$_2$H$^+$ (1--0) line, is indicated by the vertical green dashed line. The derived infall velocity is annotated in the top-left corner of the panel. The residuals are plotted with a vertical offset of -0.1 K along the $y$-axis for clarity.}
    \label{fig:C16_infall}
\end{figure}

To quantitatively analyze the infall motions within the dense cores, we selected a number of dense cores whose averaged CS (2–1) spectra\footnote{Note that we found CS to be brighter and less contaminated by outflow activities than HCO$^{+}$. Hence, we chose the CS line only to derive the infall speed.}, extracted within the 30\% peak intensity contour, exhibit SNR $\gtrsim7$. The 30\% level contour was adopted as it effectively includes the innermost region of the dense cores and contains a sufficient number of blue-asymmetric profiles to obtain an average profile with an adequate SNR. The blue-asymmetric CS profiles were fitted using the two-layer radiative transfer model, Hill5 \citep{2005ApJ...620..800D}, to derive the infall velocity ($v_{\rm in}$). The fitting was performed through a Markov Chain Monte Carlo (MCMC) approach using the \texttt{emcee}\footnote{\href{https://emcee.readthedocs.io/en/stable/}{https://emcee.readthedocs.io/en/stable/}} Python package \citep{2013PASP..125..306F}. Figure \ref{fig:C16_infall} shows the best-fit Hill5 profile over the averaged CS spectrum for core C16. The derived $v_{\rm in}$ values range from 0.11 km s$^{-1}$ (C52) to 0.73 km s$^{-1}$ (C6), with a median value of 0.42 km s$^{-1}$. The working principle of the Hill5 model is described in Appendix \ref{app:infall}, where Figure \ref{fig:corner} presents the posterior probability distributions of all fitted parameters for core C3, and Table \ref{tab:infall_summary} summarises the derived $v_{\rm in}$ values for all the selected cores.
 
Comparison of $v_{\rm in}$ with the free-fall velocity, $v_{\rm ff} = \sqrt{2GM_{\rm in}/R_{\rm in}}$, shows that on average $v_{\rm ff} \sim 3\times v_{\rm infall}$, where $M_{\rm in}$ and $R_{\rm in}$ are the mass and radius of the core within the 30\% peak intensity contour. This indicates that the infall motions within these cores are not purely gravitational but rather regulated by additional support mechanisms such as turbulence and magnetic fields.

Further, assuming a roughly constant $v_{\rm in}$ across $R_{\rm in}$, we estimated the infall timescale as $t_{\rm in} = R_{\rm in} / {v_{\rm in}}$. Comparison with the free-fall timescale, $t_{\rm ff} = R_{\rm in} / v_{\rm ff}$, revealed that $t_{\rm in} \sim 4\times t_{\rm ff}$, suggesting that the collapse occurs over a period several times longer than the free-fall time, consistent with the scenario explained above.

We also calculated the accretion timescale, $t_{\rm acc} = M_{\rm core}/\dot{M}_{\parallel}$, defined as the time required for a core to accumulate its present mass through filamentary accretion. A comparison of these timescales shows that $t_{\rm acc} \sim 5\times t_{\rm in}$ and $t_{\rm acc} \sim 20\times t_{\rm ff}$, indicating that the inflow of material from the surrounding filamentary network proceeds much more slowly than the inward motion within the core itself. This suggests that filamentary accretion likely began well before the present stage of core evolution and continues to feed material into the core but on a much longer timescale than the core’s regulated internal collapse.

\subsection{Evolution of Dense Cores within the Filaments}
To discuss how the dense cores evolve within their surrounding filaments, we define the relative abundance of dense cores as 
\begin{equation} \label{eq:rel_ab}
X=\frac{\left[\rm C^{18}O\right]}{\left[\rm N_{2}H^{+}\right]}\approx\frac{\int T_{\rm A,\: C^{18}O}^{*}\:dv}{\int T_{\rm A,\:  N_{2}H^{+}}^{*}\:dv},
\end{equation}
where the denominator represents the integrated intensity of N$_{2}$H$^{+}$ at the peak intensity position of a dense core and the numerator corresponds to the integrated intensity of C$^{18}$O at that position. A higher value of $X$ suggests that N$_{2}$H$^{+}$ emission is comparatively weak relative to C$^{18}$O, indicating a less evolved core. Conversely, a lower $X$ value implies a more evolved dense core, characterized by stronger N$_{2}$H$^{+}$ emission relative to C$^{18}$O. This is likely due to the depletion of C$^{18}$O in the central regions of dense cores, as described in Section \ref{sec:column}, a well-known  phenomenon associated with dense core evolution, wherein CO freezes out onto dust grains under the cold temperatures typical of dense core interiors.

\begin{figure}[!htb]
    \centering
    \includegraphics[width=0.47\textwidth]{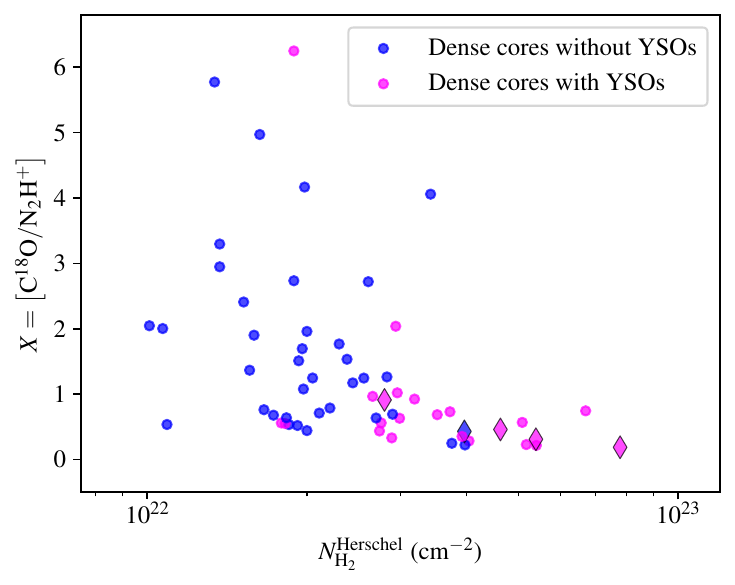}
    \caption{$\left[\rm C^{18}O\right]/\left[\rm N_{2}H^{+}\right]$ as a function of Herschel H$_{2}$ column density (at peak intensity position) for all the dense cores. The magenta and blue dots represent dense cores with and without YSOs. The diamond symbols represent the dense cores where NH$_{2}$D is detected.}
    \label{fig:rel_ab}
\end{figure}

Figure \ref{fig:rel_ab} shows the variation of $X$ as a function of Herschel H$_{2}$ column density ($N_{\rm H_{2}}$) for each dense core. $N_{\rm H_{2}}$ serves as an effective tracer of the evolutionary stage of a dense core, with higher values indicating more evolved cores and lower values corresponding to less evolved ones. From Figure \ref{fig:rel_ab}, it is apparent that dense cores exhibiting lower $X$ values tend to have higher $N_{\rm H_{2}}$ values, whereas those with higher $X$ values generally display lower $N_{\rm H_{2}}$ values. Furthermore, nearly all dense cores associated with YSOs are located at the lower-right extreme of the plot, consistent with their high evolutionary status. However, in the presence of a YSO, the freeze-out of C$^{18}$O is expected to be reversed, with the molecule being evaporated back to the gaseous state due to the photons emitted by the YSO. This process could contribute to the destruction of N$_{2}$H$^{+}$ \citep{1999ApJ...523L.165C,2002ApJ...572..238C,2002ApJ...570L.101B}, potentially increasing the relative abundance of dense cores associated with YSOs. However, no such effect is observed in this case. This may be due to the localized nature of the evaporation process, which might not be detectable given the coarse beam size of C$^{18}$O (49$^{\prime\prime}$), which likely does not resolve the small-scale variations induced by this effect.

Additionally, NH$_{2}$D emission, a well-known tracer of highly evolved dense cores, has been detected slightly above the $3\times\sigma_{\rm RMS\_I}$ level in the case of five dense cores (4 with and 1 without YSOs). These dense cores, denoted with diamond symbols in Figure \ref{fig:rel_ab}, are predominantly located at the lower-right extreme of the plot, further supporting their evolutionary status.

\subsection{Dynamical Properties of Filaments and Dense Cores: A Comparison between W40 and Serpens South}
Section \ref{sec:intro} provides a brief overview of the stellar content, morphology, and evolutionary status of the W40 and Serpens South regions. In this section, we present a comparative analysis of the two regions based on the dynamical properties of the velocity-coherent filaments and dense cores identified in this study. Although no systematic difference in V$_{\rm LSR}$ is there between the two regions, we roughly separate them using a vertical boundary at RA = 18:30:37.2 (see Figures \ref{fig:boundary}, \ref{fig:filvel}, \ref{fig:cores}), with the eastern side corresponding to W40 and the western side to Serpens South. Out of the 24 velocity-coherent filaments identified in the Aquila region, 15 are found within W40 and 9 within Serpens South, with $\sim$60\% of the filaments in both regions hosting $\rm N_{2}H^{+}$ dense cores.

The filamentary structures in W40 are relatively simple in velocity structure and well separated, whereas those in Serpens South display complex velocity patterns, including spatial overlap between multiple filaments (e.g., F4, F13, F15, F20; Figure \ref{fig:filvel}). While most filament properties analysed in this work are broadly similar across the two regions, filaments within the Serpens South region are more massive, with a median line mass $\sim$20\% higher, and also exhibit stronger accretion activity, with an average mass accretion rate $\sim$40\% greater than those in W40. Regarding the dense cores, W40 hosts 33 identified $\rm N_{2}H^{+}$ cores, which are typically clumpy and compact, with a median radius of 0.12 pc. In contrast, Serpens South contains the remaining 40 $\rm N_{2}H^{+}$ cores, which tend to be more extended, with a median radius of 0.16 pc. A striking difference is also observed in the YSO population: only 17 YSOs are associated with W40, compared to 56 in Serpens South. Furthermore, Mach number analysis indicates that a few dense cores in Serpens South exhibit supersonic turbulence ($\mathcal{M}>1$) (see Figure \ref{fig:mach}, Appendix \ref{app:mach}), particularly within the hub region, likely driven by ongoing infall and outflow activities \citep{2015Natur.527...70P} associated with the large YSO population. In contrast, dense cores within the W40 region exhibit subsonic to transonic motions only.

The differences we find in the properties of filaments and dense cores between W40 and Serpens South are consistent with, and provide further support for, the evolutionary distinctions highlighted in previous studies summarised in Section \ref{sec:intro}. Specifically, the relatively simple velocity structures, compact dense cores, and lower YSO count in W40 align with its characterisation as a more evolved H {\sc ii} region powered by massive O- and B-type stars, where strong stellar feedback has already shaped the environment and dispersed part of the parental material, thereby limiting the formation of dense cores and the emergence of new YSOs. In contrast, the more massive filaments, extended dense cores, higher YSO population, and enhanced turbulence in Serpens South are in agreement with its identification as a younger, highly active hub-filament system, where filamentary accretion and interactions continue to drive vigorous star formation.

\subsection{Relation with other studies from the TRAO-FUNS Project}
The primary objective of the TRAO-FUNS project is to obtain kinematical and chemical information of filaments and embedded dense cores to understand the physical process involved in their formation within the molecular clouds belonging to the Gould Belt. Table \ref{tab:trao} lists the derived physical properties of four molecular clouds analyzed so far.

The range of filament lengths is found to be consistent across the four molecular clouds, except for L1478, which has relatively shorter filaments compared to other regions. Regarding the range of filament widths, Aquila exhibits a similar range to that of L1478 and Orion B, whose distances from Earth are also comparable to that of Aquila. However, these observed widths are smaller than the filament width of IC 5146, which is at a distance of $\sim$700 pc. This result aligns with recent findings reported by \cite{2022A&A...657L..13P}, highlighting the effect of distance on filament widths. However, \cite{2022A&A...667L...1A} have suggested that 0.1 pc is a characteristic filamentary width in nearby molecular clouds. Recent interferometric studies have challenged this characteristic value by detecting much narrower, velocity-coherent, fiber-like structures ($\sim$0.03 pc) within filamentary regions (\citealt{2011ApJ...739L...2P,2014ApJ...790L..19F,2017MNRAS.464L..31H,2018A&A...610A..77H}). Note that the TRAO beam size is larger compared to those of Herschel and interferometric studies, which may have influenced our derived filament widths. We plan to revisit this issue in future studies, with the analysis of additional targets observed in the TRAO-FUNS survey and high angular resolution data for the filamentary clouds.

Upon revisiting previous TRAO-FUNS studies, we find that all velocity-coherent filaments hosting dense cores across the four regions are both thermally supercritical and gravitationally bound, indicating that these conditions are a prerequisite for core formation. Herschel observations of the Aquila region also revealed that nearly 80\% of identified cores reside within supercritical filaments \citep{2015A&A...584A..91K}. Nevertheless, several filaments that satisfy these criteria are not associated with dense cores, indicating that supercriticality and boundedness are necessary but not sufficient conditions and likely represent an evolutionary precursor to active star formation. 

The overall efficiency of core and YSO formation appears to be further modulated by the H$_{2}$ column density. In L1478, all ten filaments are bound and supercritical, yet the numbers of dense cores and YSOs remain comparatively low, consistent with the region’s lowest average H$_{2}$ column density compared to the others. By contrast, Aquila exhibits the highest H$_{2}$ column density and simultaneously hosts the largest numbers of dense cores and YSOs, in addition to all filaments being gravitationally bound and thermally supercritical. Orion B and IC 5146 occupy intermediate regimes, where significant fractions of the filaments are bound and supercritical ($\sim$60\% and 80\%, respectively), and correspondingly have moderate numbers of dense cores and YSOs reported. These results highlight the combined importance of filament criticality and the H$_{2}$ column density. While criticality and gravitational boundedness are essential for dense core formation, the H$_{2}$ column density provides the environmental regulation that determines how efficiently supercritical filaments evolve to form dense cores and YSOs, ultimately giving rise to stars.

Future studies on the rest of the molecular clouds observed in the TRAO-FUNS survey (Polaris flare, Taurus, Perseus, Serpens Main, and Cepheus) will be able to provide definitive conclusions regarding the physical processes involved in the role of filamentary structures in the formation of dense cores. Further studies will also focus on the chemical differentiation between the filaments and dense cores using our chemical tracers, NH$_{2}$D and SO, in various star-forming regions.

\begin{deluxetable*}{lcccccccccc}
\tablecaption{Comparison of Physical Properties of Molecular Clouds Observed by the TRAO-FUNS Project. \label{tab:trao}}
\setlength{\tabcolsep}{7pt}
\tablewidth{0pt}
\tabletypesize{\scriptsize}
\tablehead{
\colhead{Cloud} & \colhead{Distance} & \colhead{Area} & \colhead{Fil.\#} & \colhead{Length} & \colhead{Width} & \colhead{Core\#} & \colhead{YSO\#} & \colhead{Bound$+$Supercritical Fil.\#} & \colhead{Fil.\# with cores} & \colhead{$\left<N_{\rm H_{2}}\right>$}\\
\colhead{} & \colhead{(pc)} & \colhead{(pc$^{2}$)} & \colhead{} & \colhead{(pc)} & \colhead{(pc)} & \colhead{} & \colhead{} & \colhead{} & \colhead{} & \colhead{($\times$10$^{20}$ cm$^{-2}$)}\\
\colhead{(1)} & \colhead{(2)} & \colhead{(3)} & \colhead{(4)} & \colhead{(5)} & \colhead{(6)} & \colhead{(7)} & \colhead{(8)} & \colhead{(9)} & \colhead{(10)} & \colhead{(11)}
}
\startdata
L1478 & 450 & 63 & 10 (0.16) & 0.4--1.1 & 0.08--0.19 & 8 (0.13) & 3 (0.05) & 10 (0.16) & 3 (0.05) & 7--40 \\
IC 5146 & 700 & 140 & 24 (0.17) & 0.2--4.2 & 0.18--0.67 & 22 (0.16) & 40 (0.29) & 21 (0.15) & 10 (0.07) & 22--88 \\
Orion B & 423 & 74 & 32 (0.43) & 0.2--6.5 & 0.05--0.26 & 48 (0.65) & 51 (0.69) & 19 (0.26) & 12 (0.16) & 28--145 \\
Aquila & 455 & 40 & 24 (0.60) & 0.6--4.0 & 0.05--0.32 & 73 (1.83) & 73 (1.83) & 24 (0.60) & 16 (0.40) & 84--270 \\
\enddata
\tablecomments{(1) Molecular clouds analysed in the TRAO-FUNS study so far. Refer to \cite{2019ApJ...877..114C,2021ApJ...919....3C} and \cite{2023ApJ...957...94Y} for details on L1478, IC 5146, and Orion B, respectively. (2) Distance to each molecular cloud. (3) Total observed area of C$^{18}$O. (4) Number of identified velocity-coherent filaments. (5) and (6) Range of measured lengths and widths of filaments. (7) and (8) Number of identified dense cores and YSOs. (9) Number of gravitationally bound and thermally supercritical filaments. (10) Number of filaments that contain dense cores. (11) Range of average H$_{2}$ column densities of filaments. The values within parenthesis in columns of (4), (7)--(10) represent the number normalised to 1 pc$^{2}$ observed area.}
\end{deluxetable*}

\section{Summary} \label{sec:summary}
This paper presents the key findings obtained from observations of multiple molecular lines towards the W40 and Serpens South regions within the Aquila molecular cloud complex using the TRAO 14 m single-dish telescope as part of the `TRAO-FUNS' project to study the velocity-coherent filamentary structures in the cloud and their role in the formation and evolution of dense cores. A summary of the results is provided below:
\begin{enumerate}
    \item A total of 24 velocity-coherent filaments were identified by applying the `FoF' algorithm to the Gaussian-decomposed velocity components of the C$^{18}$O spectra, with some filaments being further separated into smaller ones using PCA and DBSCAN. The lengths and widths of the filaments were derived using the \texttt{FilFinder} algorithm. The other physical parameters, such as mass, line mass, total velocity dispersion, etc., were also measured and listed in Table \ref{tab:prop}. Many of the filaments previously identified from the H$_{2}$ column density map obtained from the Herschel survey are found to contain multiple velocity-coherent filaments, thus highlighting the necessity of velocity information.
    \item We identified 64 N$_{2}$H$^{+}$ dense cores within the observed region by applying the `FellWalker' algorithm to the N$_{2}$H$^{+}$ integrated intensity map. The core C1 is further divided into 10 cores by applying the FoF algorithm to the decomposed velocity components of the N$_{2}$H$^{+}$ spectra present in the C1 region. The physical properties, such as radius, mass, total velocity dispersion, virial parameters, etc., were derived and listed in Table \ref{tab:cores} for each of the dense cores.
    \item A tight correlation was observed between the systematic velocities of dense cores and the surrounding filaments, indicating the close kinematical relation between the two entities. Also, end-to-end velocity gradients along with local variations at sub-parsec scales, particularly near the dense cores, were observed in most of the filaments harboring dense cores.
    \item Our analysis revealed that velocity gradients along filaments are present near the vicinity of the embedded dense cores, indicating longitudinal gas flows that converge toward core centers. The corresponding mass flow rates span 7–178 M$_\odot$ Myr$^{-1}$, with a median of 35 M$_\odot$ Myr$^{-1}$, and are notably higher in Serpens South, consistent with its hub-filamentary environment. However, comparison of accretion and free-fall timescales shows that cores collapse much faster than they are replenished by filamentary inflows. This demonstrates that while filamentary accretion plays a critical supporting role, gravitational collapse dominates the mass assembly at the core scale.
    \item The distribution of non-thermal velocity dispersions extracted from N$_{2}$H$^{+}$ and C$^{18}$O for dense cores and filaments, respectively, suggested that almost all the dense cores have subsonic/transonic motions, whereas the surrounding filaments are primarily supersonic. This may indicate that the gas turbulent motions in the filaments around the dense cores have been dissipated at the core scale to help in the formation of dense cores.
    \item All the identified filaments are found to be gravitationally bound and thermally supercritical, regardless of the presence of N$_{2}$H$^{+}$ dense cores within them. This implies that even filaments without dense cores provide a suitable environment for the potential formation of dense cores in the future. Furthermore, it is observed that filaments with larger $M_{\rm line}/M_{\rm line}^{\rm th,\: crit}$ tend to host a greater number of dense cores, the majority of which are associated with YSOs. Therefore, it can be inferred that the gravitational criticality of a filament plays a crucial role in its fragmentation and the subsequent formation of dense cores and YSOs. Virial analysis of the dense cores indicates that nearly 90\% of the identified dense cores, including those with YSOs, have $\alpha_{\rm vir}^{\rm core}$ values below the critical threshold of 2, confirming their gravitationally bound status.
    \item Nearly 60\% of the identified dense cores, based on the analysis of the CS, HCO$^{+}$ and H$^{13}$CO$^{+}$ spectra, display a characteristic blue asymmetry profile, indicative of ongoing infall motions within these cores. The infall velocity for a subset of these cores was derived by fitting the Hill5 model to the CS spectra averaged within the 30\% peak intensity contour. A comparison between the derived infall and free-fall velocities, as well as free-fall, infall, and accretion timescales, revealed that the contraction motions within the cores are not purely gravitational but rather regulated by possible support from turbulence and magnetic fields.
    \item We compared the velocity-coherent filaments and dense cores between W40 and Serpens South. Out of 24 filaments, 15 lie in W40 and 9 in Serpens South, with $\sim$60\% in both regions hosting N$_{2}$H$^{+}$ dense cores. W40 shows simpler velocity structures in filaments, compact dense cores (median radius 0.12 pc), and only 17 YSOs, consistent with a more evolved H {\sc ii} region shaped by stellar feedback. Serpens South, by contrast, has more massive filaments, larger dense cores (median radius 0.16 pc), higher accretion rates ($\sim$40\% greater compared to W40), and 56 YSOs, consistent with the presence of the active hub-filament system.
    \item Among the four regions studied in the TRAO-FUNS survey so far, all core-hosting velocity-coherent filaments are found to be gravitationally bound and thermally supercritical, confirming these properties as prerequisites for core formation. At the same time, the total numbers of dense cores and YSOs in each region correlate with the clouds’ H$_2$ column densities. Aquila, which shows the highest H$_2$ column density, hosts the most number of dense cores and YSOs and only slightly fewer filaments than Orion B. Moreover, all filaments in Aquila, irrespective of the presence of dense cores, are gravitationally bound and thermally supercritical. These results highlight that both filament criticality and ambient column density jointly regulate the efficiency of core and star formation within a molecular cloud.
\end{enumerate}

\section{Acknowledgments}
This work is supported by the Basic Science Research Program through the NRF funded by the Ministry of Education, Science and Technology (grant No. NRF-2019R1A2C1010851) and by the Korea Astronomy and Space Science Institute grant funded by the Korea government (MSIT; project No. 2024-1-841-00). S. K. was  supported by the National Research Foundation of Korea(NRF) funded by Korea government(KASA, Korea AeroSpace Administration) (grant number RS-2024-00509838). W.K. is supported by the National Research Foundation of Korea (NRF) grant funded by the Korea government (MSIT) (RS-2024-00342488 \& RS-2024-00416859). This research has made use of data from the Herschel Gould Belt survey (HGBS) project (\href{http://gouldbelt-herschel.cea.fr}{http://gouldbelt-herschel.cea.fr}). The HGBS is a Herschel Key Programme jointly carried out by SPIRE Specialist Astronomy Group 3 (SAG 3), scientists of several institutes in the PACS Consortium (CEA Saclay, INAF-IFSI Rome and INAF-Arcetri, KU Leuven, MPIA Heidelberg), and scientists of the Herschel Science Center (HSC). This research has made use of the NASA/IPAC Infrared Science Archive, which is funded by the National Aeronautics and Space Administration and operated by the California Institute of Technology. The Starlink software \citep{2014ASPC..485..391C} is currently supported by the East Asian Observatory.

\facilities{TRAO, Herschel (PACS, SPIRE), IRSA (eHOPS)}
\software{Astropy \citep{2013A&A...558A..33A,2018AJ....156..123A}, APLpy \citep{2012ascl.soft08017R}, FilFinder \citep{2015MNRAS.452.3435K}, FellWalker \citep{2015A&C....10...22B}, Starlink \citep{2014ASPC..485..391C}
          }

\clearpage

\appendix
\section{Separation of Filament 14A} \label{app:fil14a}
\setcounter{figure}{0} \renewcommand{\thefigure}{A.\arabic{figure}}
\setcounter{table}{0} \renewcommand{\thetable}{A.\arabic{table}}

\begin{figure*}[!htb]
    \centering
    \includegraphics[width=\textwidth]{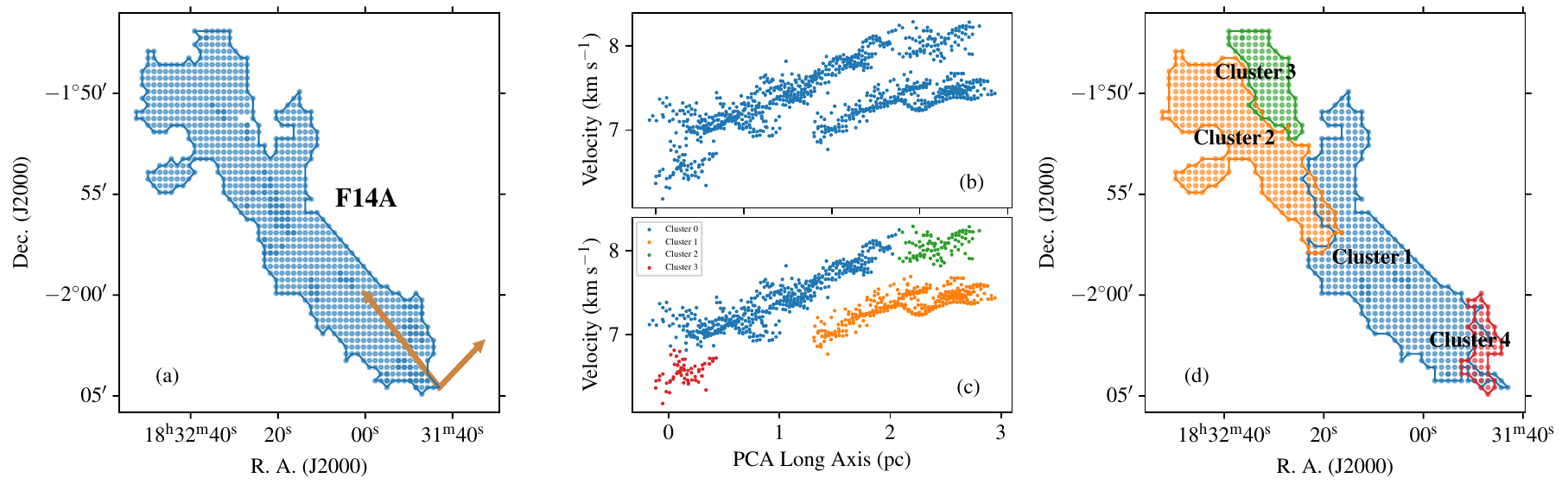}
    \caption{(a): The structure F14A, obtained from the FoF algorithm. The axes shown at the bottom of the figures represent the orientation of PCA axes, with the longer one being the PCA long axis. (b): Distribution of components in the PCA long-axis \& velocity frame. (c): Separation of substructures using the \texttt{DBSCAN} algorithm. (d): Spatial distribution of the identified distinct velocity-coherent structures.}
    \label{fig:pca_db}
\end{figure*}

A closer inspection of F14A in the PPV space suggested that it contains a number of sub-structures that were not easily separable using the FoF algorithm due to the very small velocity difference between them. To resolve this issue, we employed principal component analysis (PCA) on the spatial coordinates of each of these structures to determine the principal axis of elongation (PCA long axis; see Figure \ref{fig:pca_db}(a)). Then, we mapped the PCA long axis coordinates with the velocity values to get the distributions of sub-structures along this axis (Figure \ref{fig:pca_db}(b)). Subsequently, we applied the \texttt{DBSCAN} algorithm in the \texttt{sklearn.cluster} package iteratively on the PCA-velocity dataset to separate the sub-structures (see Figure \ref{fig:pca_db}(c)). The sub-structures were named in the following way: Cluster 1: F14a, Cluster 2: F14c, Cluster 3: F14d, and Cluster 4: F14e. A similar velocity distribution was observed with the structure F10, and it was split into two sub-structures named F10a and F10b (not shown) with the above-mentioned approach.

\section{Separation of Dense Core C1} \label{app:core_c1}
\begin{figure*}[!htb]
    \centering
    \includegraphics[width=\textwidth]{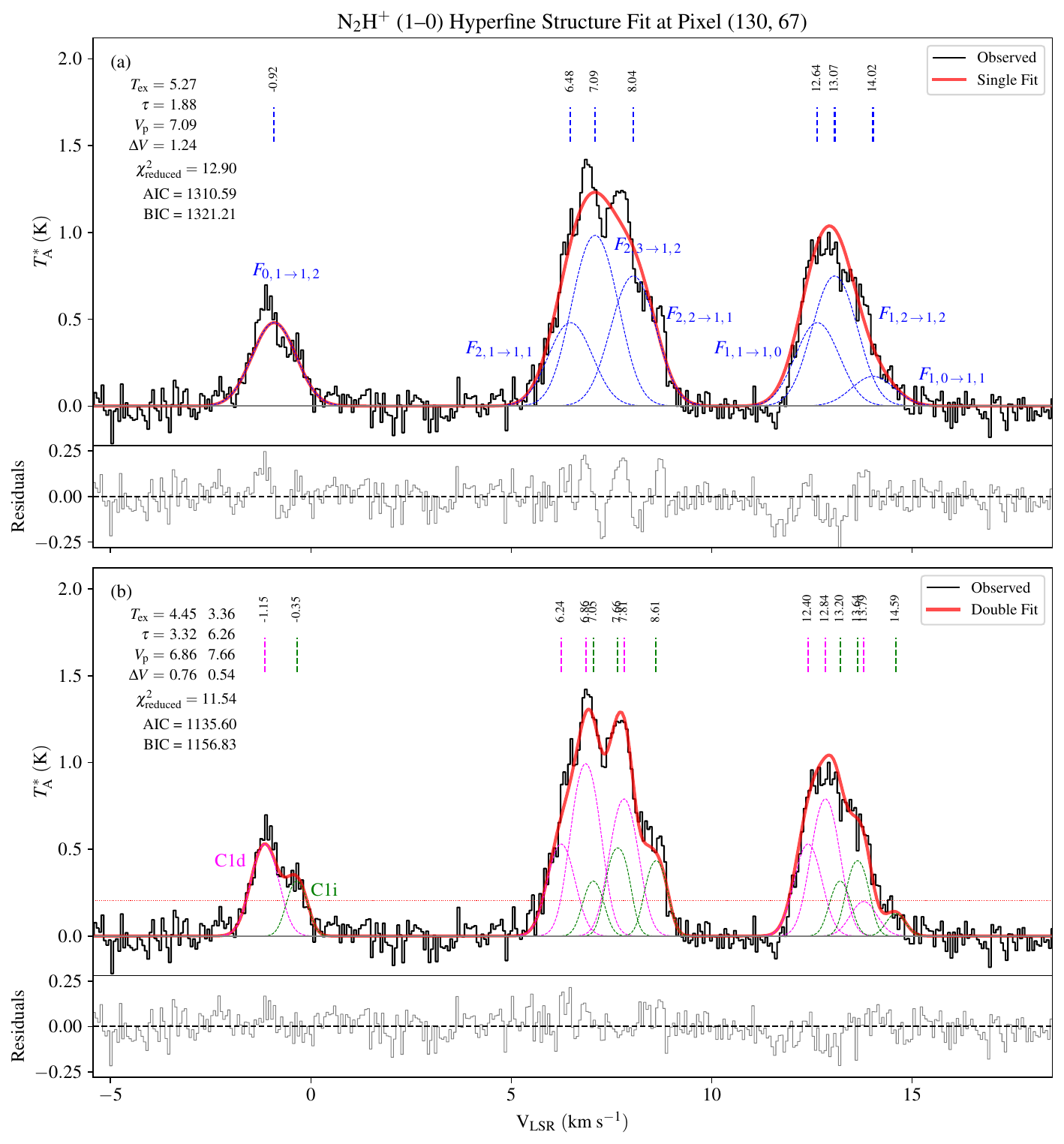}
    \caption{Hyperfine structure fitting to the N$_{2}$H$^{+}$ spectra at a reference pixel position (130, 67). The black line refers to the observed spectra, and the thick red line refers to the overall fitting in both the panels. Panel (a) shows a single-component fit with 7 Gaussian functions to the hyperfine spectra, with each of the fitted hyperfine components indicated by dashed blue lines. Each of the seven hyperfine components is labeled accordingly, alongside their respective velocity centroid values. Panel (b) presents a two-component fit with 14 Gaussian functions to the hyperfine spectra, with the two sets of the fitted hyperfine components shown in green and magenta colored dashed lines. The magenta component is associated with core C1d, and the green component corresponds to core C1i, as determined by the FoF algorithm. The best-fit parameters, as well as $\chi^{2}_{\rm reduced}$, AIC and BIC values, are also listed in both the panels, and the residuals are also plotted at the bottom of each panel.}
    \label{fig:n2hp_fit}
\end{figure*}

To separate the dense core C1 based on kinematic information, we performed hyperfine fitting of the N$_{2}$H$^{+}$ spectra on a pixel-by-pixel basis using seven Gaussian components, corresponding to its seven hyperfine transitions. The velocity offsets and relative intensities of these components were adopted from \citet{1995ApJ...455L..77C}. The resulting distribution of velocity centroids across the C1 region is presented in Figure \ref{fig:nh2_compare}(a). However, detailed inspection of the residuals revealed that, at several pixel positions, the single-component fit fails to adequately reproduce the observed spectra, with significant residuals remaining (see Figure \ref{fig:n2hp_fit}(a)). This inconsistency likely arises from the presence of multiple velocity components along the line-of-sight. To investigate this, we compared the isolated (optically thin) hyperfine component of the N$_{2}$H$^{+}$ spectra—shifted to align with the main component—with spectra of H$^{13}$CO$^{+}$ and C$^{18}$O. Both of these tracers exhibit clear evidence for two distinct velocity components, supporting the presence of multiple kinematic features.

\begin{figure}[!htb]
    \centering
    \includegraphics[width=\textwidth]{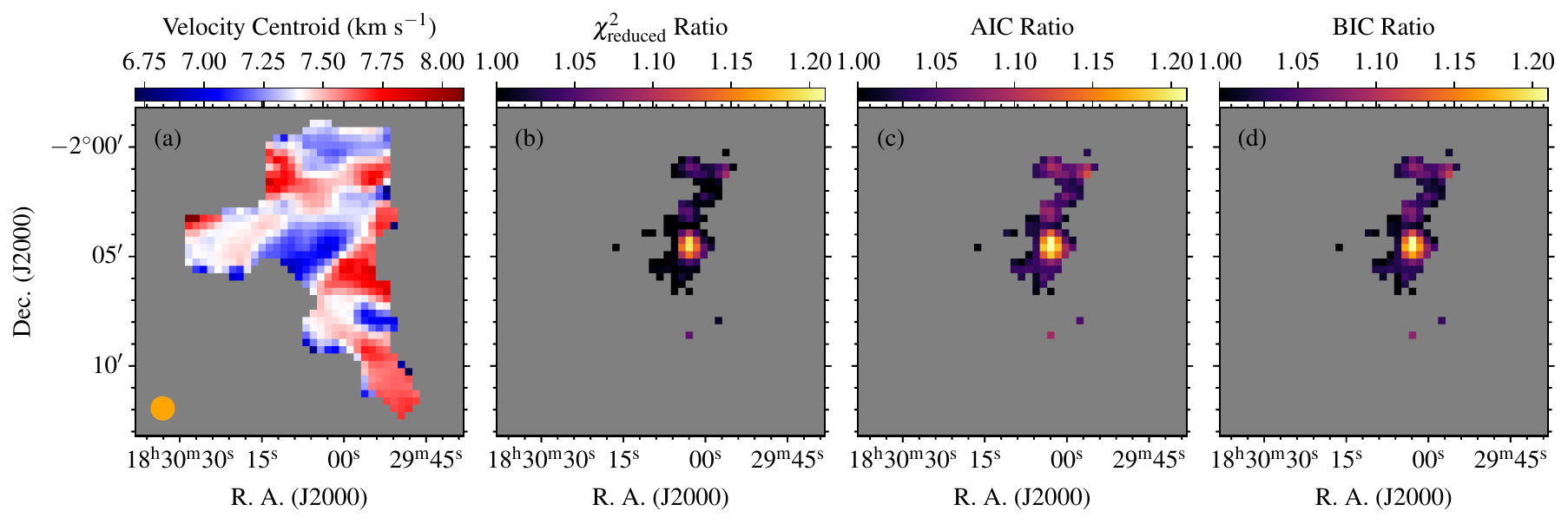}
    \caption{(a): Distribution of velocity centroid values obtained by single-component hyperfine fitting to the N$_{2}$H$^{+}$ spectra across the C1 region. (b), (c), and (d): Distribution of the ratios of reduced $\chi^{2}_{\rm reduced}$, AIC, and BIC values obtained from single-component and double-component hyperfine structure fitting.}
    \label{fig:c1_summary}
\end{figure}

Subsequently, we fitted the N$_{2}$H$^{+}$ spectra with a two-component hyperfine model, comprising 14 Gaussian components, across the entire C1 region. We retained only those pixels where the peak intensities of both fitted components exceeded a signal-to-noise ratio (SNR) of 3. For these pixels, we quantitatively compared the quality of the single- and double-component fits using the reduced chi-square ($\chi^{2}_{\rm reduced}$), Akaike Information Criterion (AIC), and Bayesian Information Criterion (BIC)\footnote{$\chi^{2}_{\rm reduced}=\frac{\chi^{2}}{N-k}, \: {\rm AIC} = \chi^{2}+2k, \: {\rm BIC} = \chi^{2}+k\ln N,$\\where $N$ is the number of data points and $k$ is the number of free parameters.}, by computing the ratios of their respective values (see Figures \ref{fig:c1_summary}(b)–(d)). Pixels were identified as exhibiting two velocity components if at least two out of these three diagnostic ratios were greater than unity, indicating an improved fit from the two-component model (see Figure \ref{fig:n2hp_fit}(b)). We additionally tested a three-component hyperfine fit using 21 Gaussian components in regions where the two-component fit was favored over the single-component model. However, in all such cases, the two-component fit remained superior, suggesting that the spectra are best described by two distinct velocity components.

Following the derivation of velocity information from both single- and two-component hyperfine fits, we applied the FoF algorithm to segment the entire C1 region into distinct velocity-coherent structures. This analysis revealed ten such structures, which we designate as dense cores C1a, C1b, $\cdots$, C1j. Notably, C1d overlaps with C1i, and C1c overlaps with C1e in position space (see Figure \ref{fig:cores}), but all of these have completely different velocity distributions in the PPV space\footnote{Velocity distribution of C1a to C1j dense cores in the PPV space: \href{https://imbunty575.github.io/Aquila-Filaments/C1_PPV.html}{https://imbunty575.github.io/Aquila-Filaments/C1\_PPV.html}}. Note that cores C1a and C1h exhibit relatively large aspect ratios, making their classification as compact dense cores somewhat ambiguous. These elongated structures may contain multiple dense cores within them, which could not be resolved due to the coarse beam size of TRAO observations.

\clearpage

\section{Comparison of N$_{2}$H$^{+}$ emission with C$^{18}$O Velocity-Coherent Filaments} \label{app:n2hp_dist}
\begin{figure*}[h]
    \centering
    \includegraphics[width=\textwidth]{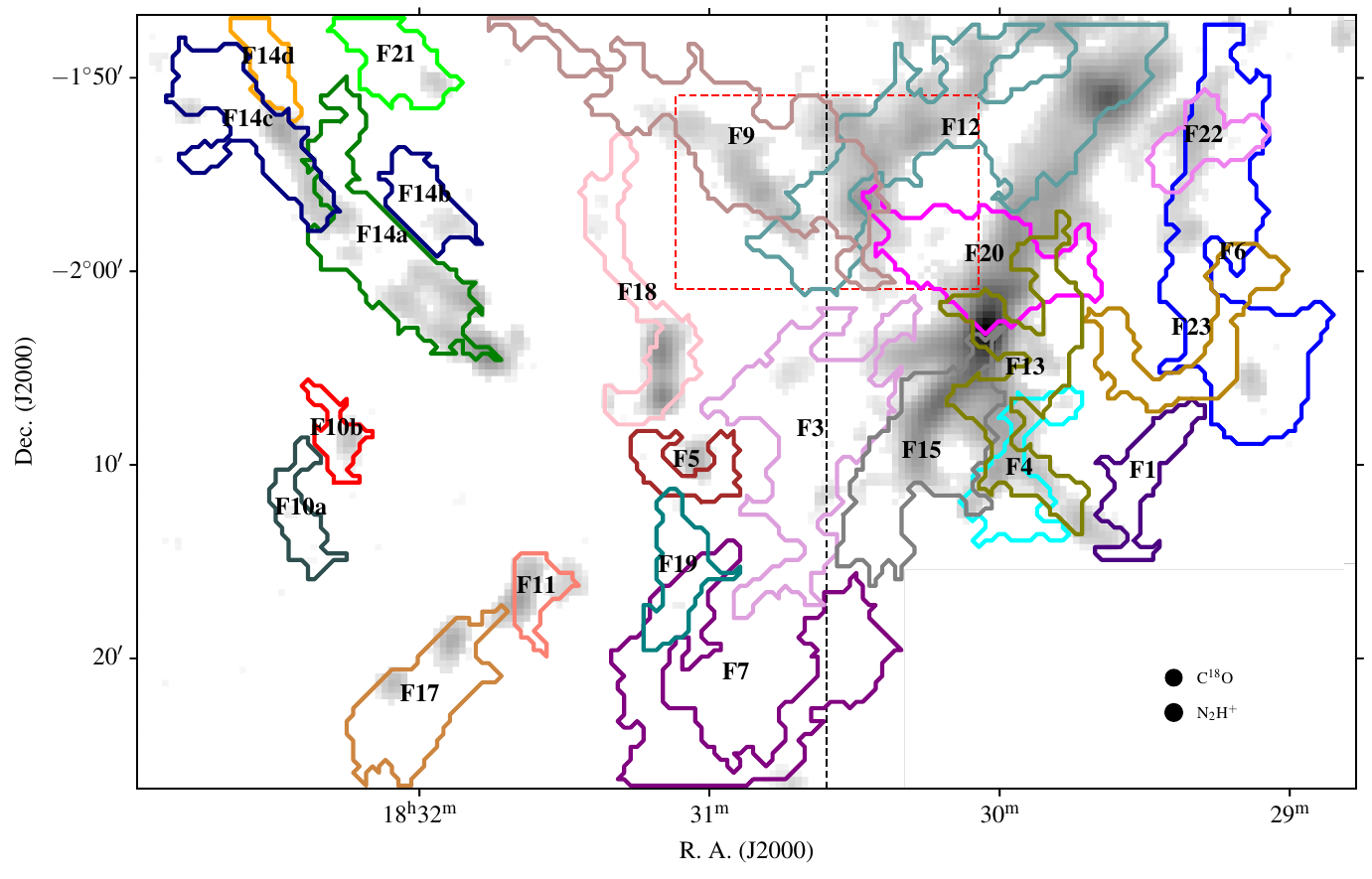}
    \caption{Comparison of N$_{2}$H$^{+}$ emission ($\mathbf{>3\times\sigma_{\rm RMS\_II}}$) with respect to the C$^{18}$O velocity-coherent filaments over the Aquila region. The filament boundaries and names are the same as shown in Figure \ref{fig:filvel}. The background represents N$_{2}$H$^{+}$ integrated intensity map. The vertical black dashed line (RA = 18:30:37.2) roughly marks the boundary between W40 and Serpens South regions. The red-dashed rectangle represents the zoomed-in region shown in Figure \ref{fig:fil_core_spectra}. The beam sizes of C$^{18}$O and N$_{2}$H$^{+}$ emission maps are shown as black dots in the lower-right corner of the figure.}
    \label{fig:fil_core}
\end{figure*}

Here we present the comparison of N$_{2}$H$^{+}$ emission with respect to the C$^{18}$O velocity-coherent filaments (Figure \ref{fig:fil_core}). The entire Serpens South region, as traced with N$_{2}$H$^{+}$, extends over filaments F12, F20, F15, F13, and F4. The widespread emission of N$_{2}$H$^{+}$ highlights the dense environment characteristic of the Serpens South molecular cloud.

\begin{figure*}[!htb]
    \centering
    \includegraphics[width=\textwidth]{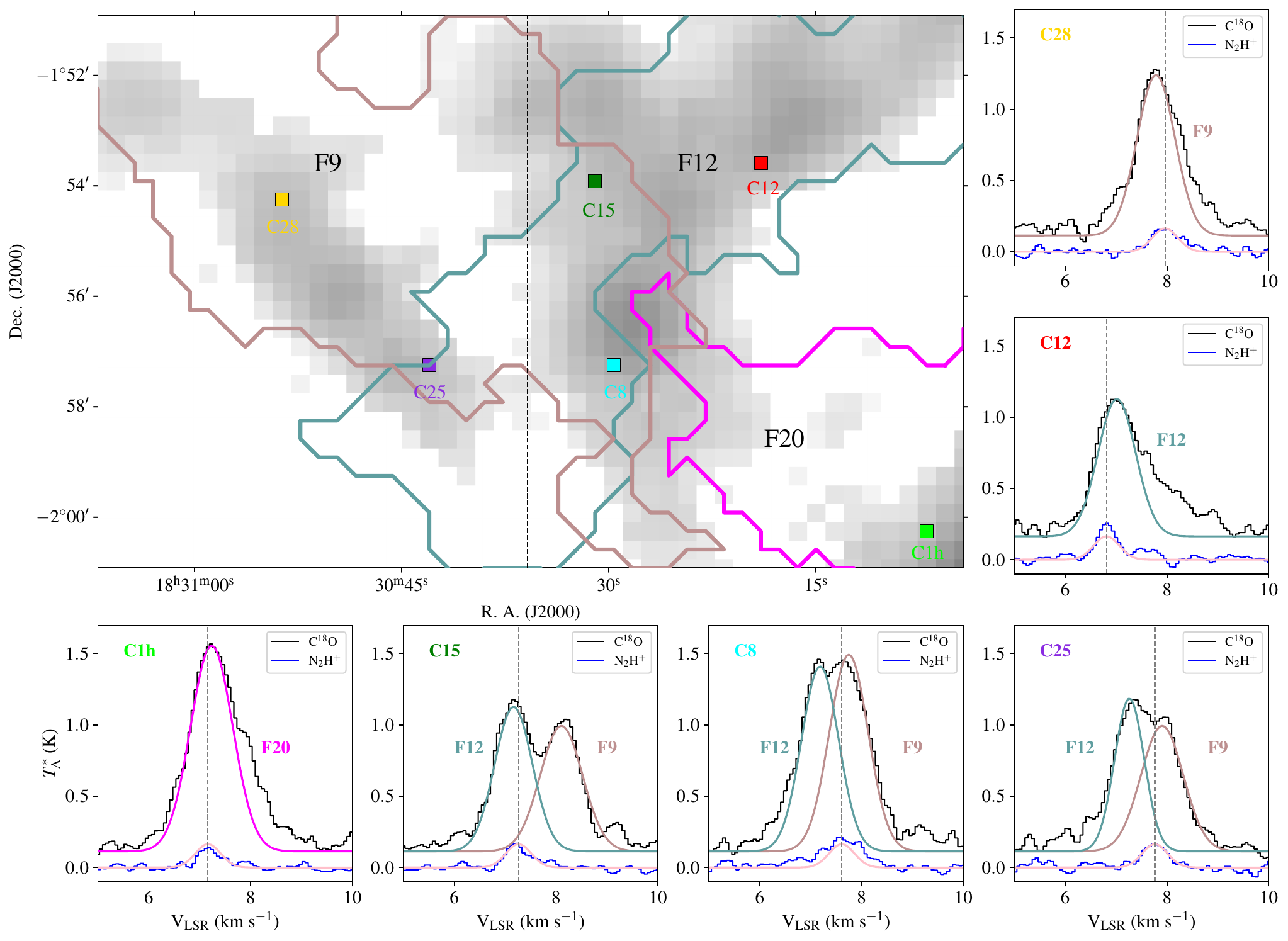}
    \caption{The top-left panel presents a zoomed-in view of the overlapping regions of filaments F9, F12, and F20, with the background showing the integrated intensity map of N$_{2}$H$^{+}$ ($\mathbf{>3\times\sigma_{\rm RMS\_II}}$). The vertical black dashed line (RA = 18:30:37.2) roughly marks the boundary between W40 and Serpens South regions. The colored squares on this map indicate specific pixel positions of the dense cores, which are further detailed in the smaller panels. Each small panel displays the C$^{18}$O spectra (black) and the isolated N$_{2}$H$^{+}$ spectra (blue) corresponding to the indicated pixel positions. The C$^{18}$O spectra is shifted vertically based on the peak position of the N$_{2}$H$^{+}$ spectra. The color coding of the filaments in the top-left panel is consistent with that used for the decomposed C$^{18}$O line profiles and the filament names in the smaller panels. The pink line overlaid on the N$_{2}$H$^{+}$ spectra represents the hyperfine structure fit result for the isolated component, and the gray dashed line represents the centroid velocity obtained from the fit.}
    \label{fig:fil_core_spectra}
\end{figure*}

Figure \ref{fig:fil_core_spectra} presents a comparison of the C$^{18}$O and N$_{2}$H$^{+}$ spectra at the positions marked by filled squares. Some of these positions exhibit two distinct velocity components, as observed in the C$^{18}$O spectra (represented by the green, cyan, and violet squares). Each velocity component corresponds to a different filament, distinguished by their respective velocities. However, the dense cores at these positions are assigned to the filament that exhibits a velocity consistent with that of the N$_{2}$H$^{+}$ emission.

\clearpage
\section{Comparison of N$_{2}$H$^{+}$ Dense Cores with Herschel-identified Cores} \label{app:n2hp_herschel}
\cite{2015A&A...584A..91K} have identified a total of 292 prestellar and 58 protostellar cores within the entire Aquila region, consisting of W40, Serpens South, and MWC297 regions, using both PACS and SPIRE images at five wavelengths from 70 $\mu$m to 500 $\mu$m based on multi-scale, multi-wavelength core extraction with the $getsources$ algorithm.

\begin{figure*}[!htb]
    \centering
    \includegraphics[width=\textwidth]{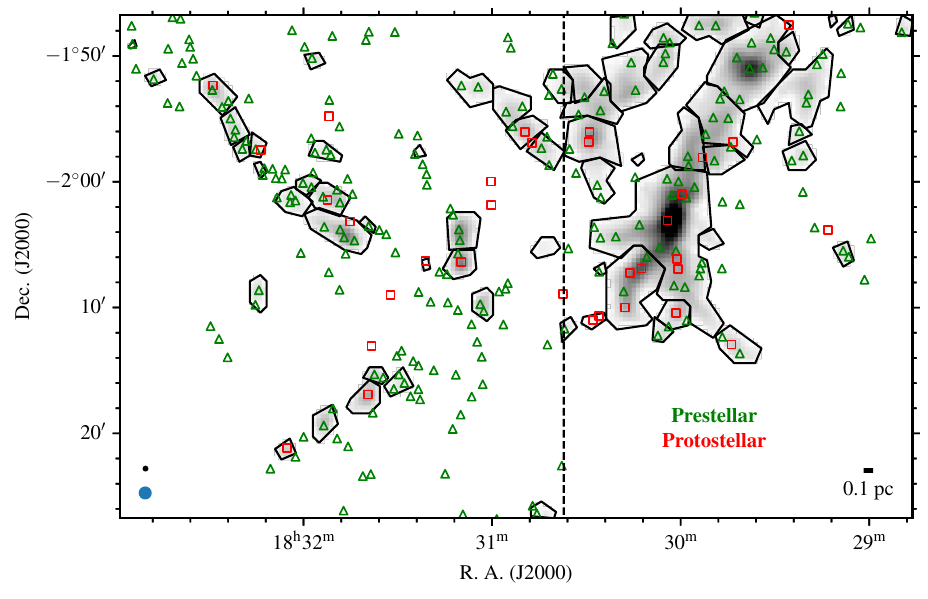}
    \caption{Herschel-identified prestellar (green triangles) and protostellar (red squares) core positions \citep{2015A&A...584A..91K} overlaid on the N$_{2}$H$^{+}$ dense cores identified in this study. The vertical black dashed line (RA = 18:30:37.2) roughly marks the boundary between W40 and Serpens South regions. The blue and black circles in the left-bottom side of the figure represent the beam sizes of TRAO N$_{2}$H$^{+}$ integrated intensity and Herschel H$_{2}$ column density maps, respectively.}
    \label{fig:herschel_cores}
\end{figure*}

While most of the identified cores are located within the W40 and Serpens South regions, here we have presented a spatial correlation between these cores and the N$_{2}$H$^{+}$ dense cores identified in this study in Figure \ref{fig:herschel_cores}. From the figure, it can be clearly seen that nearly all of the Herschel-identified protostellar cores (red squares) and a large number of prestellar cores (green triangles) spatially overlap with our N$_{2}$H$^{+}$ dense cores. This confirms that N$_{2}$H$^{+}$ emission reliably traces the dense, cold regions where stars are forming (or about to form). However, there is a small fraction of Herschel-identified prestellar cores that do not show corresponding N$_{2}$H$^{+}$ emission. This discrepancy may arise from sensitivity or beam dilution effects, as TRAO has a much coarser beam compared to that of Herschel.

\clearpage
\section{Variation of Velocity along the Filaments} \label{app:velocity}
\begin{figure*}[h]
    \centering
    \includegraphics[width=0.97\textwidth]{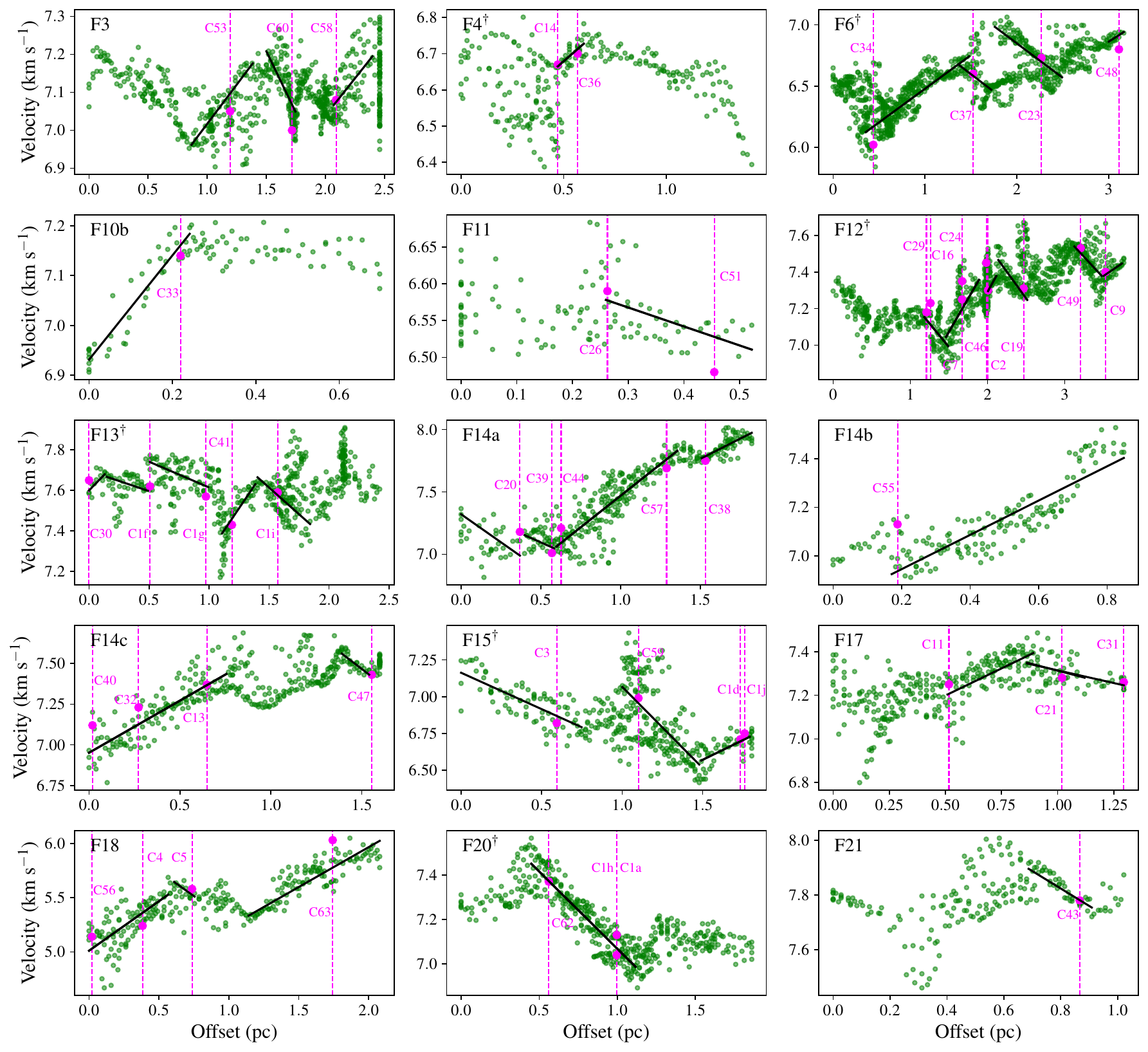}
    \caption{Variation of velocity centroid values along the filaments' skeletons. The 0 pc offset for each filament starts from the dot marks shown in Figure \ref{fig:filvel}. The filament IDs are labeled at the top-left corner of each panel. The vertical magenta lines indicate the representative position of the dense cores along the filaments' skeletons. The core IDs are mentioned next to these magenta lines. The average velocity centroid values of each dense core (see Table \ref{tab:cores}) are presented with magenta dots along the vertical magenta line. The black solid lines mark the least-squares fit to the velocity centroid values near the centroid positions of each dense core. $^{\dagger}$: Filaments within Serpens South.}
    \label{fig:velocity}
\end{figure*}

Here we present the variation in velocity centroid values of the filaments, derived from Gaussian decomposition of the C$^{18}$O spectra, along the filamentary skeletons in Figure \ref{fig:velocity}. The centroid positions of the dense cores embedded within each filament are indicated by vertical magenta lines, while the corresponding average centroid velocities, obtained from hyperfine structure fitting of the N$_{2}$H$^{+}$ spectra, are marked by magenta dots placed along these lines. To quantify the kinematic properties, we performed least-squares fits to the filament velocity centroid values in the vicinity of each dense core position. From these fits, we derived the local velocity gradients $(\nabla V_{\parallel})$ and subsequently estimated the associated mass flow rates $(\dot{M}_{\parallel})$ along the filaments.

\begin{deluxetable}{lcccc}
\tablecaption{Summary of Velocity Gradients and Mass Flow Rates along the Filaments \label{tab:vel_grad}}
\setlength{\tabcolsep}{33pt}
\tablehead{
\colhead{Core--Filament} & \colhead{$L_{\rm fil}$} & \colhead{$M_{\rm fil}$} & \colhead{$|\nabla V_{\parallel}|$} & \colhead{$\dot{M}_{\parallel}$}\\
\colhead{} & \colhead{(pc)} & \colhead{(M$_{\odot}$)} & \colhead{(km s$^{-1}$ pc$^{-1}$)} & \colhead{(M$_{\odot}$ Myr$^{-1}$)}\\
\colhead{(1)} & \colhead{(2)} & \colhead{(3)} & \colhead{(4)} & \colhead{(5)}
}
\startdata
C1a--F20 & 0.67 & 251 $\pm$ 55 & 0.69 $\pm$ 0.02 & 178 $\pm$ 39\\
C1d--F15 & 0.30 & 29 $\pm$ 6 & 0.54 $\pm$ 0.12 & 16 $\pm$ 5\\
C1f--F13 & 0.34 & 23 $\pm$ 5 & 0.20 $\pm$ 0.10 & 5 $\pm$ 2\\
C1g--F13 & 0.48 & 47 $\pm$ 10 & 0.25 $\pm$ 0.07 & 12 $\pm$ 4\\
C1h--F20 & 0.67 & 251 $\pm$ 55 & 0.69 $\pm$ 0.02 & 178 $\pm$ 39\\
C1i--F13 & 0.44 & 76 $\pm$ 17 & 0.53 $\pm$ 0.10 & 41 $\pm$ 12\\
C1j--F15 & 0.30 & 29 $\pm$ 6 & 0.54 $\pm$ 0.12 & 16 $\pm$ 5\\
C2--F12 & 0.11 & 33 $\pm$ 7 & 0.77 $\pm$ 0.28 & 26 $\pm$ 11\\
C3--F15 & 0.75 & 35 $\pm$ 8 & 0.49 $\pm$ 0.05 & 18 $\pm$ 4\\
C4--F18 & 0.57 & 166 $\pm$ 36 & 0.91 $\pm$ 0.10 & 155 $\pm$ 38\\
C5--F18 & 0.15 & 18 $\pm$ 4 & 0.93 $\pm$ 0.74 & 17 $\pm$ 14\\
C7--F12 & 0.43 & 94 $\pm$ 21 & 0.73 $\pm$ 0.08 & 70 $\pm$ 17\\
C8--F9 & 0.46 & 29 $\pm$ 6 & 0.44 $\pm$ 0.09 & 13 $\pm$ 4\\
C9--F12 & 0.24 & 38 $\pm$ 8 & 0.28 $\pm$ 0.16 & 11 $\pm$ 7\\
C11--F17 & 0.38 & 61 $\pm$ 13 & 0.50 $\pm$ 0.06 & 31 $\pm$ 8\\
C13--F14c & 0.15 & 11 $\pm$ 3 & 0.86 $\pm$ 0.12 & 10 $\pm$ 3\\
C14--F4 & 0.13 & 29 $\pm$ 6 & 0.49 $\pm$ 0.12 & 15 $\pm$ 5\\
C16--F12 & 0.31 & 45 $\pm$ 10 & 0.51 $\pm$ 0.11 & 24 $\pm$ 7\\
C17--F9 & 0.36 & 58 $\pm$ 13 & 0.59 $\pm$ 0.10 & 35 $\pm$ 10\\
C19--F12 & 0.37 & 64 $\pm$ 14 & 0.58 $\pm$ 0.08 & 38 $\pm$ 10\\
C20--F14a & 0.19 & 36 $\pm$ 8 & 1.01 $\pm$ 0.05 & 145 $\pm$ 33\\
C21--F17 & 0.26 & 28 $\pm$ 6 & 0.27 $\pm$ 0.13 & 8 $\pm$ 4\\
C23--F6 & 0.73 & 133 $\pm$ 29 & 0.56 $\pm$ 0.05 & 76 $\pm$ 18\\
C24--F12 & 0.43 & 94 $\pm$ 21 & 0.73 $\pm$ 0.08 & 70 $\pm$ 17\\
C25--F9 & 0.32 & 74 $\pm$ 16 & 0.85 $\pm$ 0.09 & 65 $\pm$ 16\\
C26--F11 & 0.26 & 27 $\pm$ 6 & 0.26 $\pm$ 0.07 & 7 $\pm$ 2\\
C28--F9 & 0.32 & 74 $\pm$ 16 & 0.85 $\pm$ 0.09 & 65 $\pm$ 16\\
C29--F12 & 0.31 & 45 $\pm$ 10 & 0.51 $\pm$ 0.11 & 24 $\pm$ 7\\
C30--F13 & 0.13 & 9 $\pm$ 2 & 0.61 $\pm$ 0.23 & 5 $\pm$ 2\\
C31--F17 & 0.43 & 35 $\pm$ 8 & 0.23 $\pm$ 0.05 & 8 $\pm$ 3\\
C32--F14a & 0.16 & 17 $\pm$ 4 & 1.30 $\pm$ 0.50 & 23 $\pm$ 10\\
C33--F10b & 0.24 & 16 $\pm$ 4 & 1.05 $\pm$ 0.06 & 17 $\pm$ 4\\
C34--F6 & 1.13 & 223 $\pm$ 49 & 0.55 $\pm$ 0.02 & 125 $\pm$ 28\\
C35--F9 & 0.57 & 125 $\pm$ 28 & 0.50 $\pm$ 0.04 & 64 $\pm$ 15\\
C36--F4 & 0.13 & 29 $\pm$ 6 & 0.49 $\pm$ 0.12 & 15 $\pm$ 5\\
C37--F6 & 0.35 & 45 $\pm$ 10 & 0.57 $\pm$ 0.21 & 26 $\pm$ 11\\
C39--F14a & 0.76 & 140 $\pm$ 31 & 1.01 $\pm$ 0.05 & 145 $\pm$ 33\\
C41--F13 & 0.28 & 35 $\pm$ 8 & 0.87 $\pm$ 0.11 & 32 $\pm$ 8\\
C42--F9 & 0.32 & 95 $\pm$ 21 & 1.52 $\pm$ 0.08 & 148 $\pm$ 33\\
C43--F21 & 0.22 & 29 $\pm$ 6 & 0.64 $\pm$ 0.07 & 19 $\pm$ 5\\
C44--F14a & 0.31 & 42 $\pm$ 9 & 0.65 $\pm$ 0.08 & 28 $\pm$ 7\\
C46--F12 & 0.11 & 33 $\pm$ 7 & 0.77 $\pm$ 0.28 & 26 $\pm$ 11\\
C47--F14c & 0.76 & 79 $\pm$ 17 & 0.64 $\pm$ 0.04 & 52 $\pm$ 12\\
C48--F6 & 0.16 & 10 $\pm$ 2 & 0.48 $\pm$ 0.14 & 5 $\pm$ 2\\
C49--F12 & 0.35 & 129 $\pm$ 28 & 0.45 $\pm$ 0.07 & 60 $\pm$ 16\\
C51--F11 & 0.26 & 27 $\pm$ 6 & 0.26 $\pm$ 0.07 & 7 $\pm$ 2\\
C53--F3 & 0.52 & 49 $\pm$ 11 & 0.41 $\pm$ 0.04 & 21 $\pm$ 5\\
C55--F14b & 0.68 & 73 $\pm$ 16 & 0.70 $\pm$ 0.03 & 52 $\pm$ 12\\
C56--F18 & 0.57 & 166 $\pm$ 36 & 0.91 $\pm$ 0.10 & 155 $\pm$ 38\\
C57--F14a & 0.14 & 17 $\pm$ 4 & 0.40 $\pm$ 0.20 & 7 $\pm$ 4\\
C58--F3 & 0.33 & 35 $\pm$ 8 & 0.39 $\pm$ 0.07 & 14 $\pm$ 4\\
C59--F15 & 0.48 & 139 $\pm$ 31 & 1.10 $\pm$ 0.10 & 157 $\pm$ 37\\
C60--F3 & 0.25 & 63 $\pm$ 14 & 0.65 $\pm$ 0.06 & 42 $\pm$ 10\\
C62--F20 & 0.67 & 251 $\pm$ 55 & 0.69 $\pm$ 0.02 & 178 $\pm$ 39\\
C63--F18 & 0.94 & 56 $\pm$ 12 & 0.74 $\pm$ 0.04 & 42 $\pm$ 10\\
\enddata
\tablecomments{(1) Core IDs and the corresponding Filament IDs. (2) Length of the filment along which the velocity gradient is measured. (3) Mass of the filament within $L_{\rm fil}$. (4) Absolute velocity gradient measured along $L_{\rm fil}$ using least-squares fit. (5) Mass flow rate along the filament assuming $\alpha=45^{\circ}$.}
\end{deluxetable}

\section{Variation of Mach Numbers along the Filaments} \label{app:mach}
\begin{figure*}[h]
    \centering
    \includegraphics[width=\textwidth]{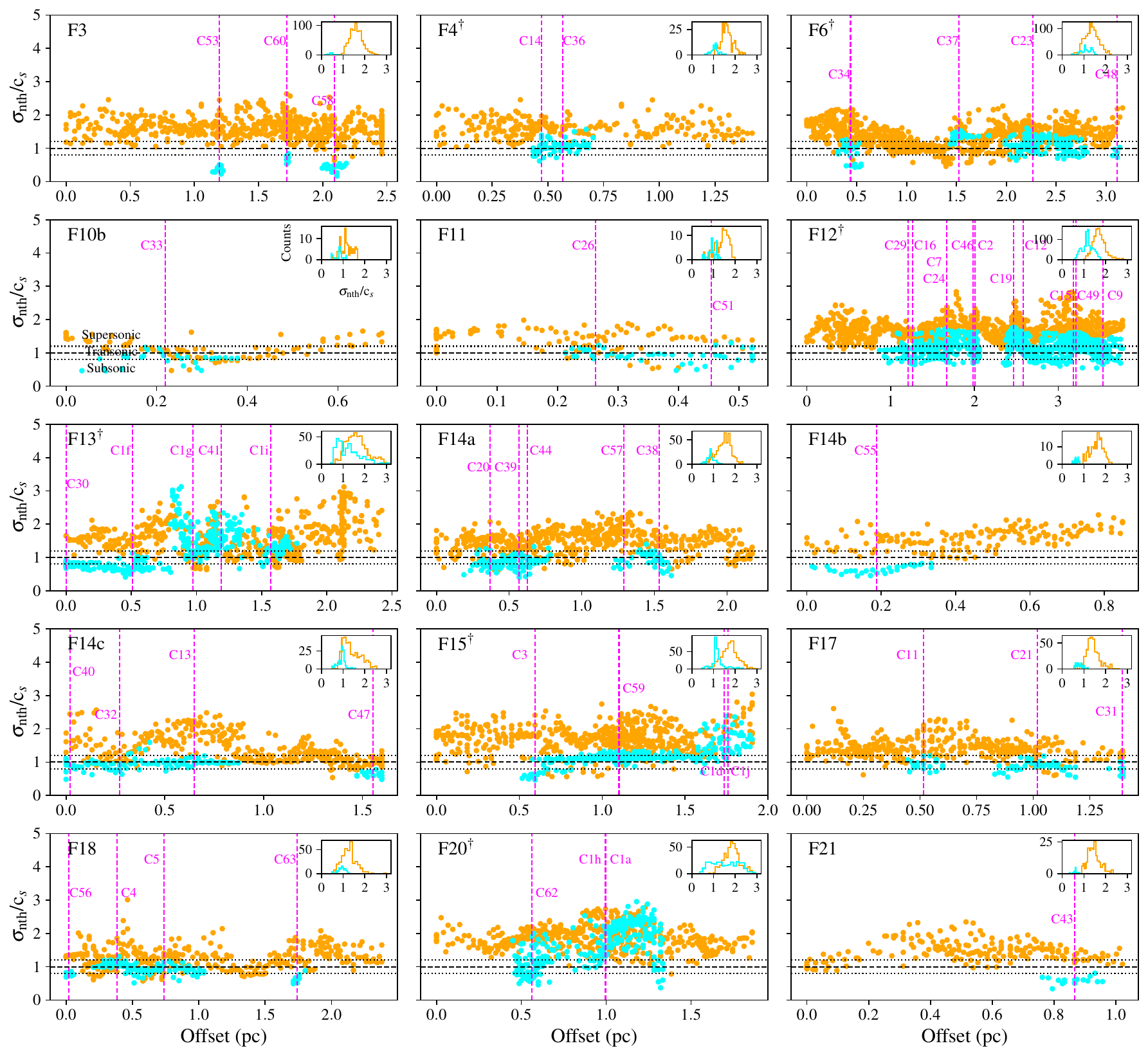}
    \caption{Variation of $\mathcal{M}$ along the filaments' skeletons. The 0 pc offset for each filament starts from the dot marks shown in Figure \ref{fig:filvel}. The filament IDs are labeled at the top-left corner of each panel. The vertical magenta lines indicate the representative position of the dense cores along the filaments' skeletons. The core IDs are mentioned next to these magenta lines. The horizontal dashed line in each panel at 1 marks the boundary between the subsonic and supersonic motions, while the two horizontal dotted lines mark the range of transonic motions. The inset within each panel depicts the histogram of this distribution. $^{\dagger}$: Filaments within Serpens South.}
    \label{fig:mach}
\end{figure*}

The variation of Mach numbers ($\mathcal{M}$), i.e., the non-thermal velocity dispersion divided by the speed of sound along the skeleton of each filament, is presented in Figure \ref{fig:mach}. The orange and blue dots are estimated from Gaussian decomposition of C$^{18}$O spectra (for filaments) and hyperfine structure fitting to the N$_{2}$H$^{+}$ spectra (for dense cores), respectively.

\clearpage

\section{Hill5 Model Fit to Infall Profiles} \label{app:infall}
The Hill5 model is an analytic radiative transfer model proposed by \cite{2005ApJ...620..800D} to explain the blue-asymmetric line profiles indicative of contracting motions. In this model, the contracting cloud is represented by two approaching layers of gas moving inward with the same infall velocity, with excitation temperatures of both the layers linearly increasing toward the inner region as a function of optical depth \citep[see][for a detailed description]{2005ApJ...620..800D}. Considering radiative transfer effects between two layers, the resulting brightness temperature that one can measure is given by,
\begin{equation}
    \Delta T_{\rm B} (v) = J(T_{P})\left[\frac{1-e^{-\tau_{f} (v)}}{\tau_{f} (v)} - \frac{e^{-\tau_{f} (v)}(1-e^{-\tau_{r} (v)})}{\tau_{r} (v)}\right] - J(T_{\rm bg})\left[1-e^{-\tau_{r} (v)-\tau_{f} (v)}\right],
\end{equation}
where $J(T)=\frac{T_{0}}{e^{T_{0}/T}-1}$, $T_{0} = h\nu/k_{\rm B}$, $\nu$ is the frequency, $h$ and $k_{\rm B}$ are Planck and Boltzmann constants, respectively, $T_{P}$ is the peak brightness temperature at the core center, $\tau_{f}=\tau_{0}\exp{\left[-(v- {\rm V_{ LSR}}-v_{\rm in})^{2}/2\sigma^{2}\right]}$ and $\tau_{r}=\tau_{0}\exp{\left[-(v- {\rm V_{ LSR}}+v_{\rm in})^{2}/2\sigma^{2}\right]}$ are the optical depths of the front and rear layers along the line-of-sight, respectively, where $\tau_{0}$ is the optical depth at the core center, V$_{\rm LSR}$ is the systematic velocity of the core, $v_{\rm in}$ is the infall velocity, and $\sigma$ is the velocity dispersion of the gas in both layers.

\begin{figure*}
    \centering
    \includegraphics[width=\textwidth]{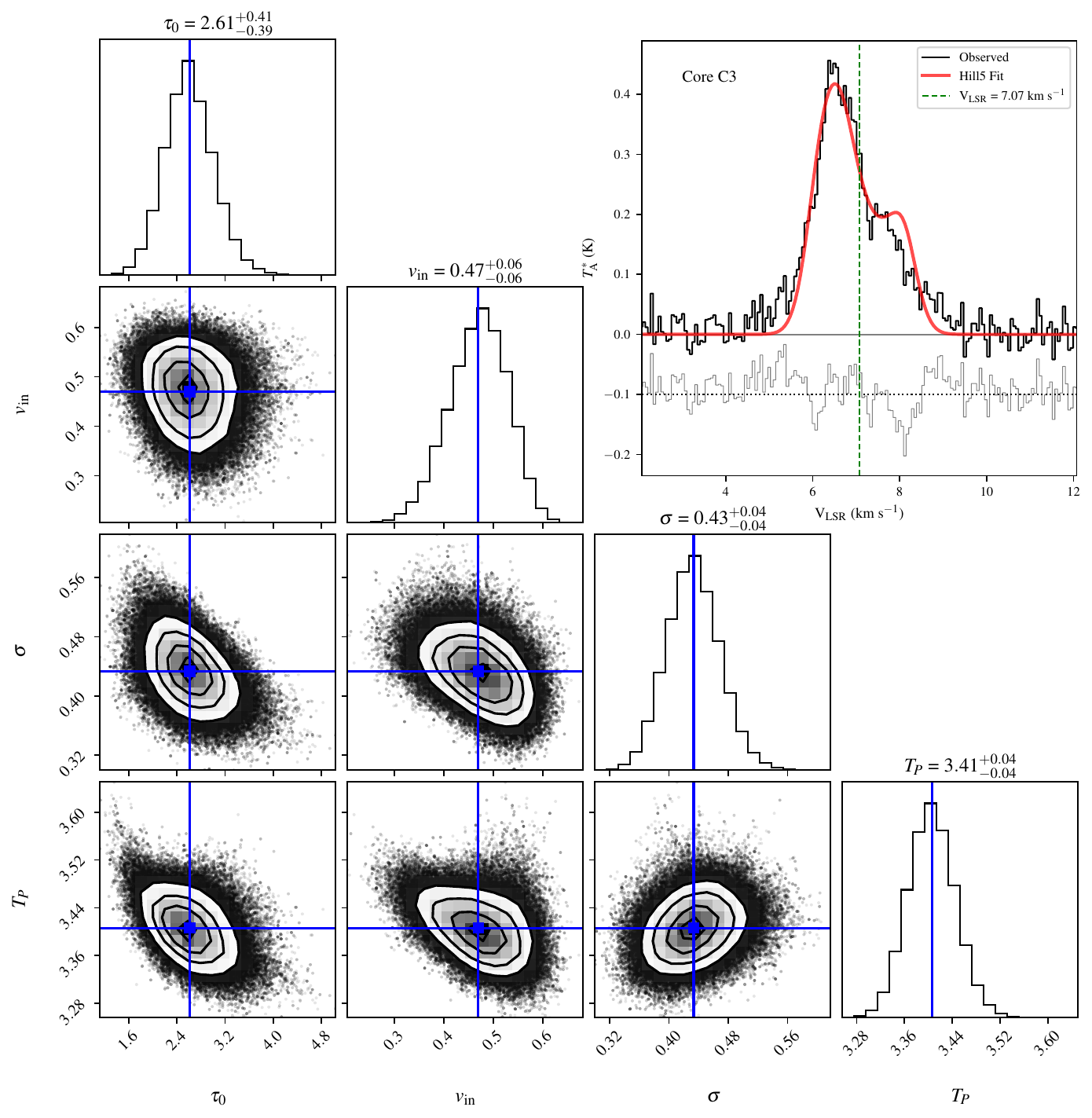}
    \caption{Determination of the infall velocity for core C3. The corner plots display the posterior probability distributions of the free parameters obtained from the MCMC fitting. The blue lines indicate the median values of the parameters. The best-fit Hill5 model (red line) is overplotted on the averaged CS spectrum (black line) in the upper-right panel. The vertical green dashed line marks V$_{\rm LSR}$ derived from the N$_2$H$^+$ line. The residual is shown with a vertical offset of –0.1 K for clarity.}
    \label{fig:corner}
\end{figure*}

$T_{P}$, $\tau_{0}$, $v_{\rm in}$, and $\sigma$ were treated as free parameters during our MCMC sampling. $V_{\rm LSR}$ for each dense core was fixed to the value obtained from hyperfine fitting of the N$_{2}$H$^{+}$ spectra. In the initial MCMC trial, broad parameter ranges were adopted to explore the full parameter space and identify the region containing the global minimum. A subsequent trial was then performed with narrower, physically motivated bounds to refine the fit around the most probable solution. Figure \ref{fig:corner} presents the resulting posterior probability distributions of all parameters from the final MCMC run, along with the best-fit model to the CS line profile of core C3.

\begin{deluxetable}{lccccccc}[h]
\tablecaption{Summary of Infall Velocities and Timescales \label{tab:infall_summary}}
\setlength{\tabcolsep}{17pt}
\tablehead{
\colhead{Core ID} & \colhead{$R_{\rm in}$} & \colhead{$M_{\rm in}$} & \colhead{$v_{\rm in}$} & \colhead{$v_{\rm ff}$} & \colhead{$t_{\rm in}$} & \colhead{$t_{\rm ff}$} & \colhead{$t_{\rm acc}$}\\
\colhead{} & \colhead{(pc)} & \colhead{(M$_{\odot}$)} & \colhead{(km s$^{-1}$)} & \colhead{(km s$^{-1}$)} & \colhead{(Myr)} & \colhead{(Myr)} & \colhead{(Myr)}\\
\colhead{(1)} & \colhead{(2)} & \colhead{(3)} & \colhead{(4)} & \colhead{(5)} & \colhead{(6)} & \colhead{(7)} & \colhead{(8)}
}
\startdata
C2 & 0.22 $\pm$ 0.02 & 114 $\pm$ 25 & $0.43^{+0.04}_{-0.04}$ & 2.12 $\pm$ 0.25 & $0.50^{+0.07}_{-0.07}$ & 0.10 $\pm$ 0.01 & 9.50 $\pm$ 4.61\\
C3 & 0.23 $\pm$ 0.02 & 114 $\pm$ 25 & $0.47^{+0.06}_{-0.06}$ & 2.07 $\pm$ 0.24 & $0.48^{+0.07}_{-0.07}$ & 0.11 $\pm$ 0.02 & 10.09 $\pm$ 3.32\\
C6 & 0.14 $\pm$ 0.02 & 37 $\pm$ 8 & $0.73^{+0.02}_{-0.02}$ & 1.51 $\pm$ 0.20 & $0.19^{+0.03}_{-0.03}$ & 0.09 $\pm$ 0.02 & $\cdots$\\
C7 & 0.19 $\pm$ 0.02 & 129 $\pm$ 28 & $0.45^{+0.11}_{-0.12}$ & 2.42 $\pm$ 0.29 & $0.41^{+0.11}_{-0.12}$ & 0.08 $\pm$ 0.01 & 2.13 $\pm$ 0.71\\
C8 & 0.18 $\pm$ 0.02 & 66 $\pm$ 15 & $0.27^{+0.08}_{-0.08}$ & 1.78 $\pm$ 0.23 & $0.65^{+0.21}_{-0.16}$ & 0.10 $\pm$ 0.02 & 8.47 $\pm$ 3.22\\
C11 & 0.12 $\pm$ 0.01 & 23 $\pm$ 5 & $0.59^{+0.08}_{-0.09}$ & 1.29 $\pm$ 0.15 & $0.20^{+0.03}_{-0.03}$ & 0.09 $\pm$ 0.01 & 0.97 $\pm$ 0.33\\
C14 & 0.13 $\pm$ 0.01 & 31 $\pm$ 7 & $0.41^{+0.18}_{-0.06}$ & 1.44 $\pm$ 0.17 & $0.31^{+0.14}_{-0.05}$ & 0.09 $\pm$ 0.01 & 2.58 $\pm$ 1.02\\
C16 & 0.19 $\pm$ 0.02 & 47 $\pm$ 10 & $0.32^{+0.07}_{-0.06}$ & 1.46 $\pm$ 0.17 & $0.58^{+0.14}_{-0.12}$ & 0.13 $\pm$ 0.02 & 2.25 $\pm$ 0.85\\
C21 & 0.12 $\pm$ 0.01 & 24 $\pm$ 5 & $0.22^{+0.05}_{-0.04}$ & 1.32 $\pm$ 0.15 & $0.53^{+0.13}_{-0.11}$ & 0.09 $\pm$ 0.01 & 3.92 $\pm$ 2.27\\
C28 & 0.16 $\pm$ 0.02 & 43 $\pm$ 10 & $0.33^{+0.11}_{-0.10}$ & 1.52 $\pm$ 0.20 & $0.47^{+0.17}_{-0.16}$ & 0.10 $\pm$ 0.02 & 0.78 $\pm$ 0.26\\
C32 & 0.12 $\pm$ 0.01 & 23 $\pm$ 5 & $0.27^{+0.09}_{-0.10}$ & 1.29 $\pm$ 0.15 & $0.43^{+0.15}_{-0.17}$ & 0.09 $\pm$ 0.01 & 1.21 $\pm$ 0.6\\
C36 & 0.08 $\pm$ 0.01 & 12 $\pm$ 3 & $0.48^{+0.04}_{-0.09}$ & 1.14 $\pm$ 0.16 & $0.16^{+0.02}_{-0.04}$ & 0.07 $\pm$ 0.01 & 0.91 $\pm$ 0.36\\
C40 & 0.06 $\pm$ 0.01 & 11 $\pm$ 2 & $0.47^{+0.09}_{-0.12}$ & 1.26 $\pm$ 0.16 & $0.12^{+0.03}_{-0.04}$ & 0.05 $\pm$ 0.01 & $\cdots$\\
C48 & 0.07 $\pm$ 0.01 & 3 $\pm$ 1 & $0.42^{+0.02}_{-0.03}$ & 0.61 $\pm$ 0.11 & $0.16^{+0.02}_{-0.03}$ & 0.11 $\pm$ 0.04 & 0.47 $\pm$ 0.21\\
C52 & 0.03 $\pm$ 0.01 & 7 $\pm$ 1 & $0.11^{+0.05}_{-0.05}$ & 1.42 $\pm$ 0.26 & $0.27^{+0.15}_{-0.15}$ & 0.02 $\pm$ 0.01 & $\cdots$\\
\enddata
\tablecomments{(1) Core IDs. (2) and (3) Radius and mass of each dense core within the 30\% intensity contour, respectively. (4) and (5) Infall and free-fall velocities, respectively. (6), (7) and (8) Infall, free-fall and accretion timescales. The uncertainties in $v_{\rm in}$ are obtained from the Hill5 model fitting.}
\end{deluxetable}

\clearpage
\section{Spectral Profiles of Dense Cores} \label{app:profiles}
\begin{figure*}[b]
    \centering
    \includegraphics[width=\textwidth]{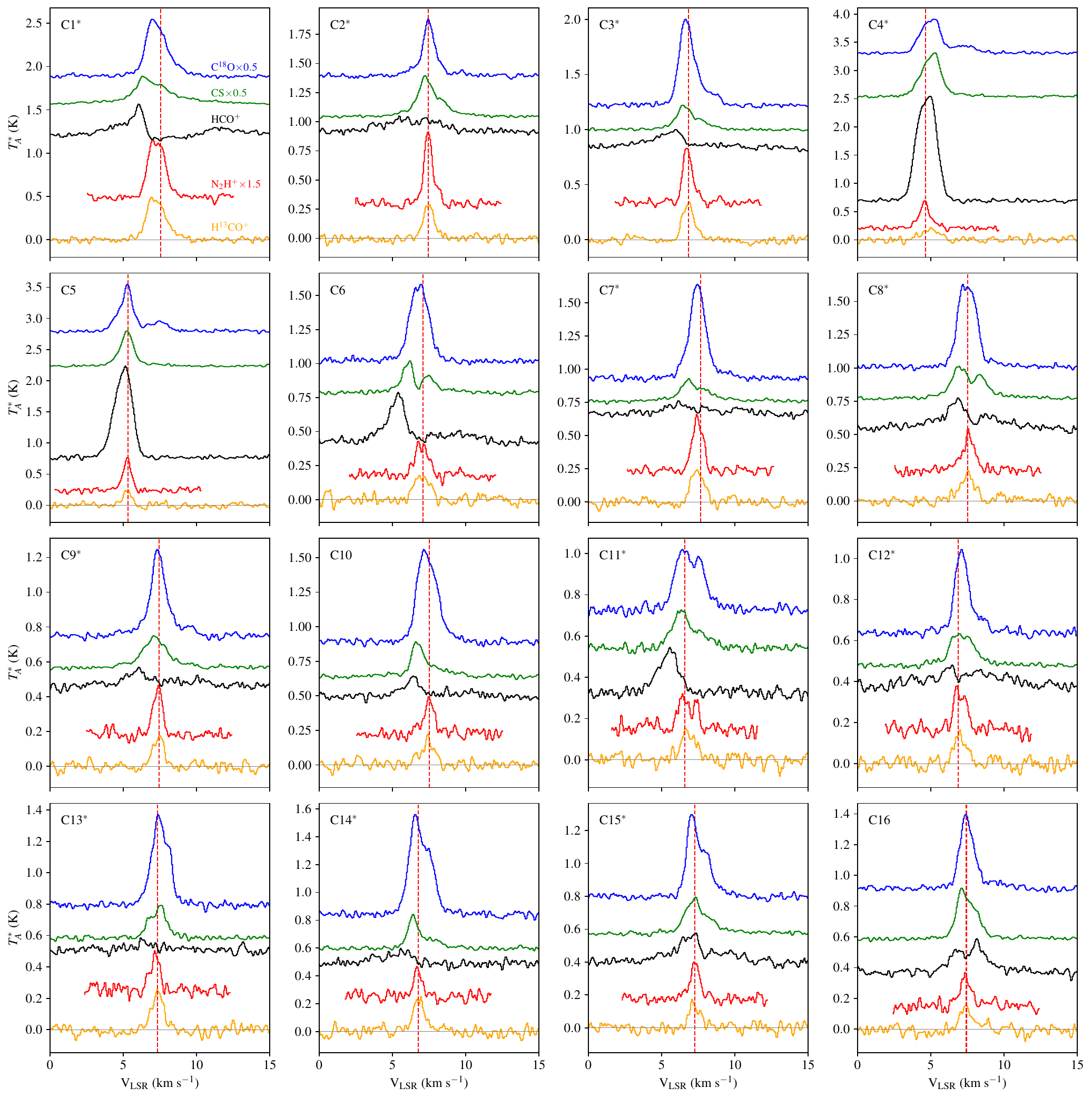}
    \caption{The spectral profiles of various molecules across the identified dense cores (C1-C16). For N$_{2}$H$^{+}$, only the isolated hyperfine component is shown. The C$^{18}$O and CS spectra are multiplied with a factor of 0.5, while N$_{2}$H$^{+}$ spectra is multiplied with 1.5 for better visualisation. The vertical red dashed line marks the systematic velocity of the dense core obtained from the hyperfine structure fitting to the N$_{2}$H$^{+}$ spectra.}
    \label{fig:spectral_profiles1}
\end{figure*}

\begin{figure*}
    \centering
    \includegraphics[width=\textwidth]{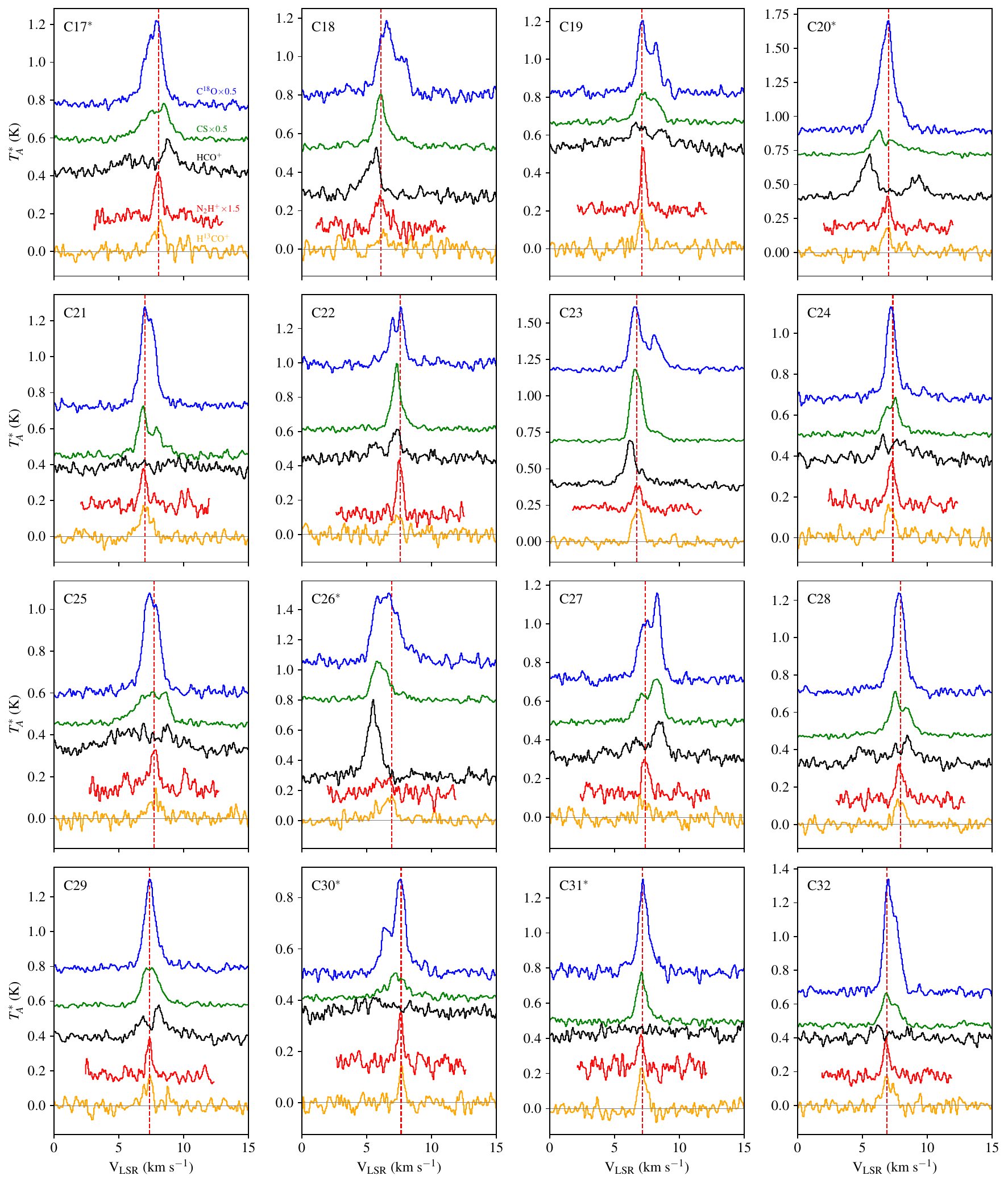}
    \caption{Same as Figure \ref{fig:spectral_profiles1} for dense cores C17-C32.}
    \label{fig:spectral_profiles2}
\end{figure*}

\begin{figure*}
    \centering
    \includegraphics[width=\textwidth]{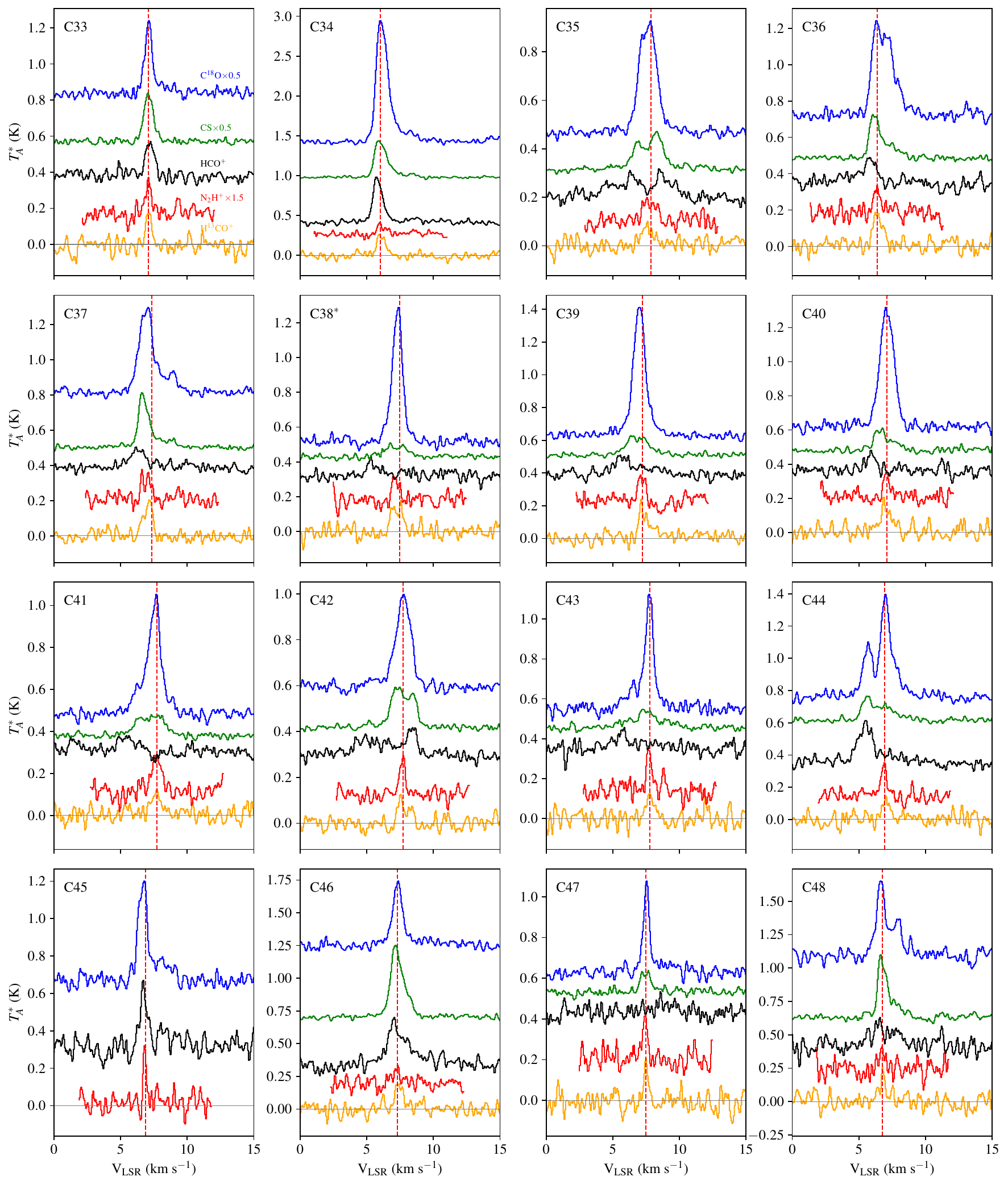}
    \caption{Same as Figure \ref{fig:spectral_profiles1} for dense cores C33-C48.}
    \label{fig:spectral_profiles3}
\end{figure*}

\begin{figure*}
    \centering
    \includegraphics[width=\textwidth]{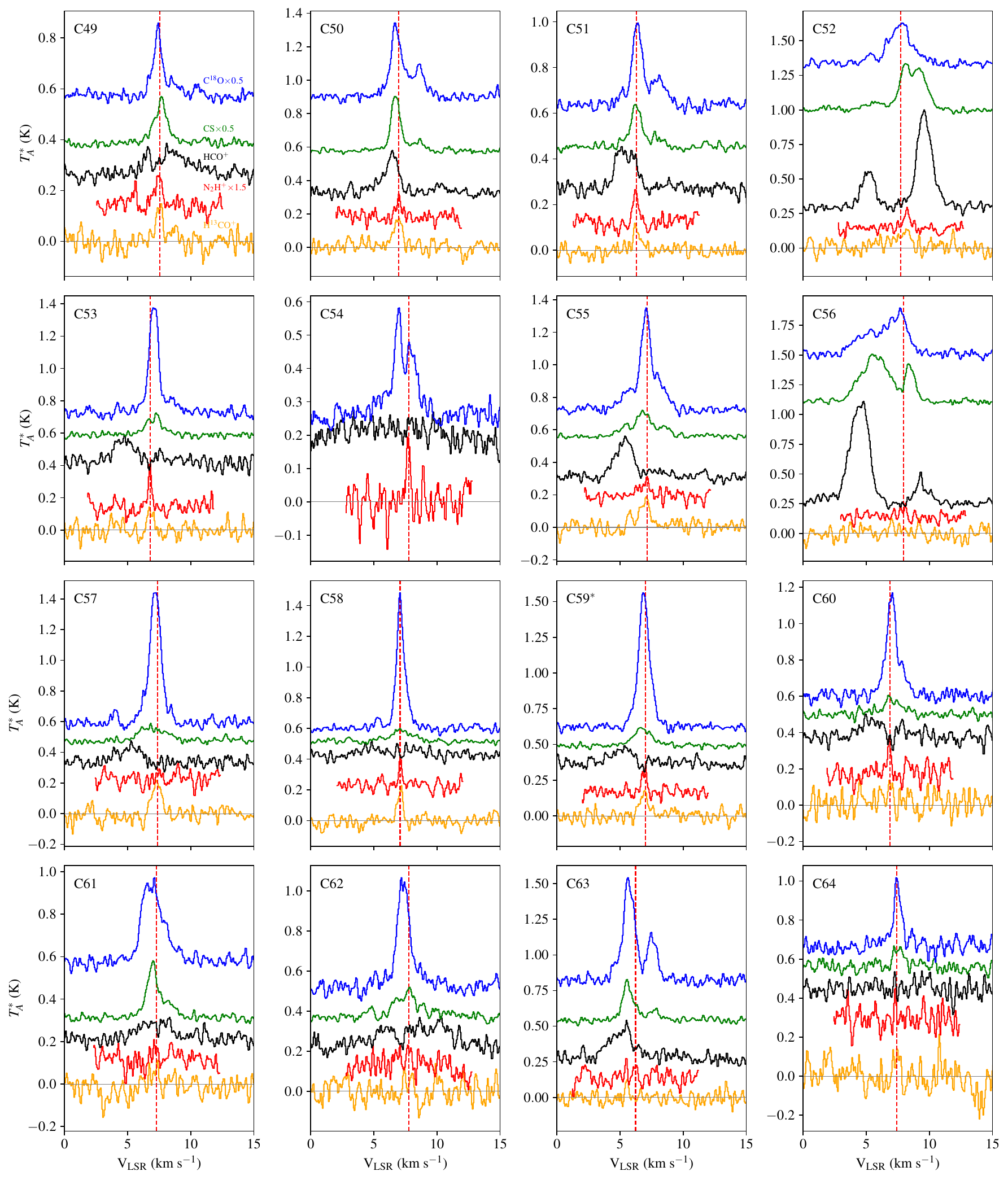}
    \caption{Same as Figure \ref{fig:spectral_profiles1} for dense cores C49-C64.}
    \label{fig:spectral_profiles4}
\end{figure*}

The spectral profiles of N$_{2}$H$^{+}$, H$^{13}$CO$^{+}$, HCO$^{+}$, CS, and C$^{18}$O, averaged over the 30\% intensity contours of the 64 identified dense cores, are shown in Figures \ref{fig:spectral_profiles1}, \ref{fig:spectral_profiles2}, \ref{fig:spectral_profiles3}, and \ref{fig:spectral_profiles4}. Each spectrum has been smoothed by a factor of 2 along the velocity axis to enhance the signal-to-noise ratio. The names of the cores are provided in the top-left corner of each panel. The names with an asterisk symbol next to them represent the dense cores with YSOs.

\clearpage

\bibliography{References}{}

@ARTICLE{2019ApJ...877..114C,
       author = {{Chung}, Eun Jung and {Lee}, Chang Won and {Kim}, Shinyoung and {Kim}, Gwanjeong and {Caselli}, Paola and {Tafalla}, Mario and {Myers}, Philip C. and {Soam}, Archana and {Liu}, Tie and {Gopinathan}, Maheswar and {Kim}, Miryang and {Kim}, Kyoung Hee and {Kwon}, Woojin and {Kang}, Hyunwoo and {Lee}, Changhoon},
        title = "{TRAO Survey of Nearby Filamentary Molecular Clouds, the Universal Nursery of Stars (TRAO FUNS). I. Dynamics and Chemistry of L1478 in the California Molecular Cloud}",
      journal = {\apj},
         year = 2019,
        month = jun,
       volume = {877},
       number = {2},
          eid = {114},
        pages = {114},
          doi = {10.3847/1538-4357/ab12d1},
archivePrefix = {arXiv},
       eprint = {1904.10594},
 primaryClass = {astro-ph.GA},
       adsurl = {https://ui.adsabs.harvard.edu/abs/2019ApJ...877..114C}
}

@ARTICLE{2021ApJ...919....3C,
       author = {{Chung}, Eun Jung and {Lee}, Chang Won and {Kim}, Shinyoung and {Gopinathan}, Maheswar and {Tafalla}, Mario and {Caselli}, Paola and {Myers}, Philip C. and {Liu}, Tie and {Yoo}, Hyunju and {Kim}, Kyoung Hee and {Kim}, Mi-Ryang and {Soam}, Archana and {Cho}, Jungyeon and {Kwon}, Woojin and {Lee}, Changhoon and {Kang}, Hyunwoo},
        title = "{TRAO Survey of the Nearby Filamentary Molecular Clouds, the Universal Nursery of Stars (TRAO FUNS). II. Filaments and Dense Cores in IC 5146}",
      journal = {\apj},
         year = 2021,
        month = sep,
       volume = {919},
       number = {1},
          eid = {3},
        pages = {3},
          doi = {10.3847/1538-4357/ac0881},
archivePrefix = {arXiv},
       eprint = {2106.03897},
 primaryClass = {astro-ph.GA},
       adsurl = {https://ui.adsabs.harvard.edu/abs/2021ApJ...919....3C}
}

@ARTICLE{2001ApJS..136..703L,
       author = {{Lee}, Chang Won and {Myers}, Philip C. and {Tafalla}, Mario},
        title = "{A Survey for Infall Motions toward Starless Cores. II. CS (2-1) and N$_{2}$H$^{+}$ (1-0) Mapping Observations}",
      journal = {\apjs},
         year = 2001,
        month = oct,
       volume = {136},
       number = {2},
        pages = {703-734},
          doi = {10.1086/322534},
archivePrefix = {arXiv},
       eprint = {astro-ph/0105515},
 primaryClass = {astro-ph},
       adsurl = {https://ui.adsabs.harvard.edu/abs/2001ApJS..136..703L}
}

@ARTICLE{2023ApJ...957...94Y,
       author = {{Yoo}, Hyunju and {Lee}, Chang Won and {Chung}, Eun Jung and {Kim}, Shinyoung and {Tafalla}, Mario and {Caselli}, Paola and {Myers}, Philip C. and {Kim}, Kyoung Hee and {Liu}, Tie and {Kwon}, Woojin and {Soam}, Archana and {Kim}, Jongsoo},
        title = "{TRAO Survey of Nearby Filamentary Molecular Clouds, the Universal Nursery of Stars (TRAO-FUNS). III. Filaments and Dense Cores in the NGC 2068 and NGC 2071 Regions of Orion B}",
      journal = {\apj},
         year = 2023,
        month = nov,
       volume = {957},
       number = {2},
          eid = {94},
        pages = {94},
          doi = {10.3847/1538-4357/acf8c2},
archivePrefix = {arXiv},
       eprint = {2309.09683},
 primaryClass = {astro-ph.GA},
       adsurl = {https://ui.adsabs.harvard.edu/abs/2023ApJ...957...94Y}
}

@ARTICLE{2015A&A...584A..91K,
       author = {{K{\"o}nyves}, V. and {Andr{\'e}}, Ph. and {Men'shchikov}, A. and {Palmeirim}, P. and {Arzoumanian}, D. and {Schneider}, N. and {Roy}, A. and {Didelon}, P. and {Maury}, A. and {Shimajiri}, Y. and {Di Francesco}, J. and {Bontemps}, S. and {Peretto}, N. and {Benedettini}, M. and {Bernard}, J. -Ph. and {Elia}, D. and {Griffin}, M.~J. and {Hill}, T. and {Kirk}, J. and {Ladjelate}, B. and {Marsh}, K. and {Martin}, P.~G. and {Motte}, F. and {Nguy{\^e}n Luong}, Q. and {Pezzuto}, S. and {Roussel}, H. and {Rygl}, K.~L.~J. and {Sadavoy}, S.~I. and {Schisano}, E. and {Spinoglio}, L. and {Ward-Thompson}, D. and {White}, G.~J.},
        title = "{A census of dense cores in the Aquila cloud complex: SPIRE/PACS observations from the Herschel Gould Belt survey}",
      journal = {\aap},
         year = 2015,
        month = dec,
       volume = {584},
          eid = {A91},
        pages = {A91},
          doi = {10.1051/0004-6361/201525861},
archivePrefix = {arXiv},
       eprint = {1507.05926},
 primaryClass = {astro-ph.GA},
       adsurl = {https://ui.adsabs.harvard.edu/abs/2015A&A...584A..91K}
}

@ARTICLE{2015MNRAS.452.3435K,
       author = {{Koch}, Eric W. and {Rosolowsky}, Erik W.},
        title = "{Filament identification through mathematical morphology}",
      journal = {\mnras},
         year = 2015,
        month = oct,
       volume = {452},
       number = {4},
        pages = {3435-3450},
          doi = {10.1093/mnras/stv1521},
archivePrefix = {arXiv},
       eprint = {1507.02289},
 primaryClass = {astro-ph.GA},
       adsurl = {https://ui.adsabs.harvard.edu/abs/2015MNRAS.452.3435K}
}

@INPROCEEDINGS{2014prpl.conf...27A,
       author = {{Andr{\'e}}, P. and {Di Francesco}, J. and {Ward-Thompson}, D. and {Inutsuka}, S. -I. and {Pudritz}, R.~E. and {Pineda}, J.~E.},
        title = "{From Filamentary Networks to Dense Cores in Molecular Clouds: Toward a New Paradigm for Star Formation}",
    booktitle = {Protostars and Planets VI},
         year = 2014,
       editor = {{Beuther}, Henrik and {Klessen}, Ralf S. and {Dullemond}, Cornelis P. and {Henning}, Thomas},
        month = jan,
        pages = {27-51},
          doi = {10.2458/azu_uapress_9780816531240-ch002},
archivePrefix = {arXiv},
       eprint = {1312.6232},
 primaryClass = {astro-ph.GA},
       adsurl = {https://ui.adsabs.harvard.edu/abs/2014prpl.conf...27A}
}

@ARTICLE{2008A&A...487..993K,
       author = {{Kauffmann}, J. and {Bertoldi}, F. and {Bourke}, T.~L. and {Evans}, II, N.~J. and {Lee}, C.~W.},
        title = "{MAMBO mapping of Spitzer c2d small clouds and cores}",
      journal = {\aap},
         year = 2008,
        month = sep,
       volume = {487},
       number = {3},
        pages = {993-1017},
          doi = {10.1051/0004-6361:200809481},
archivePrefix = {arXiv},
       eprint = {0805.4205},
 primaryClass = {astro-ph},
       adsurl = {https://ui.adsabs.harvard.edu/abs/2008A&A...487..993K}
}

@ARTICLE{2015A&C....10...22B,
       author = {{Berry}, D.~S.},
        title = "{FellWalker-A clump identification algorithm}",
      journal = {Astronomy and Computing},
         year = 2015,
        month = apr,
       volume = {10},
        pages = {22-31},
          doi = {10.1016/j.ascom.2014.11.004},
archivePrefix = {arXiv},
       eprint = {1411.6267},
 primaryClass = {astro-ph.IM},
       adsurl = {https://ui.adsabs.harvard.edu/abs/2015A&C....10...22B}
}

@INPROCEEDINGS{2007ASPC..376..425B,
       author = {{Berry}, D.~S. and {Reinhold}, K. and {Jenness}, T. and {Economou}, F.},
        title = "{CUPID: A Clump Identification and Analysis Package}",
    booktitle = {Astronomical Data Analysis Software and Systems XVI},
         year = 2007,
       editor = {{Shaw}, R.~A. and {Hill}, F. and {Bell}, D.~J.},
       series = {Astronomical Society of the Pacific Conference Series},
       volume = {376},
        month = oct,
        pages = {425},
       adsurl = {https://ui.adsabs.harvard.edu/abs/2007ASPC..376..425B}
}

@ARTICLE{2020ApJ...895..137S,
       author = {{Shimoikura}, Tomomi and {Dobashi}, Kazuhito and {Hatano}, Yoshiko and {Nakamura}, Fumitaka},
        title = "{A Detailed Analysis of the Cloud Structure and Dynamics in Aquila Rift}",
      journal = {\apj},
         year = 2020,
        month = jun,
       volume = {895},
       number = {2},
          eid = {137},
        pages = {137},
          doi = {10.3847/1538-4357/ab8c4f},
       adsurl = {https://ui.adsabs.harvard.edu/abs/2020ApJ...895..137S}
}

@ARTICLE{2011A&A...535A..77M,
       author = {{Maury}, A.~J. and {Andr{\'e}}, P. and {Men'shchikov}, A. and {K{\"o}nyves}, V. and {Bontemps}, S.},
        title = "{The formation of active protoclusters in the Aquila rift: a millimeter continuum view}",
      journal = {\aap},
         year = 2011,
        month = nov,
       volume = {535},
          eid = {A77},
        pages = {A77},
          doi = {10.1051/0004-6361/201117132},
archivePrefix = {arXiv},
       eprint = {1108.0668},
 primaryClass = {astro-ph.GA},
       adsurl = {https://ui.adsabs.harvard.edu/abs/2011A&A...535A..77M}
}

@ARTICLE{2013ApJ...766..115K,
       author = {{Kirk}, Helen and {Myers}, Philip C. and {Bourke}, Tyler L. and {Gutermuth}, Robert A. and {Hedden}, Abigail and {Wilson}, Grant W.},
        title = "{Filamentary Accretion Flows in the Embedded Serpens South Protocluster}",
      journal = {\apj},
         year = 2013,
        month = apr,
       volume = {766},
       number = {2},
          eid = {115},
        pages = {115},
          doi = {10.1088/0004-637X/766/2/115},
archivePrefix = {arXiv},
       eprint = {1301.6792},
 primaryClass = {astro-ph.GA},
       adsurl = {https://ui.adsabs.harvard.edu/abs/2013ApJ...766..115K}
}

@ARTICLE{2016ApJ...833..204F,
       author = {{Friesen}, R.~K. and {Bourke}, T.~L. and {Di Francesco}, J. and {Gutermuth}, R. and {Myers}, P.~C.},
        title = "{The Fragmentation and Stability of Hierarchical Structure in Serpens South}",
      journal = {\apj},
         year = 2016,
        month = dec,
       volume = {833},
       number = {2},
          eid = {204},
        pages = {204},
          doi = {10.3847/1538-4357/833/2/204},
archivePrefix = {arXiv},
       eprint = {1610.10066},
 primaryClass = {astro-ph.GA},
       adsurl = {https://ui.adsabs.harvard.edu/abs/2016ApJ...833..204F}
}

@ARTICLE{2024ApJ...969...70F,
       author = {{Friesen}, Rachel K. and {Jarvis}, Emma},
        title = "{The Stability of Dense Cores near the Serpens South Protocluster}",
      journal = {\apj},
         year = 2024,
        month = jul,
       volume = {969},
       number = {1},
          eid = {70},
        pages = {70},
          doi = {10.3847/1538-4357/ad435b},
archivePrefix = {arXiv},
       eprint = {2404.07259},
 primaryClass = {astro-ph.GA},
       adsurl = {https://ui.adsabs.harvard.edu/abs/2024ApJ...969...70F}
}

@ARTICLE{1991ApJ...374..540G,
       author = {{Garden}, R.~P. and {Hayashi}, M. and {Gatley}, I. and {Hasegawa}, T. and {Kaifu}, N.},
        title = "{A Spectroscopic Study of the DR 21 Outflow Source. III. The CO Line Emission}",
      journal = {\apj},
         year = 1991,
        month = jun,
       volume = {374},
        pages = {540},
          doi = {10.1086/170143},
       adsurl = {https://ui.adsabs.harvard.edu/abs/1991ApJ...374..540G}
}

@ARTICLE{2015MNRAS.450.1094P,
       author = {{Pattle}, K. and {Ward-Thompson}, D. and {Kirk}, J.~M. and {White}, G.~J. and {Drabek-Maunder}, E. and {Buckle}, J. and {Beaulieu}, S.~F. and {Berry}, D.~S. and {Broekhoven-Fiene}, H. and {Currie}, M.~J. and {Fich}, M. and {Hatchell}, J. and {Kirk}, H. and {Jenness}, T. and {Johnstone}, D. and {Mottram}, J.~C. and {Nutter}, D. and {Pineda}, J.~E. and {Quinn}, C. and {Salji}, C. and {Tisi}, S. and {Walker-Smith}, S. and {di Francesco}, J. and {Hogerheijde}, M.~R. and {Andr{\'e}}, Ph. and {Bastien}, P. and {Bresnahan}, D. and {Butner}, H. and {Chen}, M. and {Chrysostomou}, A. and {Coude}, S. and {Davis}, C.~J. and {Duarte-Cabral}, A. and {Fiege}, J. and {Friberg}, P. and {Friesen}, R. and {Fuller}, G.~A. and {Graves}, S. and {Greaves}, J. and {Gregson}, J. and {Griffin}, M.~J. and {Holland}, W. and {Joncas}, G. and {Knee}, L.~B.~G. and {K{\"o}nyves}, V. and {Mairs}, S. and {Marsh}, K. and {Matthews}, B.~C. and {Moriarty-Schieven}, G. and {Rawlings}, J. and {Richer}, J. and {Robertson}, D. and {Rosolowsky}, E. and {Rumble}, D. and {Sadavoy}, S. and {Spinoglio}, L. and {Thomas}, H. and {Tothill}, N. and {Viti}, S. and {Wouterloot}, J. and {Yates}, J. and {Zhu}, M.},
        title = "{The JCMT Gould Belt Survey: first results from the SCUBA-2 observations of the Ophiuchus molecular cloud and a virial analysis of its prestellar core population}",
      journal = {\mnras},
         year = 2015,
        month = jun,
       volume = {450},
       number = {1},
        pages = {1094-1122},
          doi = {10.1093/mnras/stv376},
archivePrefix = {arXiv},
       eprint = {1502.05858},
 primaryClass = {astro-ph.GA},
       adsurl = {https://ui.adsabs.harvard.edu/abs/2015MNRAS.450.1094P}
}

@ARTICLE{2008ApJ...679..481P,
       author = {{Pineda}, Jaime E. and {Caselli}, Paola and {Goodman}, Alyssa A.},
        title = "{CO Isotopologues in the Perseus Molecular Cloud Complex: the X-factor and Regional Variations}",
      journal = {\apj},
         year = 2008,
        month = may,
       volume = {679},
       number = {1},
        pages = {481-496},
          doi = {10.1086/586883},
archivePrefix = {arXiv},
       eprint = {0802.0708},
 primaryClass = {astro-ph},
       adsurl = {https://ui.adsabs.harvard.edu/abs/2008ApJ...679..481P}
}

@ARTICLE{1982ApJ...262..590F,
       author = {{Frerking}, M.~A. and {Langer}, W.~D. and {Wilson}, R.~W.},
        title = "{The relationship between carbon monoxide abundance and visual extinction in interstellar clouds.}",
      journal = {\apj},
         year = 1982,
        month = nov,
       volume = {262},
        pages = {590-605},
          doi = {10.1086/160451},
       adsurl = {https://ui.adsabs.harvard.edu/abs/1982ApJ...262..590F}
}

@ARTICLE{2019PASJ...71S...3N,
       author = {{Nakamura}, Fumitaka and {Ishii}, Shun and {Dobashi}, Kazuhito and {Shimoikura}, Tomomi and {Shimajiri}, Yoshito and {Kawabe}, Ryohei and {Tanabe}, Yoshihiro and {Hirose}, Asha and {Oyamada}, Shuri and {Urasawa}, Yumiko and {Takemura}, Hideaki and {Tsukagoshi}, Takashi and {Momose}, Munetake and {Sugitani}, Koji and {Nishi}, Ryoichi and {Okumura}, Sachiko and {Sanhueza}, Patricio and {Nguyen-Luong}, Quang and {Kusune}, Takayoshi},
        title = "{Nobeyama 45 m mapping observations toward the nearby molecular clouds Orion A, Aquila Rift, and M17: Project overview}",
      journal = {\pasj},
         year = 2019,
        month = dec,
       volume = {71},
          eid = {S3},
        pages = {S3},
          doi = {10.1093/pasj/psz057},
archivePrefix = {arXiv},
       eprint = {1909.05980},
 primaryClass = {astro-ph.GA},
       adsurl = {https://ui.adsabs.harvard.edu/abs/2019PASJ...71S...3N}
}

@ARTICLE{1999ApJ...523L.165C,
       author = {{Caselli}, P. and {Walmsley}, C.~M. and {Tafalla}, M. and {Dore}, L. and {Myers}, P.~C.},
        title = "{CO Depletion in the Starless Cloud Core L1544}",
      journal = {\apjl},
         year = 1999,
        month = oct,
       volume = {523},
       number = {2},
        pages = {L165-L169},
          doi = {10.1086/312280},
       adsurl = {https://ui.adsabs.harvard.edu/abs/1999ApJ...523L.165C}
}

@ARTICLE{2011A&A...529L...6A,
       author = {{Arzoumanian}, D. and {Andr{\'e}}, Ph. and {Didelon}, P. and {K{\"o}nyves}, V. and {Schneider}, N. and {Men'shchikov}, A. and {Sousbie}, T. and {Zavagno}, A. and {Bontemps}, S. and {di Francesco}, J. and {Griffin}, M. and {Hennemann}, M. and {Hill}, T. and {Kirk}, J. and {Martin}, P. and {Minier}, V. and {Molinari}, S. and {Motte}, F. and {Peretto}, N. and {Pezzuto}, S. and {Spinoglio}, L. and {Ward-Thompson}, D. and {White}, G. and {Wilson}, C.~D.},
        title = "{Characterizing interstellar filaments with Herschel in IC 5146}",
      journal = {\aap},
         year = 2011,
        month = may,
       volume = {529},
          eid = {L6},
        pages = {L6},
          doi = {10.1051/0004-6361/201116596},
archivePrefix = {arXiv},
       eprint = {1103.0201},
 primaryClass = {astro-ph.GA},
       adsurl = {https://ui.adsabs.harvard.edu/abs/2011A&A...529L...6A}
}

@ARTICLE{2015Natur.527...70P,
       author = {{Plunkett}, Adele L. and {Arce}, H{\'e}ctor G. and {Mardones}, Diego and {van Dokkum}, Pieter and {Dunham}, Michael M. and {Fern{\'a}ndez-L{\'o}pez}, Manuel and {Gallardo}, Jos{\'e} and {Corder}, Stuartt A.},
        title = "{Episodic molecular outflow in the very young protostellar cluster Serpens South}",
      journal = {\nat},
         year = 2015,
        month = nov,
       volume = {527},
       number = {7576},
        pages = {70-73},
          doi = {10.1038/nature15702},
archivePrefix = {arXiv},
       eprint = {1511.01100},
 primaryClass = {astro-ph.SR},
       adsurl = {https://ui.adsabs.harvard.edu/abs/2015Natur.527...70P}
}

@ARTICLE{1963AcA....13...30S,
       author = {{Stod{\'o}lkiewicz}, J.~S.},
        title = "{On the Gravitational Instability of Some Magneto-Hydrodynamical Systems of Astrophysical Interest. Part III.}",
      journal = {\actaa},
         year = 1963,
        month = jan,
       volume = {13},
        pages = {30-54},
       adsurl = {https://ui.adsabs.harvard.edu/abs/1963AcA....13...30S}
}

@ARTICLE{1964ApJ...140.1056O,
       author = {{Ostriker}, J.},
        title = "{The Equilibrium of Polytropic and Isothermal Cylinders.}",
      journal = {\apj},
         year = 1964,
        month = oct,
       volume = {140},
        pages = {1056},
          doi = {10.1086/148005},
       adsurl = {https://ui.adsabs.harvard.edu/abs/1964ApJ...140.1056O}
}

@ARTICLE{1995ApJ...455L..77C,
       author = {{Caselli}, P. and {Myers}, P.~C. and {Thaddeus}, P.},
        title = "{Radio-astronomical Spectroscopy of the Hyperfine Structure of N 2H +}",
      journal = {\apjl},
         year = 1995,
        month = dec,
       volume = {455},
        pages = {L77},
          doi = {10.1086/309805},
       adsurl = {https://ui.adsabs.harvard.edu/abs/1995ApJ...455L..77C}
}

@ARTICLE{2023ApJS..266...32P,
       author = {{Pokhrel}, Riwaj and {Megeath}, S. Thomas and {Gutermuth}, Robert A. and {Furlan}, Elise and {Fischer}, William J. and {Federman}, Samuel and {Tobin}, John J. and {Stutz}, Amelia M. and {Hartmann}, Lee and {Osorio}, Mayra and {Watson}, Dan M. and {Stanke}, Thomas and {Manoj}, P. and {Narang}, Mayank and {Atnagulov}, Prabhani and {Habel}, Nolan and {Zakri}, Wafa},
        title = "{Extension of HOPS out to 500 pc (eHOPS). I. Identification and Modeling of Protostars in the Aquila Molecular Clouds}",
      journal = {\apjs},
         year = 2023,
        month = jun,
       volume = {266},
       number = {2},
          eid = {32},
        pages = {32},
          doi = {10.3847/1538-4365/acbfac},
archivePrefix = {arXiv},
       eprint = {2209.12090},
 primaryClass = {astro-ph.GA},
       adsurl = {https://ui.adsabs.harvard.edu/abs/2023ApJS..266...32P}
}

@ARTICLE{2002ApJ...572..238C,
       author = {{Caselli}, Paola and {Benson}, Priscilla J. and {Myers}, Philip C. and {Tafalla}, Mario},
        title = "{Dense Cores in Dark Clouds. XIV. N$_{2}$H$^{+}$ (1-0) Maps of Dense Cloud Cores}",
      journal = {\apj},
         year = 2002,
        month = jun,
       volume = {572},
       number = {1},
        pages = {238-263},
          doi = {10.1086/340195},
archivePrefix = {arXiv},
       eprint = {astro-ph/0202173},
 primaryClass = {astro-ph},
       adsurl = {https://ui.adsabs.harvard.edu/abs/2002ApJ...572..238C}
}

@ARTICLE{1992ApJ...395..140B,
       author = {{Bertoldi}, Frank and {McKee}, Christopher F.},
        title = "{Pressure-confined Clumps in Magnetized Molecular Clouds}",
      journal = {\apj},
         year = 1992,
        month = aug,
       volume = {395},
        pages = {140},
          doi = {10.1086/171638},
       adsurl = {https://ui.adsabs.harvard.edu/abs/1992ApJ...395..140B}
}

@ARTICLE{2013ApJ...779..185K,
       author = {{Kauffmann}, Jens and {Pillai}, Thushara and {Goldsmith}, Paul F.},
        title = "{Low Virial Parameters in Molecular Clouds: Implications for High-mass Star Formation and Magnetic Fields}",
      journal = {\apj},
         year = 2013,
        month = dec,
       volume = {779},
       number = {2},
          eid = {185},
        pages = {185},
          doi = {10.1088/0004-637X/779/2/185},
archivePrefix = {arXiv},
       eprint = {1308.5679},
 primaryClass = {astro-ph.GA},
       adsurl = {https://ui.adsabs.harvard.edu/abs/2013ApJ...779..185K}
}

@ARTICLE{2013A&A...553A.119A,
       author = {{Arzoumanian}, D. and {Andr{\'e}}, Ph. and {Peretto}, N. and {K{\"o}nyves}, V.},
        title = "{Formation and evolution of interstellar filaments. Hints from velocity dispersion measurements}",
      journal = {\aap},
         year = 2013,
        month = may,
       volume = {553},
          eid = {A119},
        pages = {A119},
          doi = {10.1051/0004-6361/201220822},
archivePrefix = {arXiv},
       eprint = {1303.3024},
 primaryClass = {astro-ph.SR},
       adsurl = {https://ui.adsabs.harvard.edu/abs/2013A&A...553A.119A}
}

@ARTICLE{2015A&A...574A.104T,
       author = {{Tafalla}, M. and {Hacar}, A.},
        title = "{Chains of dense cores in the Taurus L1495/B213 complex}",
      journal = {\aap},
         year = 2015,
        month = feb,
       volume = {574},
          eid = {A104},
        pages = {A104},
          doi = {10.1051/0004-6361/201424576},
archivePrefix = {arXiv},
       eprint = {1412.1083},
 primaryClass = {astro-ph.GA},
       adsurl = {https://ui.adsabs.harvard.edu/abs/2015A&A...574A.104T}
}

@ARTICLE{1999ApJ...526..788L,
       author = {{Lee}, Chang Won and {Myers}, Philip C. and {Tafalla}, Mario},
        title = "{A Survey of Infall Motions toward Starless Cores. I. CS (2-1) and N$_{2}$H$^{+}$ (1-0) Observations}",
      journal = {\apj},
         year = 1999,
        month = dec,
       volume = {526},
       number = {2},
        pages = {788-805},
          doi = {10.1086/308027},
archivePrefix = {arXiv},
       eprint = {astro-ph/9906468},
 primaryClass = {astro-ph},
       adsurl = {https://ui.adsabs.harvard.edu/abs/1999ApJ...526..788L}
}

@ARTICLE{2019JKAS...52..227J,
       author = {{Jeong}, Il-Gyo and {Kang}, Hyunwoo and {Jung}, Jaehoon and {Lee}, Changhoon and {Byun}, Do-Young and {Je}, Do-Heung and {Kang}, Sung-Ju and {Lee}, Youngung and {Lee}, Chang Won},
        title = "{Performance of the TRAO 13.7-m Telescope with New Systems}",
      journal = {Journal of Korean Astronomical Society},
         year = 2019,
        month = dec,
       volume = {52},
        pages = {227-233},
       adsurl = {https://ui.adsabs.harvard.edu/abs/2019JKAS...52..227J}
}

@ARTICLE{2022A&A...657L..13P,
       author = {{Panopoulou}, G.~V. and {Clark}, S.~E. and {Hacar}, A. and {Heitsch}, F. and {Kainulainen}, J. and {Ntormousi}, E. and {Seifried}, D. and {Smith}, R.~J.},
        title = "{The width of Herschel filaments varies with distance}",
      journal = {\aap},
         year = 2022,
        month = jan,
       volume = {657},
          eid = {L13},
        pages = {L13},
          doi = {10.1051/0004-6361/202142281},
archivePrefix = {arXiv},
       eprint = {2111.08125},
 primaryClass = {astro-ph.GA},
       adsurl = {https://ui.adsabs.harvard.edu/abs/2022A&A...657L..13P}
}

@ARTICLE{2022A&A...667L...1A,
       author = {{Andr{\'e}}, P.~J. and {Palmeirim}, P. and {Arzoumanian}, D.},
        title = "{The typical width of Herschel filaments}",
      journal = {\aap},
         year = 2022,
        month = nov,
       volume = {667},
          eid = {L1},
        pages = {L1},
          doi = {10.1051/0004-6361/202244541},
archivePrefix = {arXiv},
       eprint = {2210.04736},
 primaryClass = {astro-ph.GA},
       adsurl = {https://ui.adsabs.harvard.edu/abs/2022A&A...667L...1A}
}

@ARTICLE{2018A&A...610A..77H,
       author = {{Hacar}, A. and {Tafalla}, M. and {Forbrich}, J. and {Alves}, J. and {Meingast}, S. and {Grossschedl}, J. and {Teixeira}, P.~S.},
        title = "{An ALMA study of the Orion Integral Filament. I. Evidence for narrow fibers in a massive cloud}",
      journal = {\aap},
         year = 2018,
        month = mar,
       volume = {610},
          eid = {A77},
        pages = {A77},
          doi = {10.1051/0004-6361/201731894},
archivePrefix = {arXiv},
       eprint = {1801.01500},
 primaryClass = {astro-ph.GA},
       adsurl = {https://ui.adsabs.harvard.edu/abs/2018A&A...610A..77H}
}

@ARTICLE{2011ApJ...739L...2P,
       author = {{Pineda}, Jaime E. and {Goodman}, Alyssa A. and {Arce}, H{\'e}ctor G. and {Caselli}, Paola and {Longmore}, Steven and {Corder}, Stuartt},
        title = "{Expanded Very Large Array Observations of the Barnard 5 Star-forming Core: Embedded Filaments Revealed}",
      journal = {\apjl},
         year = 2011,
        month = sep,
       volume = {739},
       number = {1},
          eid = {L2},
        pages = {L2},
          doi = {10.1088/2041-8205/739/1/L2},
archivePrefix = {arXiv},
       eprint = {1106.5474},
 primaryClass = {astro-ph.GA},
       adsurl = {https://ui.adsabs.harvard.edu/abs/2011ApJ...739L...2P}
}

@ARTICLE{2014ApJ...790L..19F,
       author = {{Fern{\'a}ndez-L{\'o}pez}, M. and {Arce}, H.~G. and {Looney}, L. and {Mundy}, L.~G. and {Storm}, S. and {Teuben}, P.~J. and {Lee}, K. and {Segura-Cox}, D. and {Isella}, A. and {Tobin}, J.~J. and {Rosolowsky}, E. and {Plunkett}, A. and {Kwon}, W. and {Kauffmann}, J. and {Ostriker}, E. and {Tassis}, K. and {Shirley}, Y.~L. and {Pound}, M.},
        title = "{CARMA Large Area Star Formation Survey: Observational Analysis of Filaments in the Serpens South Molecular Cloud}",
      journal = {\apjl},
         year = 2014,
        month = aug,
       volume = {790},
       number = {2},
          eid = {L19},
        pages = {L19},
          doi = {10.1088/2041-8205/790/2/L19},
archivePrefix = {arXiv},
       eprint = {1407.0755},
 primaryClass = {astro-ph.GA},
       adsurl = {https://ui.adsabs.harvard.edu/abs/2014ApJ...790L..19F}
}

@ARTICLE{2017MNRAS.464L..31H,
       author = {{Henshaw}, J.~D. and {Jim{\'e}nez-Serra}, I. and {Longmore}, S.~N. and {Caselli}, P. and {Pineda}, J.~E. and {Avison}, A. and {Barnes}, A.~T. and {Tan}, J.~C. and {Fontani}, F.},
        title = "{Unveiling the early-stage anatomy of a protocluster hub with ALMA}",
      journal = {\mnras},
         year = 2017,
        month = jan,
       volume = {464},
       number = {1},
        pages = {L31-L35},
          doi = {10.1093/mnrasl/slw154},
archivePrefix = {arXiv},
       eprint = {1608.00009},
 primaryClass = {astro-ph.GA},
       adsurl = {https://ui.adsabs.harvard.edu/abs/2017MNRAS.464L..31H}
}

@ARTICLE{2010A&A...518L.102A,
       author = {{Andr{\'e}}, Ph. and {Men'shchikov}, A. and {Bontemps}, S. and {K{\"o}nyves}, V. and {Motte}, F. and {Schneider}, N. and {Didelon}, P. and {Minier}, V. and {Saraceno}, P. and {Ward-Thompson}, D. and {di Francesco}, J. and {White}, G. and {Molinari}, S. and {Testi}, L. and {Abergel}, A. and {Griffin}, M. and {Henning}, Th. and {Royer}, P. and {Mer{\'\i}n}, B. and {Vavrek}, R. and {Attard}, M. and {Arzoumanian}, D. and {Wilson}, C.~D. and {Ade}, P. and {Aussel}, H. and {Baluteau}, J. -P. and {Benedettini}, M. and {Bernard}, J. -Ph. and {Blommaert}, J.~A.~D.~L. and {Cambr{\'e}sy}, L. and {Cox}, P. and {di Giorgio}, A. and {Hargrave}, P. and {Hennemann}, M. and {Huang}, M. and {Kirk}, J. and {Krause}, O. and {Launhardt}, R. and {Leeks}, S. and {Le Pennec}, J. and {Li}, J.~Z. and {Martin}, P.~G. and {Maury}, A. and {Olofsson}, G. and {Omont}, A. and {Peretto}, N. and {Pezzuto}, S. and {Prusti}, T. and {Roussel}, H. and {Russeil}, D. and {Sauvage}, M. and {Sibthorpe}, B. and {Sicilia-Aguilar}, A. and {Spinoglio}, L. and {Waelkens}, C. and {Woodcraft}, A. and {Zavagno}, A.},
        title = "{From filamentary clouds to prestellar cores to the stellar IMF: Initial highlights from the Herschel Gould Belt Survey}",
      journal = {\aap},
         year = 2010,
        month = jul,
       volume = {518},
          eid = {L102},
        pages = {L102},
          doi = {10.1051/0004-6361/201014666},
archivePrefix = {arXiv},
       eprint = {1005.2618},
 primaryClass = {astro-ph.GA},
       adsurl = {https://ui.adsabs.harvard.edu/abs/2010A&A...518L.102A}
}

@ARTICLE{2008ApJ...673L.151G,
       author = {{Gutermuth}, R.~A. and {Bourke}, T.~L. and {Allen}, L.~E. and {Myers}, P.~C. and {Megeath}, S.~T. and {Matthews}, B.~C. and {J{\o}rgensen}, J.~K. and {Di Francesco}, J. and {Ward-Thompson}, D. and {Huard}, T.~L. and {Brooke}, T.~Y. and {Dunham}, M.~M. and {Cieza}, L.~A. and {Harvey}, P.~M. and {Chapman}, N.~L.},
        title = "{The Spitzer Gould Belt Survey of Large Nearby Interstellar Clouds: Discovery of a Dense Embedded Cluster in the Serpens-Aquila Rift}",
      journal = {\apjl},
         year = 2008,
        month = feb,
       volume = {673},
       number = {2},
        pages = {L151},
          doi = {10.1086/528710},
archivePrefix = {arXiv},
       eprint = {0712.3303},
 primaryClass = {astro-ph},
       adsurl = {https://ui.adsabs.harvard.edu/abs/2008ApJ...673L.151G}
}

@ARTICLE{2017ApJ...834..143O,
       author = {{Ortiz-Le{\'o}n}, Gisela N. and {Dzib}, Sergio A. and {Kounkel}, Marina A. and {Loinard}, Laurent and {Mioduszewski}, Amy J. and {Rodr{\'\i}guez}, Luis F. and {Torres}, Rosa M. and {Pech}, Gerardo and {Rivera}, Juana L. and {Hartmann}, Lee and {Boden}, Andrew F. and {Evans}, II, Neal J. and {Brice{\~n}o}, Cesar and {Tobin}, John J. and {Galli}, Phillip A.~B.},
        title = "{The Gould{\textquoteright}s Belt Distances Survey (GOBELINS). III. The Distance to the Serpens/Aquila Molecular Complex}",
      journal = {\apj},
         year = 2017,
        month = jan,
       volume = {834},
       number = {2},
          eid = {143},
        pages = {143},
          doi = {10.3847/1538-4357/834/2/143},
archivePrefix = {arXiv},
       eprint = {1610.03128},
 primaryClass = {astro-ph.SR},
       adsurl = {https://ui.adsabs.harvard.edu/abs/2017ApJ...834..143O}
}

@ARTICLE{2018ApJ...869L..33O,
       author = {{Ortiz-Le{\'o}n}, Gisela N. and {Loinard}, Laurent and {Dzib}, Sergio A. and {Kounkel}, Marina and {Galli}, Phillip A.~B. and {Tobin}, John J. and {Evans}, II, Neal J. and {Hartmann}, Lee and {Rodr{\'\i}guez}, Luis F. and {Brice{\~n}o}, Cesar and {Torres}, Rosa M. and {Mioduszewski}, Amy J.},
        title = "{Gaia-DR2 Confirms VLBA Parallaxes in Ophiuchus, Serpens, and Aquila}",
      journal = {\apjl},
         year = 2018,
        month = dec,
       volume = {869},
       number = {2},
          eid = {L33},
        pages = {L33},
          doi = {10.3847/2041-8213/aaf6ad},
archivePrefix = {arXiv},
       eprint = {1812.02360},
 primaryClass = {astro-ph.SR},
       adsurl = {https://ui.adsabs.harvard.edu/abs/2018ApJ...869L..33O}
}

@ARTICLE{2023A&A...673L...1O,
       author = {{Ortiz-Le{\'o}n}, Gisela N. and {Dzib}, Sergio A. and {Loinard}, Laurent and {Gong}, Yan and {Pillai}, Thushara and {Plunkett}, Adele},
        title = "{The distance to the Serpens South cluster from H$_{2}$O masers}",
      journal = {\aap},
         year = 2023,
        month = may,
       volume = {673},
          eid = {L1},
        pages = {L1},
          doi = {10.1051/0004-6361/202346369},
archivePrefix = {arXiv},
       eprint = {2304.07270},
 primaryClass = {astro-ph.GA},
       adsurl = {https://ui.adsabs.harvard.edu/abs/2023A&A...673L...1O}
}

@ARTICLE{2022ApJ...938...55A,
       author = {{Anderson}, Alexa R. and {Williams}, Jonathan P. and {van der Marel}, Nienke and {Law}, Charles J. and {Ricci}, Luca and {Tobin}, John J. and {Tong}, Simin},
        title = "{Protostellar and Protoplanetary Disk Masses in the Serpens Region}",
      journal = {\apj},
         year = 2022,
        month = oct,
       volume = {938},
       number = {1},
          eid = {55},
        pages = {55},
          doi = {10.3847/1538-4357/ac8ff0},
archivePrefix = {arXiv},
       eprint = {2204.08731},
 primaryClass = {astro-ph.SR},
       adsurl = {https://ui.adsabs.harvard.edu/abs/2022ApJ...938...55A}
}

@ARTICLE{2018A&A...615A...9P,
       author = {{Plunkett}, Adele L. and {Fern{\'a}ndez-L{\'o}pez}, Manuel and {Arce}, H{\'e}ctor G. and {Busquet}, Gemma and {Mardones}, Diego and {Dunham}, Michael M.},
        title = "{Distribution of Serpens South protostars revealed with ALMA}",
      journal = {\aap},
         year = 2018,
        month = jul,
       volume = {615},
          eid = {A9},
        pages = {A9},
          doi = {10.1051/0004-6361/201732372},
archivePrefix = {arXiv},
       eprint = {1804.02405},
 primaryClass = {astro-ph.SR},
       adsurl = {https://ui.adsabs.harvard.edu/abs/2018A&A...615A...9P}
}

@ARTICLE{2022MNRAS.516.5244S,
       author = {{Sun}, Jia and {Gutermuth}, Robert A. and {Wang}, Hongchi and {Zhang}, Miaomiao and {Zhang}, Shuinai and {Ma}, Yuehui and {Du}, Xinyu and {Long}, Min},
        title = "{Deep near-infrared survey towards the W40 and Serpens South region in the Aquila Rift: A comprehensive catalogue of young stellar objects}",
      journal = {\mnras},
         year = 2022,
        month = nov,
       volume = {516},
       number = {4},
        pages = {5244-5257},
          doi = {10.1093/mnras/stac2191},
archivePrefix = {arXiv},
       eprint = {2207.09041},
 primaryClass = {astro-ph.SR},
       adsurl = {https://ui.adsabs.harvard.edu/abs/2022MNRAS.516.5244S}
}

@ARTICLE{2014ApJ...791L..23N,
       author = {{Nakamura}, Fumitaka and {Sugitani}, Koji and {Tanaka}, Tomohiro and {Nishitani}, Hiroyuki and {Dobashi}, Kazuhito and {Shimoikura}, Tomomi and {Shimajiri}, Yoshito and {Kawabe}, Ryohei and {Yonekura}, Yoshinori and {Mizuno}, Izumi and {Kimura}, Kimihiko and {Tokuda}, Kazuki and {Kozu}, Minato and {Okada}, Nozomi and {Hasegawa}, Yutaka and {Ogawa}, Hideo and {Kameno}, Seiji and {Shinnaga}, Hiroko and {Momose}, Munetake and {Nakajima}, Taku and {Onishi}, Toshikazu and {Maezawa}, Hiroyuki and {Hirota}, Tomoya and {Takano}, Shuro and {Iono}, Daisuke and {Kuno}, Nario and {Yamamoto}, Satoshi},
        title = "{Cluster Formation Triggered by Filament Collisions in Serpens South}",
      journal = {\apjl},
         year = 2014,
        month = aug,
       volume = {791},
       number = {2},
          eid = {L23},
        pages = {L23},
          doi = {10.1088/2041-8205/791/2/L23},
archivePrefix = {arXiv},
       eprint = {1407.1235},
 primaryClass = {astro-ph.SR},
       adsurl = {https://ui.adsabs.harvard.edu/abs/2014ApJ...791L..23N}
}

@ARTICLE{2009ApJS..184...18G,
       author = {{Gutermuth}, R.~A. and {Megeath}, S.~T. and {Myers}, P.~C. and {Allen}, L.~E. and {Pipher}, J.~L. and {Fazio}, G.~G.},
        title = "{A Spitzer Survey of Young Stellar Clusters Within One Kiloparsec of the Sun: Cluster Core Extraction and Basic Structural Analysis}",
      journal = {\apjs},
         year = 2009,
        month = sep,
       volume = {184},
       number = {1},
        pages = {18-83},
          doi = {10.1088/0067-0049/184/1/18},
archivePrefix = {arXiv},
       eprint = {0906.0201},
 primaryClass = {astro-ph.SR},
       adsurl = {https://ui.adsabs.harvard.edu/abs/2009ApJS..184...18G}
}

@ARTICLE{2019PASJ...71S...4S,
       author = {{Shimoikura}, Tomomi and {Dobashi}, Kazuhito and {Nakamura}, Fumitaka and {Shimajiri}, Yoshito and {Sugitani}, Koji},
        title = "{Cluster formation in the W 40 and Serpens South complex triggered by the expanding H II region}",
      journal = {\pasj},
         year = 2019,
        month = dec,
       volume = {71},
          eid = {S4},
        pages = {S4},
          doi = {10.1093/pasj/psy115},
archivePrefix = {arXiv},
       eprint = {1809.09855},
 primaryClass = {astro-ph.SR},
       adsurl = {https://ui.adsabs.harvard.edu/abs/2019PASJ...71S...4S}
}

@ARTICLE{1959ApJS....4..257S,
       author = {{Sharpless}, Stewart},
        title = "{A Catalogue of H II Regions.}",
      journal = {\apjs},
         year = 1959,
        month = dec,
       volume = {4},
        pages = {257},
          doi = {10.1086/190049},
       adsurl = {https://ui.adsabs.harvard.edu/abs/1959ApJS....4..257S}
}

@ARTICLE{2012AJ....144..116S,
       author = {{Shuping}, R.~Y. and {Vacca}, William D. and {Kassis}, Marc and {Yu}, Ka Chun},
        title = "{Spectral Classification of the Brightest Objects in the Galactic Star-forming Region W40}",
      journal = {\aj},
         year = 2012,
        month = oct,
       volume = {144},
       number = {4},
          eid = {116},
        pages = {116},
          doi = {10.1088/0004-6256/144/4/116},
archivePrefix = {arXiv},
       eprint = {1208.4648},
 primaryClass = {astro-ph.GA},
       adsurl = {https://ui.adsabs.harvard.edu/abs/2012AJ....144..116S}
}

@ARTICLE{1985ApJ...291..571S,
       author = {{Smith}, J. and {Bentley}, A. and {Castelaz}, M. and {Gehrz}, R.~D. and {Grasdalen}, G.~L. and {Hackwell}, J.~A.},
        title = "{Infrared sources and excitation of the W 40 complex.}",
      journal = {\apj},
         year = 1985,
        month = apr,
       volume = {291},
        pages = {571-580},
          doi = {10.1086/163097},
       adsurl = {https://ui.adsabs.harvard.edu/abs/1985ApJ...291..571S}
}

@ARTICLE{2013ApJ...779..113M,
       author = {{Mallick}, K.~K. and {Kumar}, M.~S.~N. and {Ojha}, D.~K. and {Bachiller}, Rafael and {Samal}, M.~R. and {Pirogov}, L.},
        title = "{The W40 Region in the Gould Belt: An Embedded Cluster and H II Region at the Junction of Filaments}",
      journal = {\apj},
         year = 2013,
        month = dec,
       volume = {779},
       number = {2},
          eid = {113},
        pages = {113},
          doi = {10.1088/0004-637X/779/2/113},
archivePrefix = {arXiv},
       eprint = {1309.7127},
 primaryClass = {astro-ph.GA},
       adsurl = {https://ui.adsabs.harvard.edu/abs/2013ApJ...779..113M}
}

@ARTICLE{2013ApJS..209...31P,
       author = {{Povich}, Matthew S. and {Kuhn}, Michael A. and {Getman}, Konstantin V. and {Busk}, Heather A. and {Feigelson}, Eric D. and {Broos}, Patrick S. and {Townsley}, Leisa K. and {King}, Robert R. and {Naylor}, Tim},
        title = "{The MYStIX Infrared-Excess Source Catalog}",
      journal = {\apjs},
         year = 2013,
        month = dec,
       volume = {209},
       number = {2},
          eid = {31},
        pages = {31},
          doi = {10.1088/0067-0049/209/2/31},
archivePrefix = {arXiv},
       eprint = {1309.4497},
 primaryClass = {astro-ph.SR},
       adsurl = {https://ui.adsabs.harvard.edu/abs/2013ApJS..209...31P}
}

@ARTICLE{2017ApJ...837..154N,
       author = {{Nakamura}, Fumitaka and {Dobashi}, Kazuhito and {Shimoikura}, Tomomi and {Tanaka}, Tomohiro and {Onishi}, Toshikazu},
        title = "{Wide-field $^{12}$CO (J=2-1) and $^{13}$CO (J=2-1) Observations toward the Aquila Rift and Serpens Molecular Cloud Complexes. I. Molecular Clouds and Their Physical Properties}",
      journal = {\apj},
         year = 2017,
        month = mar,
       volume = {837},
       number = {2},
          eid = {154},
        pages = {154},
          doi = {10.3847/1538-4357/aa5ea6},
archivePrefix = {arXiv},
       eprint = {1702.01501},
 primaryClass = {astro-ph.GA},
       adsurl = {https://ui.adsabs.harvard.edu/abs/2017ApJ...837..154N}
}

@ARTICLE{2010A&A...518L.100M,
       author = {{Molinari}, S. and {Swinyard}, B. and {Bally}, J. and {Barlow}, M. and {Bernard}, J. -P. and {Martin}, P. and {Moore}, T. and {Noriega-Crespo}, A. and {Plume}, R. and {Testi}, L. and {Zavagno}, A. and {Abergel}, A. and {Ali}, B. and {Anderson}, L. and {Andr{\'e}}, P. and {Baluteau}, J. -P. and {Battersby}, C. and {Beltr{\'a}n}, M.~T. and {Benedettini}, M. and {Billot}, N. and {Blommaert}, J. and {Bontemps}, S. and {Boulanger}, F. and {Brand}, J. and {Brunt}, C. and {Burton}, M. and {Calzoletti}, L. and {Carey}, S. and {Caselli}, P. and {Cesaroni}, R. and {Cernicharo}, J. and {Chakrabarti}, S. and {Chrysostomou}, A. and {Cohen}, M. and {Compiegne}, M. and {de Bernardis}, P. and {de Gasperis}, G. and {di Giorgio}, A.~M. and {Elia}, D. and {Faustini}, F. and {Flagey}, N. and {Fukui}, Y. and {Fuller}, G.~A. and {Ganga}, K. and {Garcia-Lario}, P. and {Glenn}, J. and {Goldsmith}, P.~F. and {Griffin}, M. and {Hoare}, M. and {Huang}, M. and {Ikhenaode}, D. and {Joblin}, C. and {Joncas}, G. and {Juvela}, M. and {Kirk}, J.~M. and {Lagache}, G. and {Li}, J.~Z. and {Lim}, T.~L. and {Lord}, S.~D. and {Marengo}, M. and {Marshall}, D.~J. and {Masi}, S. and {Massi}, F. and {Matsuura}, M. and {Minier}, V. and {Miville-Desch{\^e}nes}, M. -A. and {Montier}, L.~A. and {Morgan}, L. and {Motte}, F. and {Mottram}, J.~C. and {M{\"u}ller}, T.~G. and {Natoli}, P. and {Neves}, J. and {Olmi}, L. and {Paladini}, R. and {Paradis}, D. and {Parsons}, H. and {Peretto}, N. and {Pestalozzi}, M. and {Pezzuto}, S. and {Piacentini}, F. and {Piazzo}, L. and {Polychroni}, D. and {Pomar{\`e}s}, M. and {Popescu}, C.~C. and {Reach}, W.~T. and {Ristorcelli}, I. and {Robitaille}, J. -F. and {Robitaille}, T. and {Rod{\'o}n}, J.~A. and {Roy}, A. and {Royer}, P. and {Russeil}, D. and {Saraceno}, P. and {Sauvage}, M. and {Schilke}, P. and {Schisano}, E. and {Schneider}, N. and {Schuller}, F. and {Schulz}, B. and {Sibthorpe}, B. and {Smith}, H.~A. and {Smith}, M.~D. and {Spinoglio}, L. and {Stamatellos}, D. and {Strafella}, F. and {Stringfellow}, G.~S. and {Sturm}, E. and {Taylor}, R. and {Thompson}, M.~A. and {Traficante}, A. and {Tuffs}, R.~J. and {Umana}, G. and {Valenziano}, L. and {Vavrek}, R. and {Veneziani}, M. and {Viti}, S. and {Waelkens}, C. and {Ward-Thompson}, D. and {White}, G. and {Wilcock}, L.~A. and {Wyrowski}, F. and {Yorke}, H.~W. and {Zhang}, Q.},
        title = "{Clouds, filaments, and protostars: The Herschel Hi-GAL Milky Way}",
      journal = {\aap},
         year = 2010,
        month = jul,
       volume = {518},
          eid = {L100},
        pages = {L100},
          doi = {10.1051/0004-6361/201014659},
archivePrefix = {arXiv},
       eprint = {1005.3317},
 primaryClass = {astro-ph.GA},
       adsurl = {https://ui.adsabs.harvard.edu/abs/2010A&A...518L.100M}
}

@ARTICLE{2013A&A...554A..55H,
       author = {{Hacar}, A. and {Tafalla}, M. and {Kauffmann}, J. and {Kov{\'a}cs}, A.},
        title = "{Cores, filaments, and bundles: hierarchical core formation in the L1495/B213 Taurus region}",
      journal = {\aap},
         year = 2013,
        month = jun,
       volume = {554},
          eid = {A55},
        pages = {A55},
          doi = {10.1051/0004-6361/201220090},
archivePrefix = {arXiv},
       eprint = {1303.2118},
 primaryClass = {astro-ph.GA},
       adsurl = {https://ui.adsabs.harvard.edu/abs/2013A&A...554A..55H}
}

@ARTICLE{2019A&A...621A..42A,
       author = {{Arzoumanian}, D. and {Andr{\'e}}, Ph. and {K{\"o}nyves}, V. and {Palmeirim}, P. and {Roy}, A. and {Schneider}, N. and {Benedettini}, M. and {Didelon}, P. and {Di Francesco}, J. and {Kirk}, J. and {Ladjelate}, B.},
        title = "{Characterizing the properties of nearby molecular filaments observed with Herschel}",
      journal = {\aap},
         year = 2019,
        month = jan,
       volume = {621},
          eid = {A42},
        pages = {A42},
          doi = {10.1051/0004-6361/201832725},
archivePrefix = {arXiv},
       eprint = {1810.00721},
 primaryClass = {astro-ph.GA},
       adsurl = {https://ui.adsabs.harvard.edu/abs/2019A&A...621A..42A}
}

@ARTICLE{2013A&A...550A..38P,
       author = {{Palmeirim}, P. and {Andr{\'e}}, Ph. and {Kirk}, J. and {Ward-Thompson}, D. and {Arzoumanian}, D. and {K{\"o}nyves}, V. and {Didelon}, P. and {Schneider}, N. and {Benedettini}, M. and {Bontemps}, S. and {Di Francesco}, J. and {Elia}, D. and {Griffin}, M. and {Hennemann}, M. and {Hill}, T. and {Martin}, P.~G. and {Men'shchikov}, A. and {Molinari}, S. and {Motte}, F. and {Nguyen Luong}, Q. and {Nutter}, D. and {Peretto}, N. and {Pezzuto}, S. and {Roy}, A. and {Rygl}, K.~L.~J. and {Spinoglio}, L. and {White}, G.~L.},
        title = "{Herschel view of the Taurus B211/3 filament and striations: evidence of filamentary growth?}",
      journal = {\aap},
         year = 2013,
        month = feb,
       volume = {550},
          eid = {A38},
        pages = {A38},
          doi = {10.1051/0004-6361/201220500},
archivePrefix = {arXiv},
       eprint = {1211.6360},
 primaryClass = {astro-ph.SR},
       adsurl = {https://ui.adsabs.harvard.edu/abs/2013A&A...550A..38P}
}

@ARTICLE{2015PASP..127..299S,
       author = {{Shirley}, Yancy L.},
        title = "{The Critical Density and the Effective Excitation Density of Commonly Observed Molecular Dense Gas Tracers}",
      journal = {\pasp},
         year = 2015,
        month = mar,
       volume = {127},
       number = {949},
        pages = {299},
          doi = {10.1086/680342},
archivePrefix = {arXiv},
       eprint = {1501.01629},
 primaryClass = {astro-ph.IM},
       adsurl = {https://ui.adsabs.harvard.edu/abs/2015PASP..127..299S}
}

@ARTICLE{Wienen,
       author = {{Wienen}, M. and {Wyrowski}, F. and {Walmsley}, C.~M. and {Csengeri}, T. and {Pillai}, T. and {Giannetti}, A. and {Menten}, K.~M.},
        title = "{ATLASGAL-selected massive clumps in the inner Galaxy. IX. Deuteration of ammonia}",
      journal = {\aap},
         year = 2021,
        month = may,
       volume = {649},
          eid = {A21},
        pages = {A21},
          doi = {10.1051/0004-6361/201731208},
archivePrefix = {arXiv},
       eprint = {2102.04478},
 primaryClass = {astro-ph.GA},
       adsurl = {https://ui.adsabs.harvard.edu/abs/2021A&A...649A..21W}
}

@ARTICLE{2002ApJ...570L.101B,
       author = {{Bergin}, Edwin A. and {Alves}, Jo{\~a}o and {Huard}, Tracy and {Lada}, Charles J.},
        title = "{N$_{2}$H$^{+}$ and C$^{18}$O Depletion in a Cold Dark Cloud}",
      journal = {\apjl},
         year = 2002,
        month = may,
       volume = {570},
       number = {2},
        pages = {L101-L104},
          doi = {10.1086/340950},
archivePrefix = {arXiv},
       eprint = {astro-ph/0204016},
 primaryClass = {astro-ph},
       adsurl = {https://ui.adsabs.harvard.edu/abs/2002ApJ...570L.101B}
}

@ARTICLE{2007ApJ...657..870V,
       author = {{V{\'a}zquez-Semadeni}, Enrique and {G{\'o}mez}, Gilberto C. and {Jappsen}, A. Katharina and {Ballesteros-Paredes}, Javier and {Gonz{\'a}lez}, Ricardo F. and {Klessen}, Ralf S.},
        title = "{Molecular Cloud Evolution. II. From Cloud Formation to the Early Stages of Star Formation in Decaying Conditions}",
      journal = {\apj},
         year = 2007,
        month = mar,
       volume = {657},
       number = {2},
        pages = {870-883},
          doi = {10.1086/510771},
archivePrefix = {arXiv},
       eprint = {astro-ph/0608375},
 primaryClass = {astro-ph},
       adsurl = {https://ui.adsabs.harvard.edu/abs/2007ApJ...657..870V}
}

@ARTICLE{2011MNRAS.411...65B,
       author = {{Ballesteros-Paredes}, Javier and {Hartmann}, Lee W. and {V{\'a}zquez-Semadeni}, Enrique and {Heitsch}, Fabian and {Zamora-Avil{\'e}s}, Manuel A.},
        title = "{Gravity or turbulence? Velocity dispersion-size relation}",
      journal = {\mnras},
         year = 2011,
        month = feb,
       volume = {411},
       number = {1},
        pages = {65-70},
          doi = {10.1111/j.1365-2966.2010.17657.x},
archivePrefix = {arXiv},
       eprint = {1009.1583},
 primaryClass = {astro-ph.GA},
       adsurl = {https://ui.adsabs.harvard.edu/abs/2011MNRAS.411...65B}
}

@ARTICLE{2022MNRAS.515.2822R,
       author = {{Ram{\'\i}rez-Galeano}, Laura and {Ballesteros-Paredes}, Javier and {Smith}, Rowan J. and {Camacho}, Vianey and {Zamora-Avil{\'e}s}, Manuel},
        title = "{Why most molecular clouds are gravitationally dominated}",
      journal = {\mnras},
         year = 2022,
        month = sep,
       volume = {515},
       number = {2},
        pages = {2822-2836},
          doi = {10.1093/mnras/stac1848},
archivePrefix = {arXiv},
       eprint = {2206.09187},
 primaryClass = {astro-ph.GA},
       adsurl = {https://ui.adsabs.harvard.edu/abs/2022MNRAS.515.2822R}
}

@misc{eHOPS_dataset,
  author       = {Pokhrel, Riwaj and Megeath, S. Thomas and Gutermuth, Robert A. and Furlan, Elise and Fischer, William J. and Federman, Samuel and Tobin, John J. and Stutz, Amelia M. and Hartmann, Lee and Osorio, Mayra and Watson, Dan M. and Stanke, Thomas and Manoj, P. and Narang, Mayank and Atnagulov, Prabhani and Habel, Nolan and Zakri, Wafa},
  title        = {Extension of HOPS Out to 500 ParSecs (eHOPS)},
  year         = {2023},
  publisher    = {IPAC},
  howpublished = {\url{https://doi.org/10.26131/IRSA553}},
  note         = {Dataset, NASA/IPAC Infrared Science Archive (IRSA)},
  doi          = {10.26131/IRSA553}
}

@ARTICLE{2013A&A...558A..33A,
       author = {{Astropy Collaboration} and {Robitaille}, Thomas P. and {Tollerud}, Erik J. and {Greenfield}, Perry and {Droettboom}, Michael and {Bray}, Erik and {Aldcroft}, Tom and {Davis}, Matt and {Ginsburg}, Adam and {Price-Whelan}, Adrian M. and {Kerzendorf}, Wolfgang E. and {Conley}, Alexander and {Crighton}, Neil and {Barbary}, Kyle and {Muna}, Demitri and {Ferguson}, Henry and {Grollier}, Fr{\'e}d{\'e}ric and {Parikh}, Madhura M. and {Nair}, Prasanth H. and {Unther}, Hans M. and {Deil}, Christoph and {Woillez}, Julien and {Conseil}, Simon and {Kramer}, Roban and {Turner}, James E.~H. and {Singer}, Leo and {Fox}, Ryan and {Weaver}, Benjamin A. and {Zabalza}, Victor and {Edwards}, Zachary I. and {Azalee Bostroem}, K. and {Burke}, D.~J. and {Casey}, Andrew R. and {Crawford}, Steven M. and {Dencheva}, Nadia and {Ely}, Justin and {Jenness}, Tim and {Labrie}, Kathleen and {Lim}, Pey Lian and {Pierfederici}, Francesco and {Pontzen}, Andrew and {Ptak}, Andy and {Refsdal}, Brian and {Servillat}, Mathieu and {Streicher}, Ole},
        title = "{Astropy: A community Python package for astronomy}",
      journal = {\aap},
         year = 2013,
        month = oct,
       volume = {558},
          eid = {A33},
        pages = {A33},
          doi = {10.1051/0004-6361/201322068},
archivePrefix = {arXiv},
       eprint = {1307.6212},
 primaryClass = {astro-ph.IM},
       adsurl = {https://ui.adsabs.harvard.edu/abs/2013A&A...558A..33A}
}

@ARTICLE{2018AJ....156..123A,
       author = {{Astropy Collaboration} and {Price-Whelan}, A.~M. and {Sip{\H{o}}cz}, B.~M. and {G{\"u}nther}, H.~M. and {Lim}, P.~L. and {Crawford}, S.~M. and {Conseil}, S. and {Shupe}, D.~L. and {Craig}, M.~W. and {Dencheva}, N. and {Ginsburg}, A. and {VanderPlas}, J.~T. and {Bradley}, L.~D. and {P{\'e}rez-Su{\'a}rez}, D. and {de Val-Borro}, M. and {Aldcroft}, T.~L. and {Cruz}, K.~L. and {Robitaille}, T.~P. and {Tollerud}, E.~J. and {Ardelean}, C. and {Babej}, T. and {Bach}, Y.~P. and {Bachetti}, M. and {Bakanov}, A.~V. and {Bamford}, S.~P. and {Barentsen}, G. and {Barmby}, P. and {Baumbach}, A. and {Berry}, K.~L. and {Biscani}, F. and {Boquien}, M. and {Bostroem}, K.~A. and {Bouma}, L.~G. and {Brammer}, G.~B. and {Bray}, E.~M. and {Breytenbach}, H. and {Buddelmeijer}, H. and {Burke}, D.~J. and {Calderone}, G. and {Cano Rodr{\'\i}guez}, J.~L. and {Cara}, M. and {Cardoso}, J.~V.~M. and {Cheedella}, S. and {Copin}, Y. and {Corrales}, L. and {Crichton}, D. and {D'Avella}, D. and {Deil}, C. and {Depagne}, {\'E}. and {Dietrich}, J.~P. and {Donath}, A. and {Droettboom}, M. and {Earl}, N. and {Erben}, T. and {Fabbro}, S. and {Ferreira}, L.~A. and {Finethy}, T. and {Fox}, R.~T. and {Garrison}, L.~H. and {Gibbons}, S.~L.~J. and {Goldstein}, D.~A. and {Gommers}, R. and {Greco}, J.~P. and {Greenfield}, P. and {Groener}, A.~M. and {Grollier}, F. and {Hagen}, A. and {Hirst}, P. and {Homeier}, D. and {Horton}, A.~J. and {Hosseinzadeh}, G. and {Hu}, L. and {Hunkeler}, J.~S. and {Ivezi{\'c}}, {\v{Z}}. and {Jain}, A. and {Jenness}, T. and {Kanarek}, G. and {Kendrew}, S. and {Kern}, N.~S. and {Kerzendorf}, W.~E. and {Khvalko}, A. and {King}, J. and {Kirkby}, D. and {Kulkarni}, A.~M. and {Kumar}, A. and {Lee}, A. and {Lenz}, D. and {Littlefair}, S.~P. and {Ma}, Z. and {Macleod}, D.~M. and {Mastropietro}, M. and {McCully}, C. and {Montagnac}, S. and {Morris}, B.~M. and {Mueller}, M. and {Mumford}, S.~J. and {Muna}, D. and {Murphy}, N.~A. and {Nelson}, S. and {Nguyen}, G.~H. and {Ninan}, J.~P. and {N{\"o}the}, M. and {Ogaz}, S. and {Oh}, S. and {Parejko}, J.~K. and {Parley}, N. and {Pascual}, S. and {Patil}, R. and {Patil}, A.~A. and {Plunkett}, A.~L. and {Prochaska}, J.~X. and {Rastogi}, T. and {Reddy Janga}, V. and {Sabater}, J. and {Sakurikar}, P. and {Seifert}, M. and {Sherbert}, L.~E. and {Sherwood-Taylor}, H. and {Shih}, A.~Y. and {Sick}, J. and {Silbiger}, M.~T. and {Singanamalla}, S. and {Singer}, L.~P. and {Sladen}, P.~H. and {Sooley}, K.~A. and {Sornarajah}, S. and {Streicher}, O. and {Teuben}, P. and {Thomas}, S.~W. and {Tremblay}, G.~R. and {Turner}, J.~E.~H. and {Terr{\'o}n}, V. and {van Kerkwijk}, M.~H. and {de la Vega}, A. and {Watkins}, L.~L. and {Weaver}, B.~A. and {Whitmore}, J.~B. and {Woillez}, J. and {Zabalza}, V. and {Astropy Contributors}},
        title = "{The Astropy Project: Building an Open-science Project and Status of the v2.0 Core Package}",
      journal = {\aj},
         year = 2018,
        month = sep,
       volume = {156},
       number = {3},
          eid = {123},
        pages = {123},
          doi = {10.3847/1538-3881/aabc4f},
archivePrefix = {arXiv},
       eprint = {1801.02634},
 primaryClass = {astro-ph.IM},
       adsurl = {https://ui.adsabs.harvard.edu/abs/2018AJ....156..123A}
}

@misc{2012ascl.soft08017R,
      author        = {{Robitaille}, T. and {Bressert}, E.},
      title         = "{APLpy: Astronomical Plotting Library in Python}",
      howpublished  = {Astrophysics Source Code Library},
      year          = 2012,
      month         = aug,
      archivePrefix = "ascl",
      eprint        = {1208.017},
      adsurl        = {http://adsabs.harvard.edu/abs/2012ascl.soft08017R}
     }

@INPROCEEDINGS{2014ASPC..485..391C,
       author = {{Currie}, M.~J. and {Berry}, D.~S. and {Jenness}, T. and {Gibb}, A.~G. and {Bell}, G.~S. and {Draper}, P.~W.},
        title = "{Starlink Software in 2013}",
    booktitle = {Astronomical Data Analysis Software and Systems XXIII},
         year = 2014,
       editor = {{Manset}, N. and {Forshay}, P.},
       series = {Astronomical Society of the Pacific Conference Series},
       volume = {485},
        month = may,
        pages = {391},
       adsurl = {https://ui.adsabs.harvard.edu/abs/2014ASPC..485..391C}
}

@ARTICLE{2006JASS...23..269C,
       author = {{Chung}, Eun Jung and {Kim}, Hyoryoung and {Rhee}, Myung-Hyun},
        title = "{Algorithm of Revised-OTFTOOL}",
      journal = {Journal of Astronomy and Space Sciences},
         year = 2006,
        month = sep,
       volume = {23},
       number = {3},
        pages = {269-288},
          doi = {10.5140/JASS.2006.23.3.269},
       adsurl = {https://ui.adsabs.harvard.edu/abs/2006JASS...23..269C}
}

@ARTICLE{2006ApJ...648..461D,
       author = {{Daniel}, F. and {Cernicharo}, J. and {Dubernet}, M. -L.},
        title = "{The Excitation of N$_{2}$H$^{+}$ in Interstellar Molecular Clouds. I. Models}",
      journal = {\apj},
         year = 2006,
        month = sep,
       volume = {648},
       number = {1},
        pages = {461-471},
          doi = {10.1086/505738},
archivePrefix = {arXiv},
       eprint = {astro-ph/0606479},
 primaryClass = {astro-ph},
       adsurl = {https://ui.adsabs.harvard.edu/abs/2006ApJ...648..461D}
}

@ARTICLE{2010A&A...522A..91T,
       author = {{Tafalla}, M. and {Santiago-Garc{\'\i}a}, J. and {Hacar}, A. and {Bachiller}, R.},
        title = "{A molecular survey of outflow gas: velocity-dependent shock chemistry and the peculiar composition of the EHV gas}",
      journal = {\aap},
         year = 2010,
        month = nov,
       volume = {522},
          eid = {A91},
        pages = {A91},
          doi = {10.1051/0004-6361/201015158},
archivePrefix = {arXiv},
       eprint = {1007.4549},
 primaryClass = {astro-ph.GA},
       adsurl = {https://ui.adsabs.harvard.edu/abs/2010A&A...522A..91T}
}

@ARTICLE{2011A&A...533A..34H,
       author = {{Hacar}, A. and {Tafalla}, M.},
        title = "{Dense core formation by fragmentation of velocity-coherent filaments in L1517}",
      journal = {\aap},
         year = 2011,
        month = sep,
       volume = {533},
          eid = {A34},
        pages = {A34},
          doi = {10.1051/0004-6361/201117039},
archivePrefix = {arXiv},
       eprint = {1107.0971},
 primaryClass = {astro-ph.GA},
       adsurl = {https://ui.adsabs.harvard.edu/abs/2011A&A...533A..34H}
}

@ARTICLE{2005A&A...442..949F,
       author = {{Fuller}, G.~A. and {Williams}, S.~J. and {Sridharan}, T.~K.},
        title = "{The circumstellar environment of high mass protostellar objects. III. Evidence of infall?}",
      journal = {\aap},
         year = 2005,
        month = nov,
       volume = {442},
       number = {3},
        pages = {949-959},
          doi = {10.1051/0004-6361:20042110},
archivePrefix = {arXiv},
       eprint = {astro-ph/0508098},
 primaryClass = {astro-ph},
       adsurl = {https://ui.adsabs.harvard.edu/abs/2005A&A...442..949F}
}

@ARTICLE{2009MNRAS.392..170S,
       author = {{Sun}, Yan and {Gao}, Yu},
        title = "{A multitransition molecular line study of inward motions towards massive star-forming cores}",
      journal = {\mnras},
         year = 2009,
        month = jan,
       volume = {392},
       number = {1},
        pages = {170-180},
          doi = {10.1111/j.1365-2966.2008.13992.x},
archivePrefix = {arXiv},
       eprint = {0810.0875},
 primaryClass = {astro-ph},
       adsurl = {https://ui.adsabs.harvard.edu/abs/2009MNRAS.392..170S}
}

@ARTICLE{2017A&A...606A.123H,
       author = {{Hacar}, A. and {Tafalla}, M. and {Alves}, J.},
        title = "{Fibers in the NGC 1333 proto-cluster}",
      journal = {\aap},
         year = 2017,
        month = oct,
       volume = {606},
          eid = {A123},
        pages = {A123},
          doi = {10.1051/0004-6361/201630348},
archivePrefix = {arXiv},
       eprint = {1703.07029},
 primaryClass = {astro-ph.GA},
       adsurl = {https://ui.adsabs.harvard.edu/abs/2017A&A...606A.123H}
}

@ARTICLE{2019A&A...629A..81T,
       author = {{Trevi{\~n}o-Morales}, S.~P. and {Fuente}, A. and {S{\'a}nchez-Monge}, {\'A}. and {Kainulainen}, J. and {Didelon}, P. and {Suri}, S. and {Schneider}, N. and {Ballesteros-Paredes}, J. and {Lee}, Y. -N. and {Hennebelle}, P. and {Pilleri}, P. and {Gonz{\'a}lez-Garc{\'\i}a}, M. and {Kramer}, C. and {Garc{\'\i}a-Burillo}, S. and {Luna}, A. and {Goicoechea}, J.~R. and {Tremblin}, P. and {Geen}, S.},
        title = "{Dynamics of cluster-forming hub-filament systems. The case of the high-mass star-forming complex Monoceros R2}",
      journal = {\aap},
         year = 2019,
        month = sep,
       volume = {629},
          eid = {A81},
        pages = {A81},
          doi = {10.1051/0004-6361/201935260},
archivePrefix = {arXiv},
       eprint = {1907.03524},
 primaryClass = {astro-ph.GA},
       adsurl = {https://ui.adsabs.harvard.edu/abs/2019A&A...629A..81T}
}

@ARTICLE{2022MNRAS.514.6038Z,
       author = {{Zhou}, Jian-Wen and {Liu}, Tie and {Evans}, Neal J. and {Garay}, Guido and {Goldsmith}, Paul F. and {G{\'o}mez}, Gilberto C. and {V{\'a}zquez-Semadeni}, Enrique and {Liu}, Hong-Li and {Stutz}, Amelia M. and {Wang}, Ke and {Juvela}, Mika and {He}, Jinhua and {Li}, Di and {Bronfman}, Leonardo and {Liu}, Xunchuan and {Xu}, Feng-Wei and {Tej}, Anandmayee and {Dewangan}, L.~K. and {Li}, Shanghuo and {Zhang}, Siju and {Zhang}, Chao and {Ren}, Zhiyuan and {Tatematsu}, Ken'ichi and {Shing Li}, Pak and {Won Lee}, Chang and {Baug}, Tapas and {Qin}, Sheng-Li and {Wu}, Yuefang and {Peng}, Yaping and {Zhang}, Yong and {Liu}, Rong and {Luo}, Qiu-Yi and {Ge}, Jixing and {Saha}, Anindya and {Chakali}, Eswaraiah and {Zhang}, Qizhou and {Kim}, Kee-Tae and {Ristorcelli}, Isabelle and {Shen}, Zhi-Qiang and {Li}, Jin-Zeng},
        title = "{ATOMS: ALMA Three-millimeter Observations of Massive Star-forming regions - XI. From inflow to infall in hub-filament systems}",
      journal = {\mnras},
         year = 2022,
        month = aug,
       volume = {514},
       number = {4},
        pages = {6038-6052},
          doi = {10.1093/mnras/stac1735},
archivePrefix = {arXiv},
       eprint = {2206.08505},
 primaryClass = {astro-ph.GA},
       adsurl = {https://ui.adsabs.harvard.edu/abs/2022MNRAS.514.6038Z}
}

@ARTICLE{2022ApJ...940..112K,
       author = {{Kim}, Shinyoung and {Lee}, Chang Won and {Tafalla}, Mario and {Gophinathan}, Maheswar and {Caselli}, Paola and {Myers}, Philip C. and {Chung}, Eun Jung and {Li}, Shanghuo},
        title = "{The Role of Filamentary Structures in the Formation of Two Dense Cores, L1544 and L694-2}",
      journal = {\apj},
         year = 2022,
        month = dec,
       volume = {940},
       number = {2},
          eid = {112},
        pages = {112},
          doi = {10.3847/1538-4357/ac96e0},
archivePrefix = {arXiv},
       eprint = {2209.14943},
 primaryClass = {astro-ph.GA},
       adsurl = {https://ui.adsabs.harvard.edu/abs/2022ApJ...940..112K}
}

@ARTICLE{2025A&A...696A.202S,
       author = {{Sandoval-Garrido}, N.~A. and {Stutz}, A.~M. and {{\'A}lvarez-Guti{\'e}rrez}, R.~H. and {Galv{\'a}n-Madrid}, R. and {Motte}, F. and {Ginsburg}, A. and {Cunningham}, N. and {Reyes-Reyes}, S. and {Redaelli}, E. and {Bonfand}, M. and {Salinas}, J. and {Koley}, A. and {Bernal-Mesina}, G. and {Braine}, J. and {Bronfman}, L. and {Busquet}, G. and {Csengeri}, T. and {Di Francesco}, J. and {Fern{\'a}ndez-L{\'o}pez}, M. and {Garcia}, P. and {Gusdorf}, A. and {Liu}, H. -L. and {Sanhueza}, P.},
        title = "{ALMA-IMF: XVIII. The assembly of a star cluster: Dense N$_{2}$H$^{+}$ (1{\textendash}0) kinematics in the massive G351.77 protocluster}",
      journal = {\aap},
         year = 2025,
        month = apr,
       volume = {696},
          eid = {A202},
        pages = {A202},
          doi = {10.1051/0004-6361/202452589},
archivePrefix = {arXiv},
       eprint = {2410.09843},
 primaryClass = {astro-ph.GA},
       adsurl = {https://ui.adsabs.harvard.edu/abs/2025A&A...696A.202S}
}

@ARTICLE{2005ApJ...620..800D,
       author = {{De Vries}, Christopher H. and {Myers}, Philip C.},
        title = "{Molecular Line Profile Fitting with Analytic Radiative Transfer Models}",
      journal = {\apj},
         year = 2005,
        month = feb,
       volume = {620},
       number = {2},
        pages = {800-815},
          doi = {10.1086/427141},
archivePrefix = {arXiv},
       eprint = {astro-ph/0410748},
 primaryClass = {astro-ph},
       adsurl = {https://ui.adsabs.harvard.edu/abs/2005ApJ...620..800D}
}

@ARTICLE{2013PASP..125..306F,
       author = {{Foreman-Mackey}, Daniel and {Hogg}, David W. and {Lang}, Dustin and {Goodman}, Jonathan},
        title = "{emcee: The MCMC Hammer}",
      journal = {\pasp},
         year = 2013,
        month = mar,
       volume = {125},
       number = {925},
        pages = {306},
          doi = {10.1086/670067},
archivePrefix = {arXiv},
       eprint = {1202.3665},
 primaryClass = {astro-ph.IM},
       adsurl = {https://ui.adsabs.harvard.edu/abs/2013PASP..125..306F}
}
\bibliographystyle{aasjournal}

\end{document}